\documentclass{article}
\usepackage{arxiv}

\usepackage{tikz}
\usetikzlibrary{arrows,automata,backgrounds,calc,fit,positioning,shapes.geometric,shapes.arrows, quotes}
\usepackage[affil-it]{authblk}
\usepackage{subfig}
\usepackage{float}
\usepackage{lineno,hyperref}
\usepackage{color}
\usepackage{adjustbox}
\usepackage[T1]{fontenc}
\usepackage{lmodern}
\usepackage{graphicx}
\usepackage{pdflscape}
\usepackage{verbatim}
\usepackage{amsmath}
\usepackage{amsthm}
\usepackage{amssymb}
\usepackage{multirow}
\usepackage{tabu}
\usepackage{booktabs}
\usepackage[utf8]{inputenc} 
\usepackage{url}            
\usepackage{amsfonts}       
\usepackage{nicefrac}       
\usepackage{microtype}      
\usepackage{lipsum}
\usepackage{fourier} 
\usepackage[inline, shortlabels]{enumitem}

\newtheorem{theorem}{Theorem}
\newtheorem{lemma}{Lemma}

\title{NeuroDAVIS: A neural network model for data visualization}

\author[1]{Chayan Maitra}
\author[2]{Dibyendu B. Seal}
\author[1,*]{Rajat K. De}

\affil[1]{Machine Intelligence Unit, Indian Statistical Institute, 203 Barrackpore Trunk Road, Kolkata 700108, India.}
\affil[2]{Tatras Data Services Pvt. Ltd.,E64, Vasant Marg, Vasant Vihar, New Delhi - 110057, India.}
\affil[*]{Corresponding author: Rajat K. De, rajat@isical.ac.in}

\begin{document}
	
	\maketitle	
	\begin{abstract} The task of dimensionality reduction and visualization of high-dimensional datasets remains a challenging problem since long. Modern high-throughput technologies produce newer high-dimensional datasets having multiple views with relatively new data types. Visualization of these datasets require proper methodology that can uncover hidden patterns in the data without affecting the local and global structures, and bring out the inherent non-linearity within the data. To this end, however, very few such methodology exist, which can realise this task. In this work, we have introduced a novel unsupervised deep neural network model, called NeuroDAVIS, for data visualization. NeuroDAVIS is capable of extracting important features from the data, without assuming any data distribution, and visualize effectively in lower dimension. It has been shown theoritically that neighbourhood relationship of the data in high dimension remains preserved in lower dimension. The performance of NeuroDAVIS has been evaluated on a wide variety of synthetic and real high-dimensional datasets including numeric, textual, image and biological data. NeuroDAVIS has been highly competitive against both t-Distributed Stochastic Neighbor Embedding (t-SNE) and Uniform Manifold Approximation and Projection (UMAP) with respect to visualization quality, and preservation of data size, shape, and both local and global structure. It has outperformed Fast interpolation-based t-SNE (Fit-SNE), a variant of t-SNE, for most of the high-dimensional datasets as well. For the biological datasets, besides t-SNE, UMAP and Fit-SNE, NeuroDAVIS has also performed well compared to other state-of-the-art algorithms, like Potential of Heat-diffusion for Affinity-based Trajectory Embedding (PHATE) and the siamese neural network-based method, called IVIS. Downstream classification and clustering analyses have also revealed favourable results for NeuroDAVIS-generated embeddings.
	\end{abstract}

	\keywords{Deep learning, Unsupervised learning, Shape preservation, Global structure preservation, Single-cell transcriptomics}

\section{Introduction}
\label{sec:intro}
Machine learning-based analyses of large real-world datasets are underpinned by dimensionality reduction (DR) methods which form the basis for preprocessing and visualization of these datasets. Some commonly used DR techniques include Principal Component Analysis (PCA) \cite{PCA}, Independent Component Analysis (ICA) \cite{ICA}, Multi-dimensional Scaling (MDS) \cite{MDS}, Isomap \cite{tenenbaum2000global}, NMF \cite{NMF}, SVD \cite{SVD}, t-Distributed Stochastic Neighbor Embedding (t-SNE) \cite{van2008visualizing} and Uniform Manifold Approximation and Projection (UMAP) \cite{mcinnes2018umap}. PCA projects the data into a newer space spanned by the vectors representing the maximum variance, while ICA extracts subcomponents from a multivariate signal. MDS is a DR and visualization technique that can extract dissimilarity within structures in the data. Isomap is a non-linear DR method that combines the advantages of several methods. Both NMF and SVD are methods for matrix factorization that have significant usage in topic of modeling and signal processing. t-SNE and UMAP, though being techniques for DR, are mostly suited for visualization tasks. All the above methods fall into the category of unsupervised DR. Linear Discriminant Analysis (LDA) \cite{LDA}, on the contrary, is a supervised method used for DR and pattern recognition. 

Any DR technique should virtue preservation of both local and global structures of the data. However, we are more accustomed to see methods preferring one over the other. PCA or MDS tend to preserve pairwise distances among all observations, while t-SNE, Isomap and UMAP tend to preserve local distances over global distances. There is limited study on local and global shape preservation for the other DR techniques mentioned above.

Non-linear projection algorithms like t-SNE, however, have an edge over linear algorithms like PCA in extracting complex latent structures within the data \cite{ding2018interpretable}. Nevertheless, there are severe downsides of t-SNE. t-SNE is highly sensitive to noise, and even randomly distributed points may be transformed into spurious clusters \cite{wattenberg2016how}. Moreover, t-SNE is unpopular for preserving local distances over global distances \cite{schubert2017intrinsic, amid2018more}, which hinders drawing realistic conclusions from t-SNE visualizations. Fast interpolation-based t-SNE (Fit-SNE) \cite{fitSNE} is a variant of t-SNE, which provides accelerated performance on large datasets. However, tuning the parameters to obtain an optimal embedding requires high amount of expertise. UMAP, on the other hand, uses a manifold learning technique to reduce data dimension and scales to large datasets pretty well. However, UMAP assumes that there exists a manifold structure within the data which is not always realistic. It also gives precedence to local relationships over global relationships \cite{mcinnes2018umap}, like t-SNE. Potential of Heat-diffusion for Affinity-based Trajectory Embedding (PHATE) \cite{moon2019visualizing} is another manifold learning-based method which has found significant usage in the field of biology in recent days. However, the effectiveness of PHATE on other real-world datasets are yet to be assessed. 

Neural network (NN)-based methods have also been in use as non-linear DR tools in almost every field, more so during the last decade \cite{AE_Hinton_OriginalPaper, lin2017using, Wang_AE}. Unsupervised NNs are trained to learn a non-linear function, while features extracted from an intermediate hidden layer with relatively low cardinality serve as a low-dimensional representation of the data. The earliest usage of NN methods for projection can be mapped back to Self-Organizing Map (SOM), also known as Kohonen map \cite{kohonen1990self}. SOM is an unsupervised method that projects data into lower dimension while preserving topological structures. In recent years, Autoencoder (AE)-based methods predominantly rule the DR space \cite{AE2, AE3, AE4} where multiple variations of AE have been developed to address the problem of `Curse of Dimensionality' in several application domains including computer vision and computational biology. Most recently, a Siamese neural network architecture with triplet loss function, called IVIS \cite{szubert2019structure}, has been developed for data visualization of high-dimensional biological datasets. All these NN-based methods have shown high potentiality with respect to DR and visualization tasks. However, not all methods can simulataneously preserve local and global structures of the data.

Fascinated by the rich potentiality of NN models to capture data non-linearity, we have introduced, in this work, a novel unsupervised deep learning model, called NeuroDAVIS, which serves the purpose of data visualization while addressing the issue of both local and global structure preservation. There are several major contributions of this work. NeuroDAVIS is a general purpose NN model that can be used for dimension reduction and visualization of high-dimensional datasets. It can extract a meaningful embedding from the data, which captures significant features that are indicative of the inherent data non-linearity. NeuroDAVIS-extracted features can be used for data reconstruction as well. Moreover, NeuroDAVIS is free from assumptions about the data distribution. The performance of NeuroDAVIS has been evaluated on a wide variety of $2$D synthetic, and real high-dimensional datasets including numeric, textual, image and biological data. Despite all the limitations of t-SNE and UMAP discussed above, they are still the current state-of-the-arts in the universe of dimension reduction or visualization methods. NeuroDAVIS has been competitive against both t-SNE and UMAP, and arguably preserves data shape, size, and local and global relationships better than both of them. It has been proved mathematically that the corresponding embeddings of local neighbours in high dimension remain local at low dimension too. NeuroDAVIS is applicable to all kinds (modalities) of datasets. Furthermore, it is able to produce impressive and interpretable visualizations independently, i.e., without any kind of preprocessing. 

The remaining part of this article is organized into the following sections. Section \ref{sec:methodology} discusses the motivation behind NeuroDAVIS and develops its architecture, followed by the proof of correctness (Section \ref{sec:proof}) where mathematical justifications for some properties of NeuroDAVIS have been provided. Section \ref{sec:Results} describes the experimental results, along with comparisons on $2$D datasets (Section \ref{sec:results_2D}), their embedding, visualization and structure preservation (Section \ref{sec:results_emb_vis_pres}), global structure preservation (Section \ref{sec:results_global}), inter-cluster distance preservation (Section \ref{sec:results_interClusterdist}), cluster-size preservation (Section \ref{sec:results_cluster_size}), and finally results on high-dimensional datasets (Section \ref{sec:results_hd}) including numeric (Section \ref{sec:results_numeric}), textual (Section \ref{sec:results_textual}), image (Section \ref{sec:results_image}) and biological datasets (Section \ref{sec:results_biological}). Section \ref{sec:conc} discusses the advantages and disadvantages of NeuroDAVIS, and concludes the article.

\section{Methodology}
\label{sec:methodology}
This section develops the proposed unsupervised neural network model, called NeuroDAVIS, for visualization of high-dimensional datasets. NeuroDAVIS is a non-recurrent, feed-forward neural network, which is capable of extracting a low-dimensional embedding that can capture relevant features and provide efficient visualization of high-dimensional datasets. It can preserve both local and global relationships between clusters within the data. The motivation behind the proposed model is described below, followed by its architecture.

\subsection{Motivation}

Preservation of local and global shapes are the two main components of a visualization  challenge. Several approaches have been put forth so far, but none of them can suitably address both these issues. The majority of them try to preserve local shape rather than global shape, while some techniques accomplish the other. Thus, developing a method that visualises data at a lower dimension while successfully preserving both local and global distances, is our key goal.

The fundamental mathematical concept behind NeuroDAVIS is inspired by the well-studied regression problem. In a regression problem, a set of independent variables (regressors) is used to predict one or more dependent variables. Unlike regression, a visualization problem has only one data. In order to visualize the data in hand, a set of random regressors can be learnt in an unsupervised manner. This relaxation may make the regression task equivalent to a visualization task.

\begin{figure}[]
	\centering
	\includegraphics[width=\columnwidth]{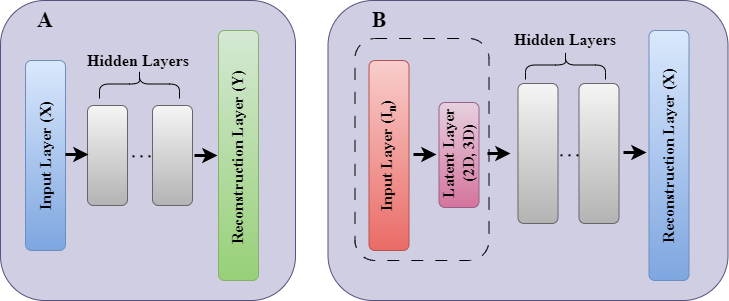}
	\caption{Block diagrams of (A) Multi-layer neural network and (B) NeuroDAVIS. The dashed part in (B) represents the modified architecture that enables the regressors to get updated.}
	\label{fig:Arch}
\end{figure}

\subsubsection*{Visualization through regression}
Let $\mathbf{X}_1 = \{\mathbf{x}_{1,i}: \mathbf{x}_{1,i} \in \mathbb{R}^{d_1}\}^{n}_{i=1}$ and $\mathbf{X}_2 = \{\mathbf{x}_{2,i}: \mathbf{x}_{2,i} \in \mathbb{R}^{d_2}\}^{n}_{i=1}$ be two datasets having $n$ samples characterized by $d_1$ and $d_2$ features respectively. The task of regressesion is to realize a continuous function $f$ from $\mathbb{R}^{d_1}$ to $\mathbb{R}^{d_2}$ such that the reconstruction loss (usually $\sum_{i=1}^n \|f(\mathbf{x}_{1,i}) - \mathbf{x}_{2,i}\|^2$) gets minimized. For this purpose, one can use a multi-layer neural network model. To ensure that the input data fit into the model appropriately, the input layer should have $d_1$ number of neurons. The input will then be processed through a number of nonlinear activation layers before being output at the output layer, which has $d_2$ number of neurons. A reconstruction loss is then determined using the expected output, and the weights and biases are updated using conventional back-propagation learning. The crucial fact about this kind of learning is that it does not support updation of the regressors. As a result, the multi-layer neural network learns a sufficiently complex function which is able to produce close predictions.

In the proposed methodology, random 2D or 3D data are prepared initially to regress the original data. Now, there are multiple objectives with some relaxations. Throughout the learning process, not only an appropriate continuous function that can effectively produce the data, has to be learnt but the regressors too. Continuity property of the learned function and its less complexity ensure that local neighbours in high-dimension are preserved on projection into low dimension (See Theorem \ref{theo:1} in Section \ref{sec:proof}).

As shown in the block diagram of NeuroDAVIS (Figure \ref{fig:Arch}B), the dashed part is used to control the regressors. Input to NeuroDAVIS is an identity matrix of size $n \times n$, where $n$ is the number of samples. During the forward pass, each column of identity matrix is fed to the input layer, thus ensuring that only certain weight values (initialized randomly) are fed as input to the latent layer. These random points generated using the first two layers are used as regressors. Similar to a multi-layer neural network, NeuroDAVIS also tries to reconstruct the data at the reconstruction layer. A reconstruction loss is then calculated and the corresponding weights and biases are updated using standard back-propagation algorithm. Thus, in every epoch, as the weight values in the first layer get updated, the regressors get modified too, making them equally distant as their corrresponding predictions. The detailed learning process has been explained in Section \ref{sec:learning}.

\subsection{Architecture}
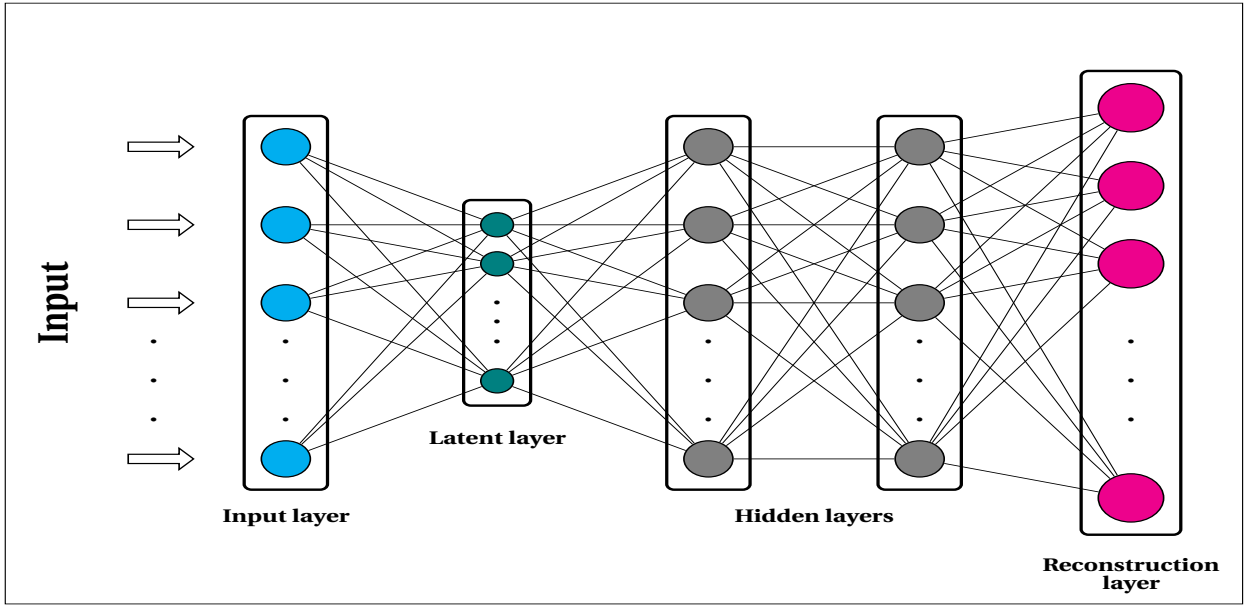
\begin{figure}
	\centering
	\resizebox{\columnwidth}{8cm}{
		\begin{tikzpicture}[framed]
			
			
			\node[draw,circle,fill=cyan,line width=0.1em,minimum size=3em] (1) at (4,4) {};
			\node[draw,circle,fill=cyan,line width=0.1em,minimum size=3em] (2) at (4,2) {};
			\node[draw,circle,fill=cyan,line width=0.1em,minimum size=3em] (3) at (4,0) {};
			
			\node (4) at (4,-3) {\textbf{$\bullet$}};
			\node (4) at (4,-2) {\textbf{$\bullet$}};
			\node (4) at (4,-1) {\textbf{$\bullet$}};
			
			\node[draw,circle,fill=cyan,line width=0.1em,minimum size=3em] (5) at (4,-4) {};
			
			\node[rectangle,rounded corners,draw=black,fit=(1) (2) (3) (4) (5),inner sep=1em,line width=0.2em]
			(6)	{};

			
			\node[draw,circle,fill=teal,line width=0.1em,minimum size=2em] (7) at (8,2) {};
			\node[draw,circle,fill=teal,line width=0.1em,minimum size=2em] (8) at (8,1) {};
			\node[draw,circle,fill=teal,line width=0.1em,minimum size=2em] (9) at (8,-2) {};

			\node (10) at (8,0) {\textbf{$\bullet$}};
			\node (10) at (8,-0.5) {\textbf{$\bullet$}};
			\node (10) at (8,-1) {\textbf{$\bullet$}};
			
			\node[rectangle,rounded corners,draw=black,fit=(7) (8) (9) (10),inner sep=1em,line width=0.2em]
			(11)	{};
			
			
			
			\node[draw,circle,fill=gray,line width=0.1em,minimum size=3em] (12) at (12,4) {};
			\node[draw,circle,fill=gray,line width=0.1em,minimum size=3em] (13) at (12,2) {};
			\node[draw,circle,fill=gray,line width=0.1em,minimum size=3em] (14) at (12,0) {};
			
			\node (15) at (12,-3) {\textbf{$\bullet$}};
			\node (15) at (12,-2) {\textbf{$\bullet$}};
			\node (15) at (12,-1) {\textbf{$\bullet$}};
			
			\node[draw,circle,fill=gray,line width=0.1em,minimum size=3em] (16) at (12,-4) {};
			
			\node[rectangle,rounded corners,draw=black,fit=(12) (16),inner sep=1em,line width=0.2em]
			(17)	{};

			\node[draw,circle,fill=gray,line width=0.1em,minimum size=3em] (18) at (16,4) {};
			\node[draw,circle,fill=gray,line width=0.1em,minimum size=3em] (19) at (16,2) {};
			\node[draw,circle,fill=gray,line width=0.1em,minimum size=3em] (20) at (16,0) {};
			
			\node (21) at (16,-3) {\textbf{$\bullet$}};
			\node (21) at (16,-2) {\textbf{$\bullet$}};
			\node (21) at (16,-1) {\textbf{$\bullet$}};
			
			\node[draw,circle,fill=gray,line width=0.1em,minimum size=3em] (22) at (16,-4) {};
			
			\node[rectangle,rounded corners,draw=black,fit=(18) (19) (20) (21) (22),inner sep=1em,line width=0.2em]
			(23)	{};
			

			\node[draw,circle,fill=magenta,line width=0.1em,minimum size=4em] (24) at (20,5) {};
			\node[draw,circle,fill=magenta,line width=0.1em,minimum size=4em] (25) at (20,3) {};
			\node[draw,circle,fill=magenta,line width=0.1em,minimum size=4em] (26) at (20,1) {};
			
			\node (27) at (20,-3) {\textbf{$\bullet$}};
			\node (27) at (20,-2) {\textbf{$\bullet$}};
			\node (27) at (20,-1) {\textbf{$\bullet$}};
			
			\node[draw,circle,fill=magenta,line width=0.1em,minimum size=4em] (28) at (20,-5) {};
			
			\node[rectangle,rounded corners,draw=black,fit=(24) (25) (26) (27) (28),inner sep=1em,line width=0.2em]
			(29)	{};
			

			
			\node[draw,line width=0.1em,single arrow,minimum height=4em,minimum width=0.5em,single arrow head extend=0.5em,	anchor=west] (33) at (1,4) {};
			
			\node[draw,line width=0.1em,single arrow,minimum height=4em,minimum width=0.5em,single arrow head extend=0.5em,	anchor=west] (34) at (1,2) {};
			
			\node[draw,line width=0.1em,single arrow,minimum height=4em,minimum width=0.5em,single arrow head extend=0.5em,	anchor=west] (35) at (1,0) {};
			
			\node[draw,line width=0.1em,single arrow,minimum height=4em,minimum width=0.5em,single arrow head extend=0.5em,	anchor=west] (36) at (1,-4) {};
			
			\node (37) at (1.5,-3) {\textbf{$\bullet$}};
			\node (37) at (1.5,-2) {\textbf{$\bullet$}};
			\node (37) at (1.5,-1) {\textbf{$\bullet$}};

			
			
			
			
			
			
			\node[text width=20em,align=center]
			(38) at (4,-5.5) {\Large\textbf{Input layer}};
			
			\node[text width=48em,align=center,text height=3em,rotate=90]
			(39) at (-0.5
			,0) {\Huge\textbf{Input}};
			
			\node[text width=20em,align=center]
			(40) at (8,-3.5) {\Large \textbf{Latent layer}};
			
			\node[text width=40em,align=center]
			(41) at (14,-5.5) {\Large\textbf{Hidden layers}};
			
			\node[text width=12em,align=center]
			(42) at (20,-7) {\Large\textbf{Reconstruction\\[0.2em]layer}};
			
			
			\draw (1) to (7);
			\draw (1) to (8);
			\draw (1) to (9);
			
			\draw (2) to (7);
			\draw (2) to (8);
			\draw (2) to (9);
			
			\draw (3) to (7);
			\draw (3) to (8);
			\draw (3) to (9);
			
			\draw (5) to (7);
			\draw (5) to (8);
			\draw (5) to (9);
			
			\draw (7) to (12);
			\draw (7) to (13);
			\draw (7) to (14);
			\draw (7) to (16);
			
			\draw (8) to (12);
			\draw (8) to (13);
			\draw (8) to (14);
			\draw (8) to (16);
			
			\draw (9) to (12);
			\draw (9) to (13);
			\draw (9) to (14);
			\draw (9) to (16);
			
			\draw (12) to (18);
			\draw (12) to (19);
			\draw (12) to (20);
			\draw (12) to (22);
			
			\draw (13) to (18);
			\draw (13) to (19);
			\draw (13) to (20);
			\draw (13) to (22);
			
			\draw (14) to (18);
			\draw (14) to (19);
			\draw (14) to (20);
			\draw (14) to (22);
			
			\draw (16) to (18);
			\draw (16) to (19);
			\draw (16) to (20);
			\draw (16) to (22);
			
			\draw (18) to (24);
			\draw (18) to (25);
			\draw (18) to (26);
			\draw (18) to (28);
			
			\draw (19) to (24);
			\draw (19) to (25);
			\draw (19) to (26);
			\draw (19) to (28);
			
			\draw (20) to (24);
			\draw (20) to (25);
			\draw (20) to (26);
			\draw (20) to (28);

			\draw (22) to (24);
			\draw (22) to (25);
			\draw (22) to (26);
			\draw (22) to (28);
			%
	\end{tikzpicture}}
	\caption{NeuroDAVIS Architecture.}
	\label{fig:NeuroDAVIS_arch}
\end{figure}

The architecture of NeuroDAVIS represents a novel neural network model as shown in Figure \ref{fig:NeuroDAVIS_arch}. The NeuroDAVIS network consists of different types of layers, viz., an \textit{Input layer}, a \textit{Latent layer}, one or more \textit{Hidden layer(s)} and a \textit{Reconstruction layer}. Let $\mathbf{X} = \{ \mathbf{x}_i: \mathbf{x}_i \in \mathbb{R}^d \}_{i=1}^{n}$ be a dataset comprising $n$ samples characterized by $d$ features. Then, the \textit{Input layer} in NeuroDAVIS consists of $n$ neurons. The number of neurons in the \textit{Latent layer} is $k$, where $k$ represents the number of dimensions to be used for visualization. In other words, $k$ is the number of dimensions at which the low-dimensional embedding is to be extracted. Usually, in real-life applications, a $2$-dimensional or $3$-dimensional visualization is possible. The \textit{Input layer} helps to create a random low-dimensional embedding of $n$ samples (observations) at the \textit{Latent layer}, which can be considered as an initial representative of the $n$ original observations. The low-dimensional embedding at the \textit{Latent layer} is projected onto the \textit{Reconstruction layer} through one or more \textit{Hidden layer(s)}. The \textit{Reconstruction layer} tries to reconstruct the original $d$ dimensional space for the $n$ samples from their random low-dimensional embedding. The number of neurons in the \textit{Reconstruction layer} is thus $d$. In this work, we have used only two hidden layers. A \textit{Hidden layer} tries to capture the non-linearity in the data and pass on the knowledge to the next layer in sequence. Thus, one can use multiple \textit{Hidden layers} based on the complexity of the data. The number of \textit{Hidden layers}, however, should not be large enough in order to avoid overfitting.


\subsection{Forward propagation}
As mentioned before, we have considered a set of $n$ samples $\mathbf{X} = \{ \mathbf{x}_i: \mathbf{x}_i \in \mathbb{R}^d \}_{i=1}^{n}$. The input to NeuroDAVIS is an identity matrix $\mathbf{I}$ of order $n \times n$. For an $i^{th}$ sample $\mathbf{x}_i$, we consider the $i^{th}$ column vector $\mathbf{e}_{i}$ of $\mathbf{I}$. That is, on presenting $\mathbf{e}_{i}$ to the \textit{Input layer}, an approximate version $\mathbf{\tilde{x}}_{i}$ of $\mathbf{x}_i$ is reconstructed at the \textit{Reconstruction layer}. Let $\mathbf{a}_{ji}$ and $\mathbf{h}_{ji}$ correspond to the input to and output from a $j^{th}$ layer on presentation of an $i^{th}$ sample; $\mathbf{W}_{j}$ be the weight matrix between ${(j-1)}^{th}$ layer and $j^{th}$ layer ($j = 1, 2, \cdots, (l+2)$); and $\mathbf{b}_{j}$ be the bias term for nodes in $j^{th}$ layer. A $j^{th}$ layer may be any of the \textit{Input layer} ($j=0$), \textit{Latent layer} ($j=1$), \textit{Hidden layer(s)} ($j=2, 3, \dots, (l+1)$) or \textit{Reconstruction layer} ($j=(l+2)$). Thus, for the \textit{Input layer}, 

\begin{equation}
	\begin{cases}
		\mathbf{a}_{0i} = \mathbf{e}_{i}, \\
		\mathbf{h}_{0i} = \mathbf{e}_{i}, 
		\hspace{1cm} \forall i=1,2,\cdots, n 
	\end{cases}
\end{equation}
For the \textit{Latent layer}, we have 

\begin{equation}
	\begin{cases}
		\mathbf{a}_{1i} = \mathbf{W}_{1}\mathbf{e}_{i} + \mathbf{b}_{1},  \\ 
		\mathbf{h}_{1i} = \mathbf{a}_{1i}, 
		\hspace{1cm} \forall i=1,2,\cdots, n 
	\end{cases}
\end{equation}
Here, $\mathbf{e}_i$ controls the weight parameters for the low-dimensional embedding of $i^{th}$ sample obtained at the \textit{Latent layer}. It ensures that only the links connected to the $i^{th}$ neuron of the \textit{Input layer} will activate neurons in \textit{Latent layer} on presentation of $i^{th}$ sample. For $l$ \textit{Hidden layer(s)}, we have 

\begin{equation}
	\begin{cases}
		\mathbf{a}_{ji} = \mathbf{W}_{j}\mathbf{h}_{(j-1)i} + \mathbf{b}_{j}, \\ 
		\mathbf{h}_{ji} = ReLU(\mathbf{a}_{ji}), 
		\hspace{1cm} \forall j=2,3, \cdots, (l+1), \forall i=1,2, \cdots, n
	\end{cases}
\end{equation}
where $ReLU(\mathbf{y}) = max(\mathbf{0},\mathbf{y})$; \textit{max} (maximum) being considered element wise.

A reconstruction of the original data is performed at the final layer, called \textit{Reconstruction layer}. For the \textit{Reconstruction layer}, we have

\begin{equation}
	\begin{cases}
		\mathbf{a}_{(l+2)i} = \mathbf{W}_{l+2}\mathbf{h}_{(l+1)i} + \mathbf{b}_{l+2}, \\
		\mathbf{h}_{(l+2)i} = \mathbf{a}_{(l+2)i}, 
		\hspace{1cm} \forall i=1,2, \cdots, n
	\end{cases}
\end{equation}
Thus, NeuroDAVIS projects the latent embedding for a sample obtained at the \textit{Latent layer} into a $d$ dimensional space through the \textit{Hidden layer(s)}.

\subsection{Learning}
\label{sec:learning}
As stated before, NeuroDAVIS has been used for dimensionality reduction and visualization of high dimensional data. For each sample $\mathbf{x}$, NeuroDAVIS tries to find an optimal reconstruction $\mathbf{\tilde{x}}$ of the input data by minimizing the reconstruction error $\|\mathbf{x}-\mathbf{\tilde{x}}\|$. $L2$ regularization has been used on the nodes' activities and edge weights to avoid overfitting. Usage of $L2$ regularization ensures minimization of model complexity. It also helps in dragging output and weight values towards zero. Regularization of the nodes' activities and weights have been controlled using regularization parameters $\alpha$ and $\beta$ respectively. The objective function thus becomes
\begin{equation}
	\label{eqn:obj_func}
	\mathcal{L}_{NeuroDAVIS} = \frac{1}{n}\sum_{i=1}^{n}\| \mathbf{x}_{i} - \mathbf{\tilde{x}}_{i}\|^2 + \alpha\sum_{j=1}^{l+1}\sum_{i=1}^{n} \|\mathbf{h}_{ji}\|_{2} + \beta\sum_{j=1}^{l+1} \|\mathbf{W}_{j}\|_{F}
\end{equation}
NeuroDAVIS has been trained using Adam optimizer \cite{kingma2014adam}. Values of $\alpha$ and $\beta$ have been set experimentally for each dataset. Additionally, the number of epochs at which $\mathcal{L}_{NeuroDAVIS}$ saturates, has been observed carefully to fix the optimal number of epochs needed for convergence.

At the onset of the forward pass, the weight values between the \textit{Input layer} and \textit{Latent layer} are initialized randomly. The learning of the NeuroDAVIS network is controlled by an identity matrix $\mathbf{I}$ fed as input to the \textit{Input layer}. The usage of the identity matrix ensures updation of the weight values solely associated with the samples in the present batch. Thus, initially, a random low-dimensional embedding of $n$ samples (observations) is created at the \textit{Latent layer}. This low-dimensional representation at the \textit{Latent layer} is then projected on to the \textit{Reconstruction layer} by NeuroDAVIS. At the \textit{Reconstruction layer}, the error value is calculated using Equation \ref{eqn:obj_func}. This error is then propagated backwards, and the weight and bias values are updated for a better reconstruction of the samples of the present batch, in the next forward pass. This process is continued till convergence. On completion of the training phase, the transformed feature set extracted from the \textit{Latent layer} serves as the low-dimensional embedding of the data, and used for visualization for $k=2$ or $3$.

\section{Proof of correctness}
\label{sec:proof}

In this section, we have established mathematically that neighbours in high dimension remain preserved after projection into low dimension.

\begin{lemma}
	\label{lem:1}
	For any real matrix $\mathbf{W}$ having Frobenius norm less than or equal to $1$ and $\eta \leq 1$, $\| \mathbf{I} - \eta \mathbf{W} \mathbf{W}^T \|_2 \leq 1 $. 
\end{lemma}

\begin{proof}
	It is well known that $$\| \mathbf{I} - \eta \mathbf{W} \mathbf{W}^T \|_2^2 = \lambda_{max} (\mathbf{I} - \eta \mathbf{W} \mathbf{W}^T )^2,  $$ where $\lambda_{max}(A)$ represents the largest eigen value of $A$. 
	
	\noindent Now, $\mathbf{W} \mathbf{W}^T$ is a positive semi-definite matrix. Hence, all eigen values of $\mathbf{W} \mathbf{W}^T$ are non-negative.
	Also, $$\lambda_{max} (\mathbf{W} \mathbf{W}^T) \leq \| \mathbf{W} \|_F \leq 1$$ \\
	Therefore, all eigen values of $\mathbf{W} \mathbf{W}^T$ lie in $[0,1]$ and $ \eta \leq 1$, which implies $\lambda_{max} (\mathbf{I} - \eta \mathbf{W} \mathbf{W}^T )^2 \leq 1$ 
\end{proof}

\begin{theorem}
\label{theo:1}
	Let $\mathbf{x}_i$ and $\mathbf{x}_j$ be two points in $d$-dimensional space $(\mathbf{x}_i, \mathbf{x}_j \in \mathbb{R}^d)$ such that they belong to a $\delta$-ball, i.e., $\|\mathbf{x}_i - \mathbf{x}_j\| < \delta$, a predefined small positive number. Their corresponding low-dimensional embedding, generated by NeuroDAVIS, will come closer in each iteration during training if the weight matrix from the final layer has a Frobenius norm less than $1$.
\end{theorem}

\begin{proof}
	Let us consider a simple NeuroDAVIS model with no hidden layer and no regularization. Let $\mathbf{y}_i$ and $\mathbf{y}_j$ be the corresponding initial low-dimensional embeddings of $\mathbf{x}_i$ and $\mathbf{x}_j$. Then $$ \mathbf{y}_i = \mathbf{W}_{1} \mathbf{e}_i + \mathbf{b}_{1} = \mathbf{W}_{1,i} + \mathbf{b}_{1}, $$ $\mathbf{e}_i$ being the $i^{th}$ column of $\mathbf{I}$.
	During each forward pass, NeuroDAVIS tries to reconstruct the original data point. Let $\mathbf{\tilde x}_i$ be a reconstruction of $\mathbf{x}_i$, i.e., $ \mathbf{\tilde x}_i = (\mathbf{W}_{1,i} + \mathbf{b}_{1}) \mathbf{W}_{2} + \mathbf{b}_{2}$.  \\
	Therefore, the loss function can be written as $$ \mathcal{L}_{NeuroDAVIS} = \frac{1}{2n} \sum_{i=1}^n \|\mathbf{x}_i - \mathbf{\tilde x}_i\|^2,$$
	The corresponding weight and bias values are updated using $ \mathbf{W}^{(t+1)} = \mathbf{W}^{(t)} - \eta \nabla \mathbf{W}^{(t)} $.\\
	Now, 
	\begin{align*}
		\frac{\partial \mathcal{L}}{\partial \mathbf{W}_{1,i}} &= (\mathbf{x}_i - \mathbf{\tilde x}_i)(- \frac{\partial \mathbf{\tilde x}_i}{\partial \mathbf{W}_{1,i}})
		= - (\mathbf{x}_i - \mathbf{\tilde x}_i) \frac{\partial }{\partial \mathbf{W}_{1,i}}( \mathbf{W}_{1,i} \mathbf{W}_{2}) = -(\mathbf{x}_i - \mathbf{\tilde x}_i) \mathbf{W}_{2}^{T}
	\end{align*}
	At $t^{th}$ iteration, the weight updation occurs as 
	$ \mathbf{W}_{1,i}^{(t+1)} = \mathbf{W}_{1,i}^{(t)} + \eta (\mathbf{x}_i - \mathbf{\tilde x}_i) \mathbf{W}_{2}^{(t)T}$\\
	Thus, 
	\begin{align*}
		\mathbf{y}_i^{(t+1)} - \mathbf{y}_j^{(t+1)}  &= \mathbf{W}_{1,i}^{(t+1)} - \mathbf{W}_{1,j}^{(t+1)} \\
		&= [ \mathbf{W}_{1,i}^{(t)} + \eta (\mathbf{x}_i - \mathbf{\tilde x}_i) \mathbf{W}_{2}^{(t)T} ] - [ \mathbf{W}_{1,j}^{(t)} + \eta (\mathbf{x}_j - \mathbf{\tilde x}_j) \mathbf{W}_{2}^{(t)T} ] \\
		&= [\mathbf{W}_{1,i}^{(t)} - \mathbf{W}_{1,j}^{(t)}] +\eta [(\mathbf{x}_i - \mathbf{x}_j)-(\mathbf{\tilde x}_i - \mathbf{\tilde x}_j)] \mathbf{W}_{2}^{(t)T} \\
		&\approx [\mathbf{W}_{1,i}^{(t)} - \mathbf{W}_{1,j}^{(t)}] -\eta(\mathbf{\tilde x}_i - \mathbf{\tilde x}_j) \mathbf{W}_{2}^{(t)T} \hspace{1.6cm}[\text{as } \|\mathbf{x}_i - \mathbf{x}_j\| < \delta]  \\
		&\approx [\mathbf{W}_{1,i}^{(t)} - \mathbf{W}_{1,j}^{(t)}] -\eta[\mathbf{W}_{1,i}^{(t)} - \mathbf{W}_{1,j}^{(t)}]\mathbf{W}_{2}^{(t)} \mathbf{W}_{2}^{(t)T} \\
		&\approx [\mathbf{W}_{1,i}^{(t)} - \mathbf{W}_{1,j}^{(t)}][\mathbf{I} - \eta \mathbf{W}_{2}^{(t)} \mathbf{W}_{2}^{(t)T}] \\
		\| \mathbf{y}_i^{(t+1)} - \mathbf{y}_j^{(t+1)} \| &\approx \| [\mathbf{W}_{1,i}^{(t)} - \mathbf{W}_{1,j}^{(t)}][\mathbf{I} - \eta \mathbf{W}_{2}^{(t)} \mathbf{W}_{2}^{(t)T}] \| \\
		&\leq \| \mathbf{W}_{1,i}^{(t)} - \mathbf{W}_{1,j}^{(t)} \| \| \mathbf{I} - \eta \mathbf{W}_{2}^{(t)} \mathbf{W}_{2}^{(t)T} \| \\
		&\leq \| \mathbf{W}_{1,i}^{(t)} - \mathbf{W}_{1,j}^{(t)} \| \hspace{2cm}[\text{By~Lemma~\ref{lem:1}}]\\
	\end{align*}
Thus,
$$ \| \mathbf{y}_i^{(t+1)} - \mathbf{y}_j^{(t+1)} \| \leq \| \mathbf{y}_i^{(t)} - \mathbf{y}_j^{(t)} \| $$
\end{proof}

\section{Results}
\label{sec:Results}
The performance of NeuroDAVIS has been evaluated on a wide variety of datasets including $2$D synthetic datasets and high-dimensional (HD) datasets of different modalities, like numeric, text, image and biological data (single-cell RNA-sequencing (scRNA-seq) data). The datasets used for evaluation have been described in Table \ref{tab:datasets}. This section explains the experiments carried out in this work to evaluate the performance of NeuroDAVIS, and analyze the results. Here we have focussed on the capability of NeuroDAVIS in low dimensional embedding, visualization, structure preservation, inter-cluster distance preservation and cluster size preservation.

\begin{table}[]
	\caption{Description of datasets used for evaluation of NeuroDAVIS}
	\centering
	\begin{tabular}{||cc||c||c||c||c||c||}
		\hline
		\multicolumn{2}{||c||}{\textbf{Category}}                                          & \textbf{Name} & \textbf{\#Samples} & \textbf{\#Features} & \textbf{\#Classes} & \textbf{Source}                                     \\ \hline
		\multicolumn{2}{||c||}{\multirow{5}{*}{2D}}                                        & EllipticRing  & 1100               & 2                   & 3                  & \multirow{5}{*}{Synthetic}                          \\ \cline{3-6}
		\multicolumn{2}{||c||}{}                                                           & Olympic       & 2500               & 2                   & 5                  &                                                     \\ \cline{3-6}
		\multicolumn{2}{||c||}{}                                                           & Spiral        & 312                & 2                   & 3                  &                                                     \\ \cline{3-6}
		\multicolumn{2}{||c||}{}                                                           & Shape         & 2000               & 2                   & 5                  &                                                     \\ \cline{3-6}
		\multicolumn{2}{||c||}{}                                                           & World Map     & 2843               & 2                   & 5                  &                                                     \\ \hline
		\multicolumn{1}{||c||}{\multirow{7}{*}{HD}} & \multirow{2}{*}{Numeric}             & Breast cancer & 569                & 30                  & 2                  & \cite{street1993nuclear}           \\ \cline{3-7} 
		\multicolumn{1}{||c||}{}                    &                                      & Wine          & 178                & 13                  & 3                  & \cite{aeberhard1992classification} \\ \cline{2-7} 
		\multicolumn{1}{||c||}{}                    & Text                                 & Spam          & 5572               & 513                 & 2                  &      \cite{almeida2011contributions}     \\ \cline{2-7} 
		\multicolumn{1}{||c||}{}                    & \multirow{2}{*}{Image}               & Coil20        & 1440               & 16385               & 20                 &  \cite{coil20}                                                   \\ \cline{3-7} 
		\multicolumn{1}{||c||}{}                    &                                      & Fashion MNIST & 60000              & 784                 & 10                 &    \cite{fmnist}                                                 \\ \cline{2-7} 
		\multicolumn{1}{||c||}{}                    & \multirow{2}{*}{scRNA-seq} & Usoskin       & 622                & 25334               & 13 &                 \cite{usoskin2015unbiased}         \\ \cline{3-7} 
		\multicolumn{1}{||c||}{}                    &                                      & Jurkat        & 3388               & 32738               & 11                 & \cite{zheng2017massively}          \\ \hline
	\end{tabular}
	\label{tab:datasets}
\end{table}

\subsection{Performance evaluation on synthetic $2$D datasets}
\label{sec:results_2D}
In order to demonstrate the effectiveness of NeuroDAVIS, we have initially applied NeuroDAVIS on four sythetic $2$D datasets, viz., Elliptic Ring, Olympic, Spiral and Shape. The results have been compared with that obtained using t-SNE and UMAP. 

\subsubsection{Embedding, visualization and structure preservation}
\label{sec:results_emb_vis_pres}
\begin{figure}
	\subfloat[Original]{
		\centering
		\includegraphics[width=0.25\columnwidth,height=4cm]{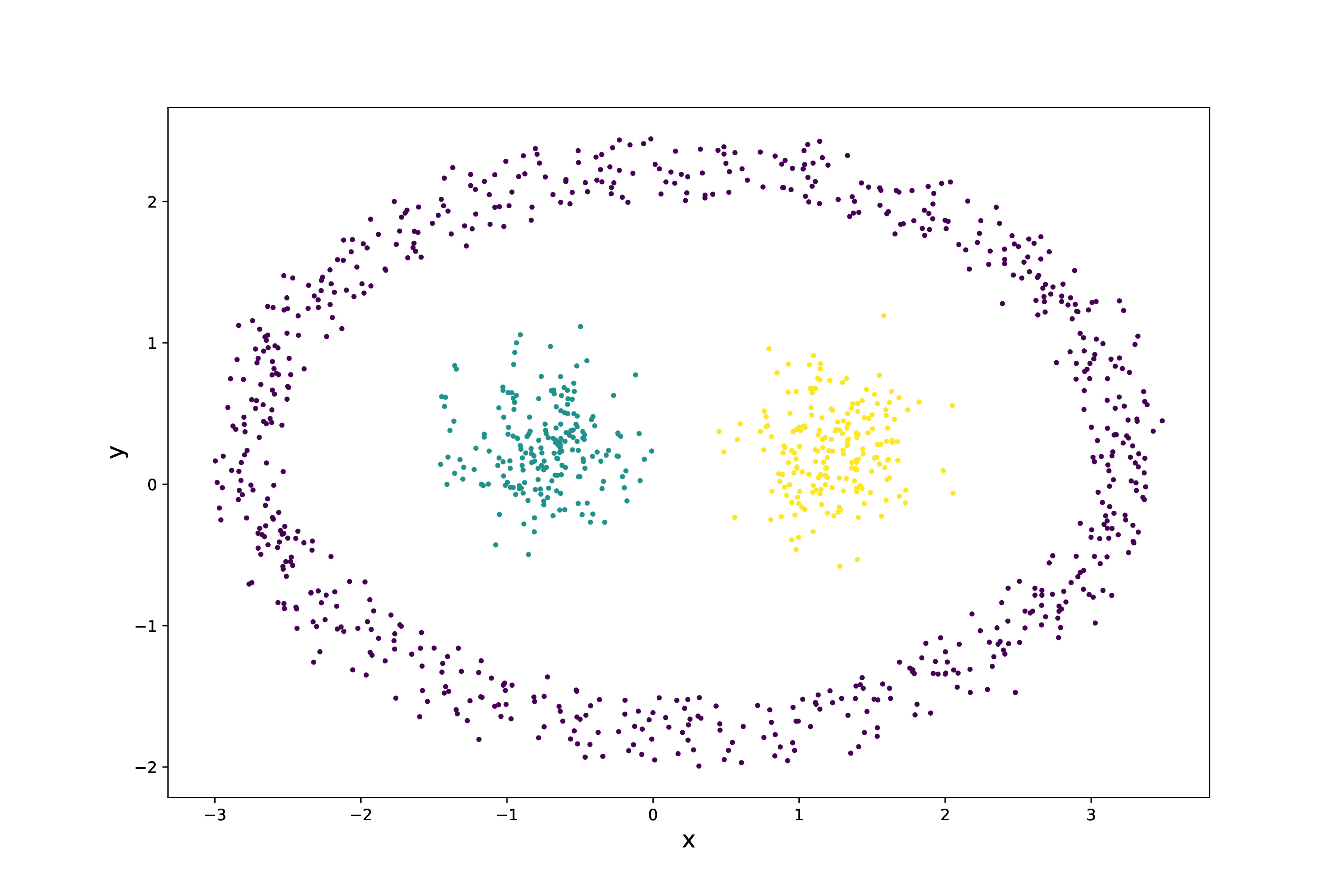}
		\label{fig:2D_EllipticRing}
	} 
	\subfloat[NeuroDAVIS]{
		\centering
		\includegraphics[width=0.25\columnwidth,height=4cm]{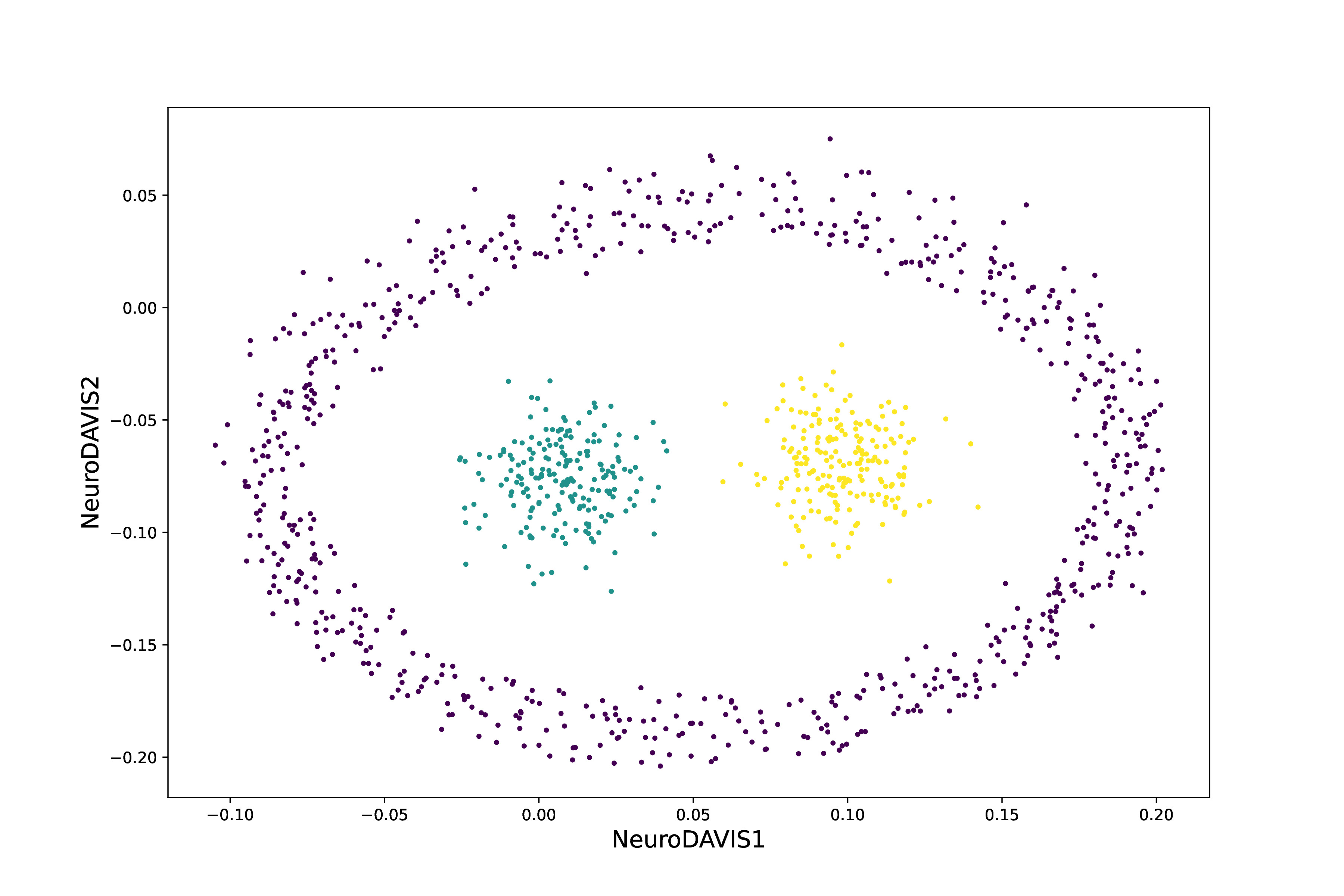}
		\label{fig:2D_EllipticRing_NeuroDAVIS}
	}
	\subfloat[t-SNE]{
		\centering
		\includegraphics[width=0.25\columnwidth,height=4cm]{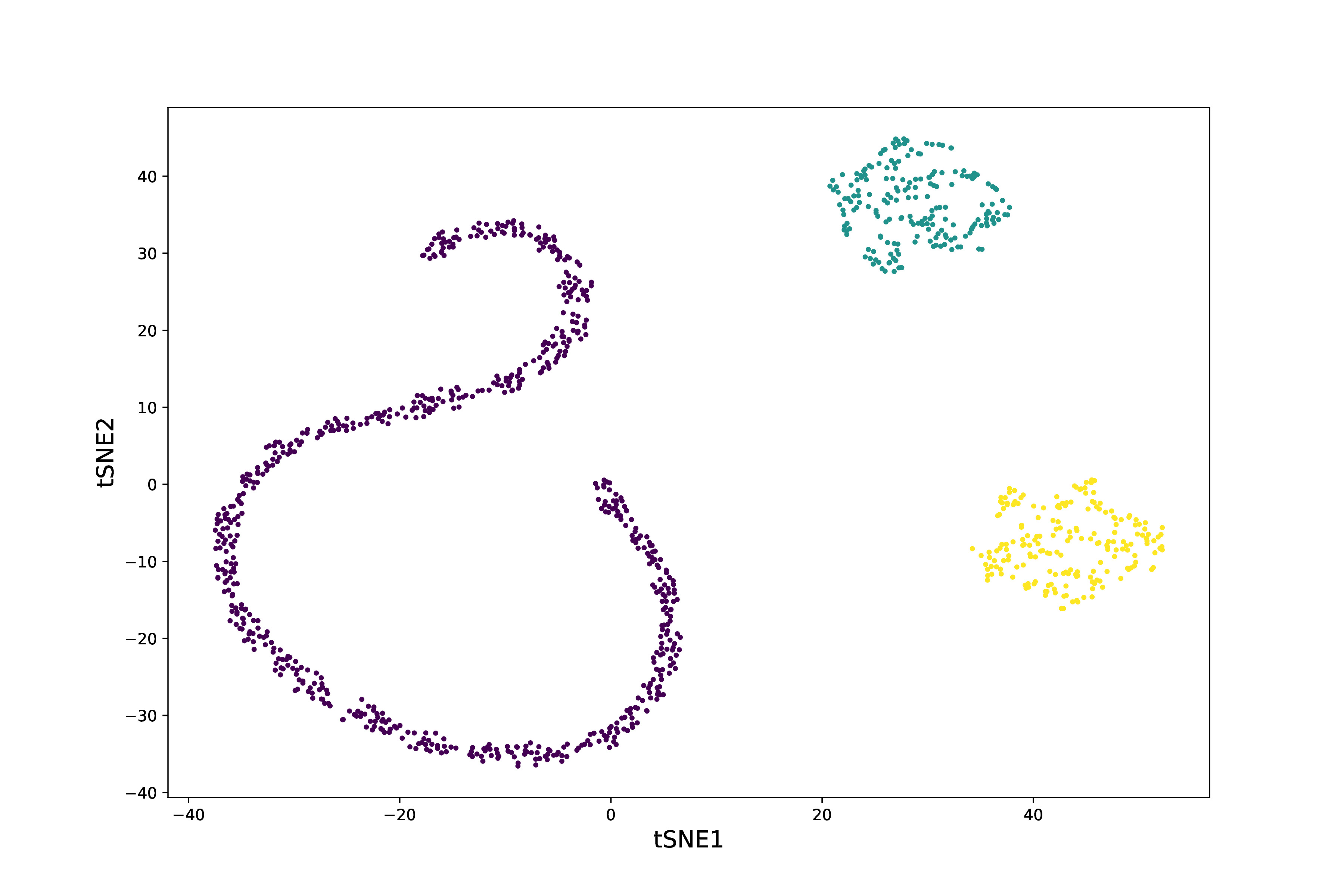}
		\label{fig:2D_EllipticRing_tSNE}
	} 
	\subfloat[UMAP]{
		\centering
		\includegraphics[width=0.25\columnwidth,height=4cm]{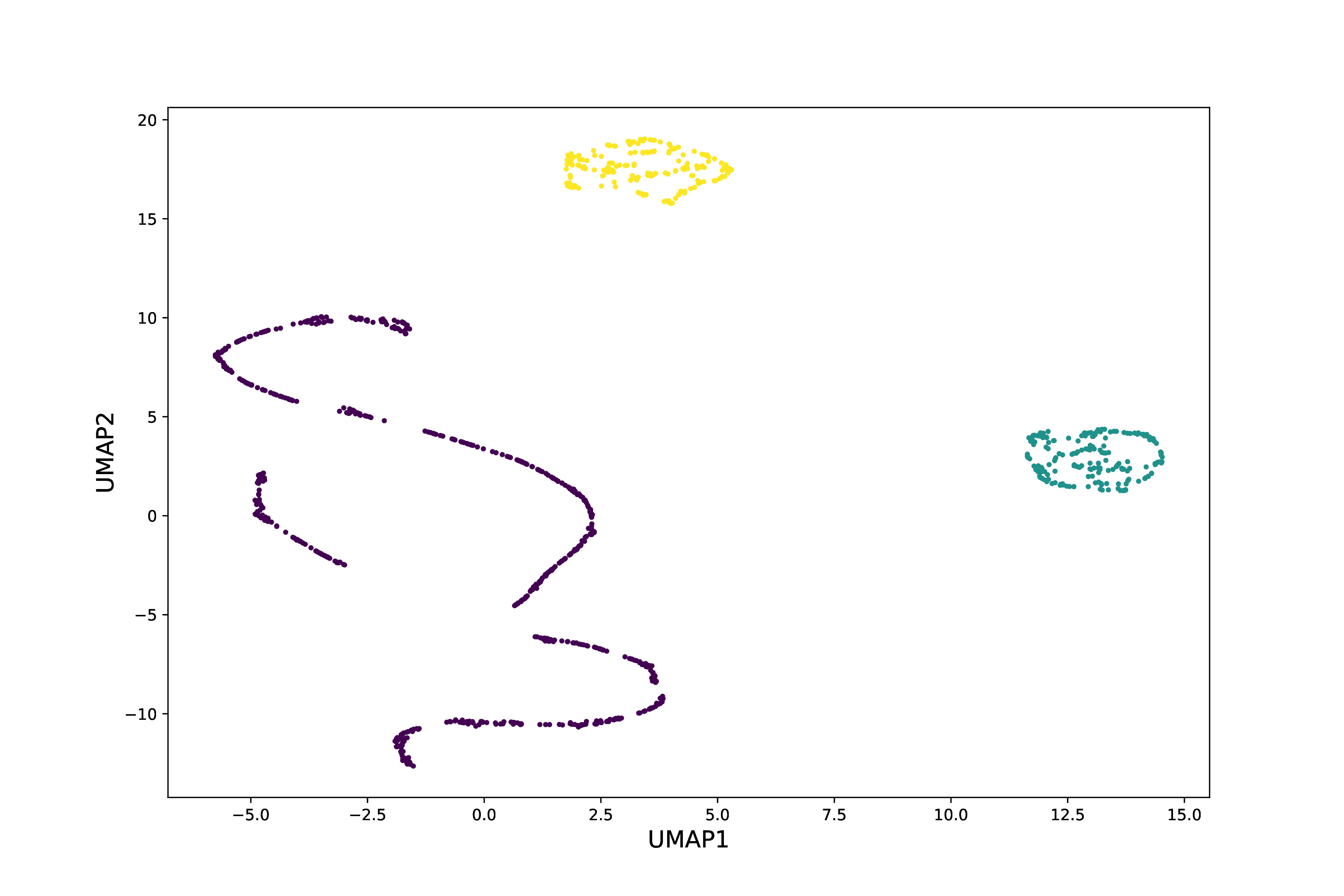}
		\label{fig:2D_EllipticRing_UMAP}
	}
	\\
	\subfloat[Original]{
		\centering
		\includegraphics[width=0.25\columnwidth,height=4cm]{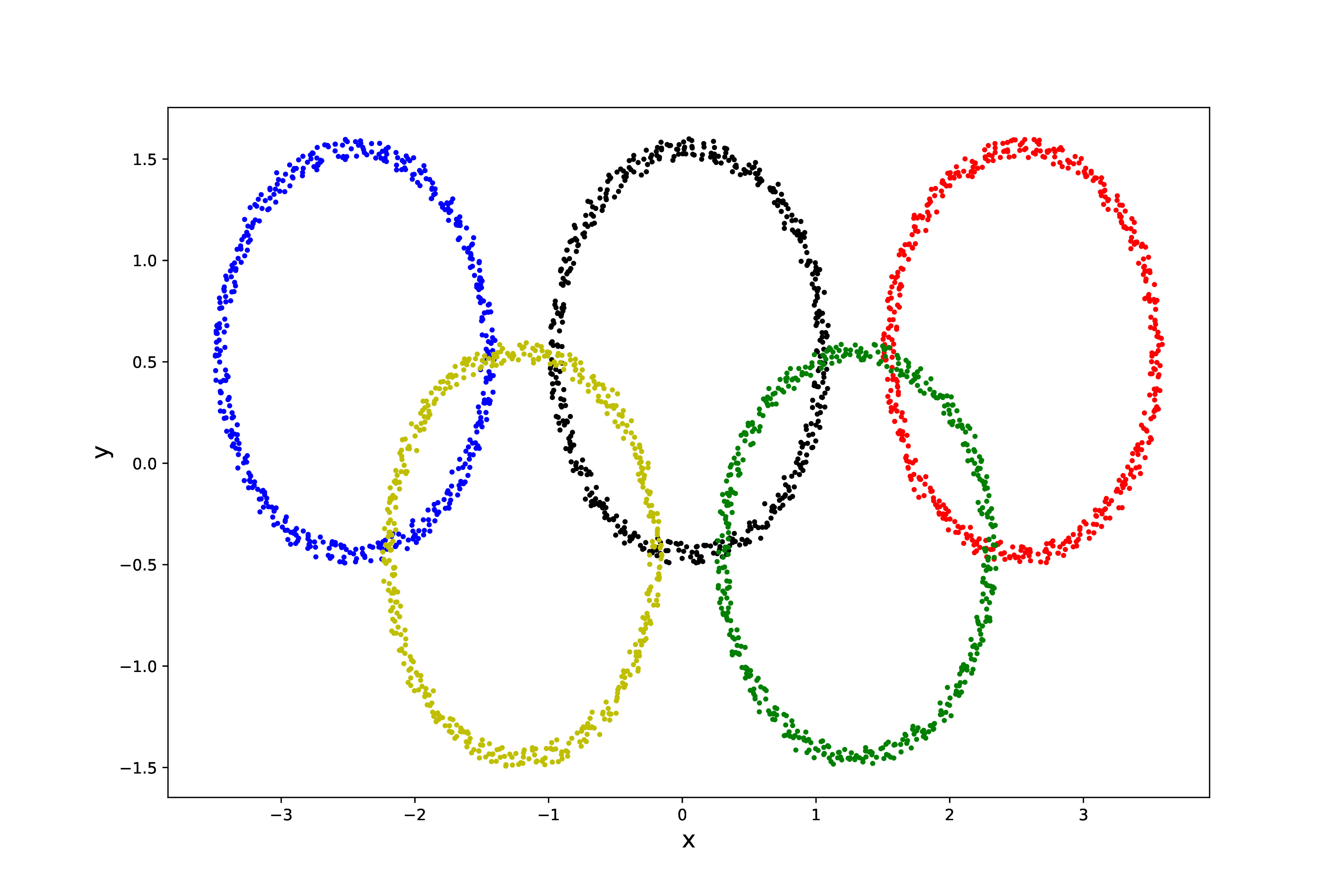}
		\label{fig:2D_Olympic}
	}
	\subfloat[NeuroDAVIS]{
		\centering
		\includegraphics[width=0.25\columnwidth,height=4cm]{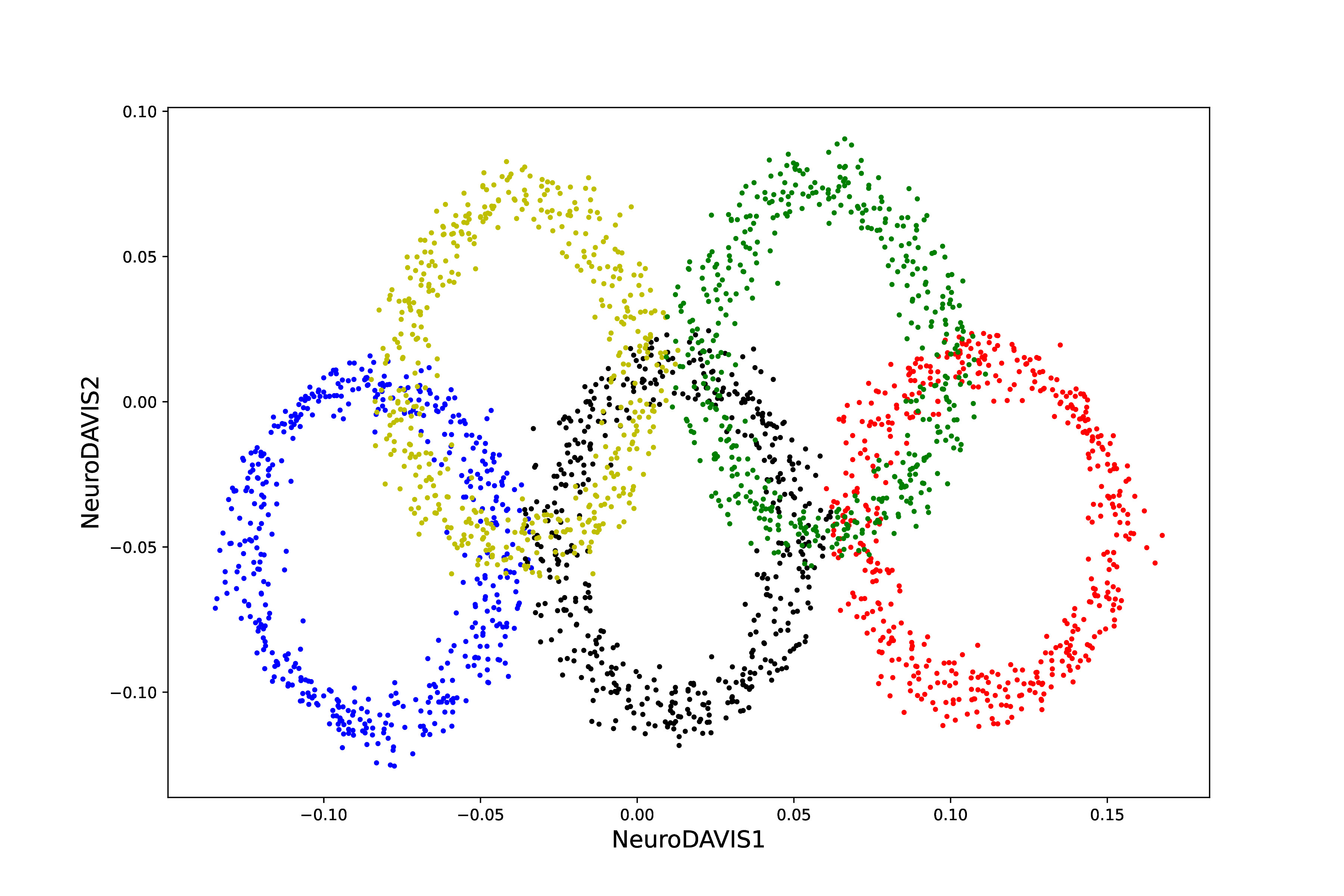}
		\label{fig:2D_Olympic_NeuroDAVIS}
	}
	\subfloat[t-SNE]{
		\centering
		\includegraphics[width=0.25\columnwidth,height=4cm]{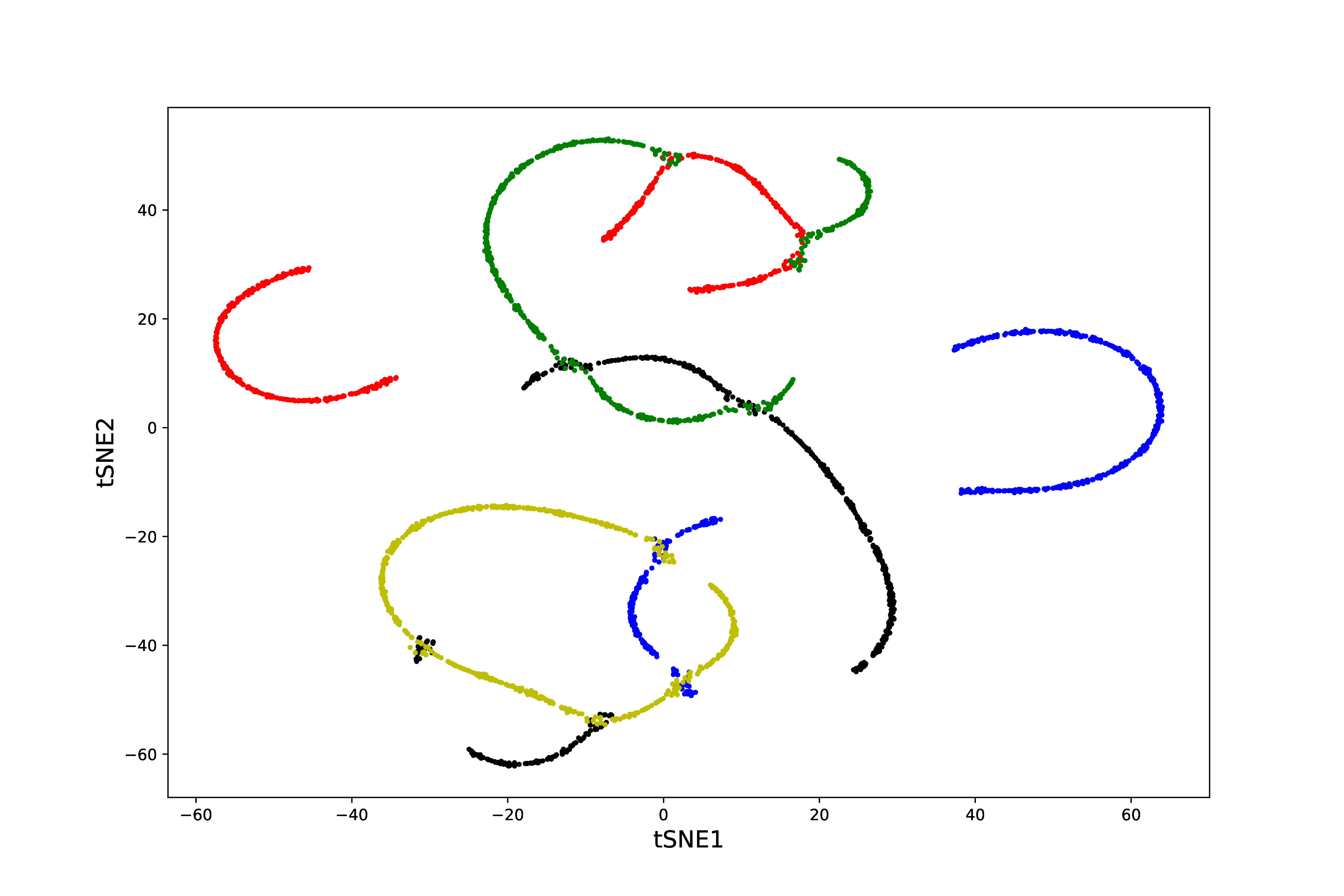}
		\label{fig:2D_Olympic_tSNE}
	}
	\subfloat[UMAP]{
		\centering
		\includegraphics[width=0.25\columnwidth,height=4cm]{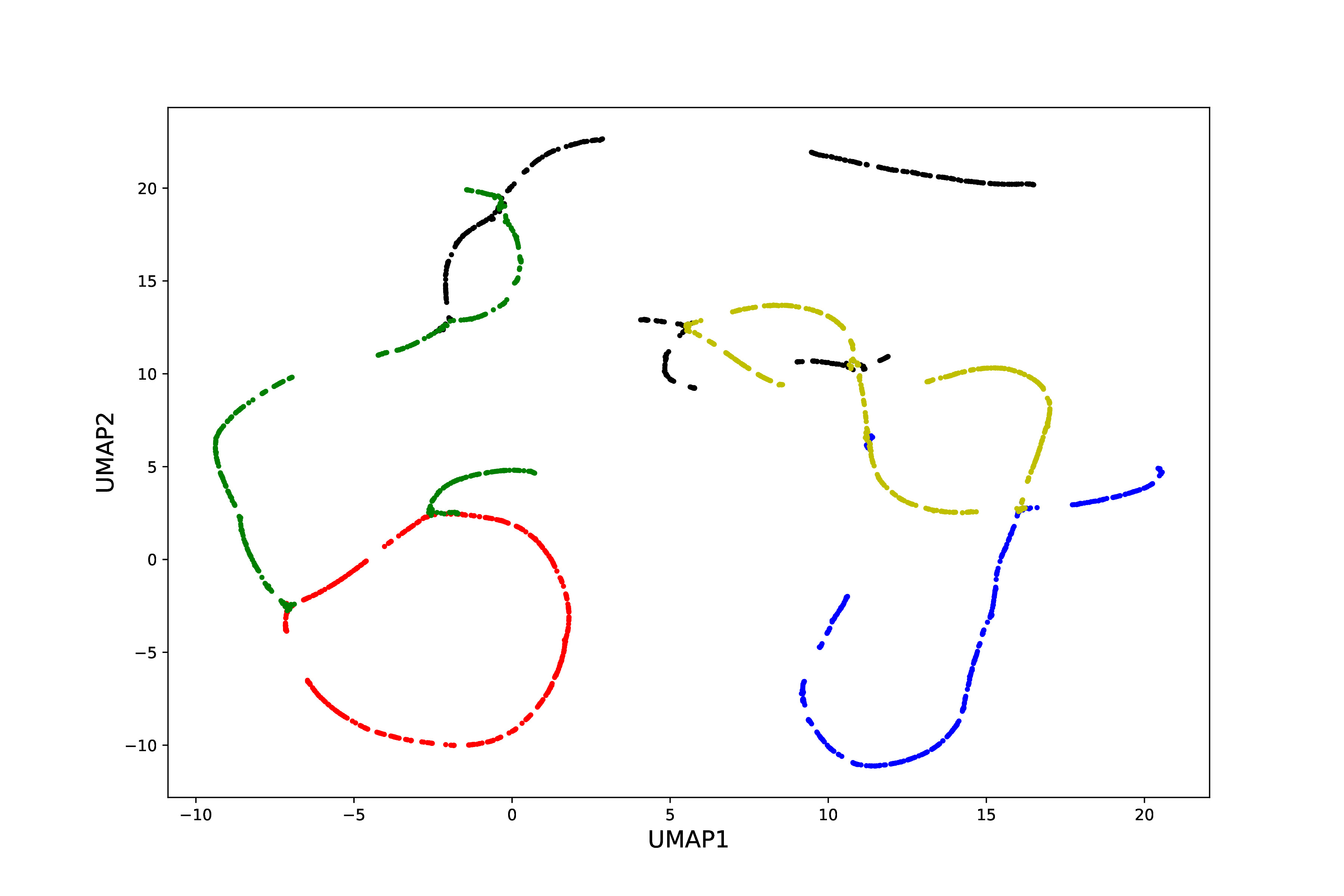}
		\label{fig:2D_Olympic_UMAP}
	}
	\\
	\subfloat[Original]{
		\centering
		\includegraphics[width=0.25\columnwidth,height=4cm]{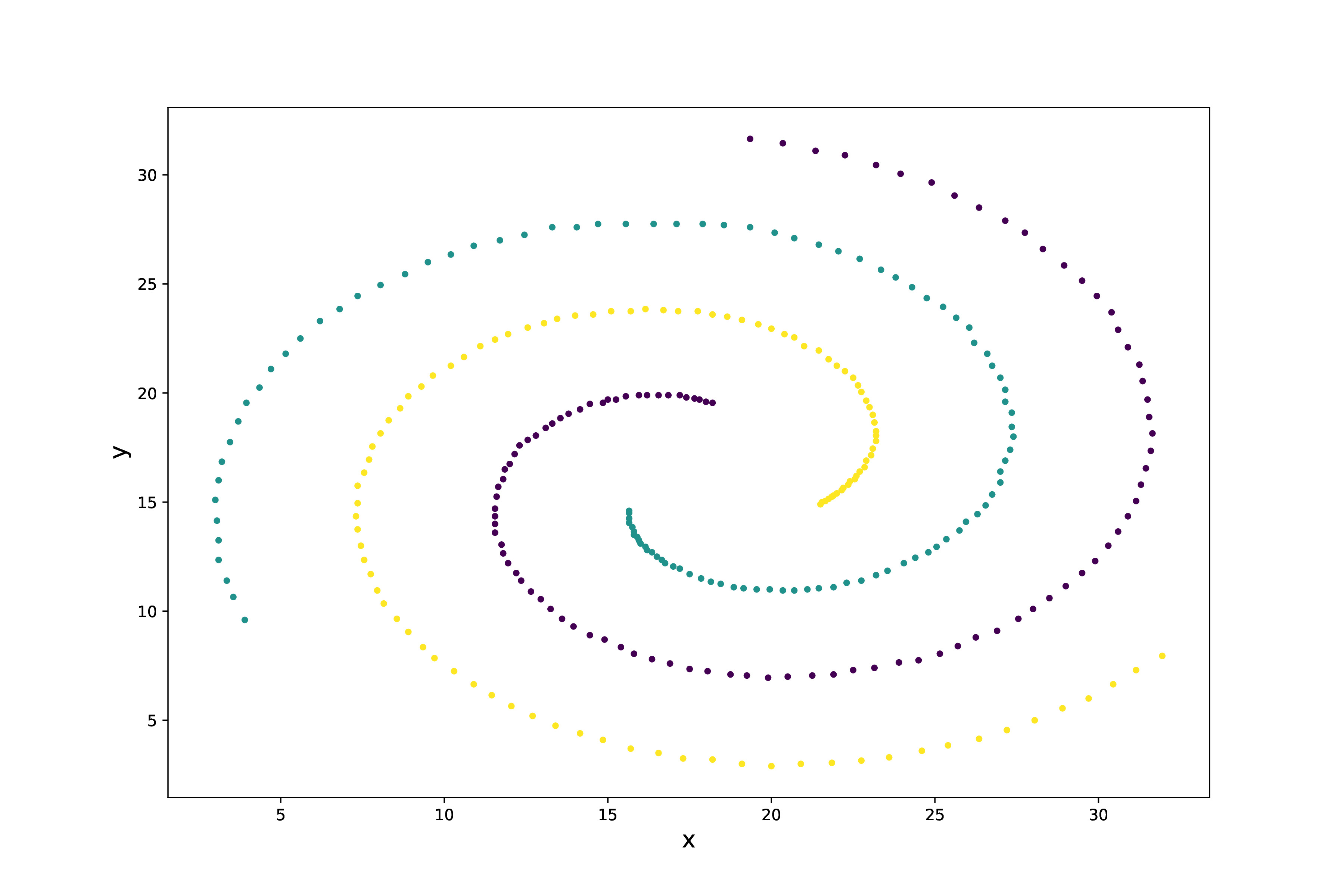}
		\label{fig:2D_Spiral}
	}
	\subfloat[NeuroDAVIS]{
		\centering
		\includegraphics[width=0.25\columnwidth,height=4cm]{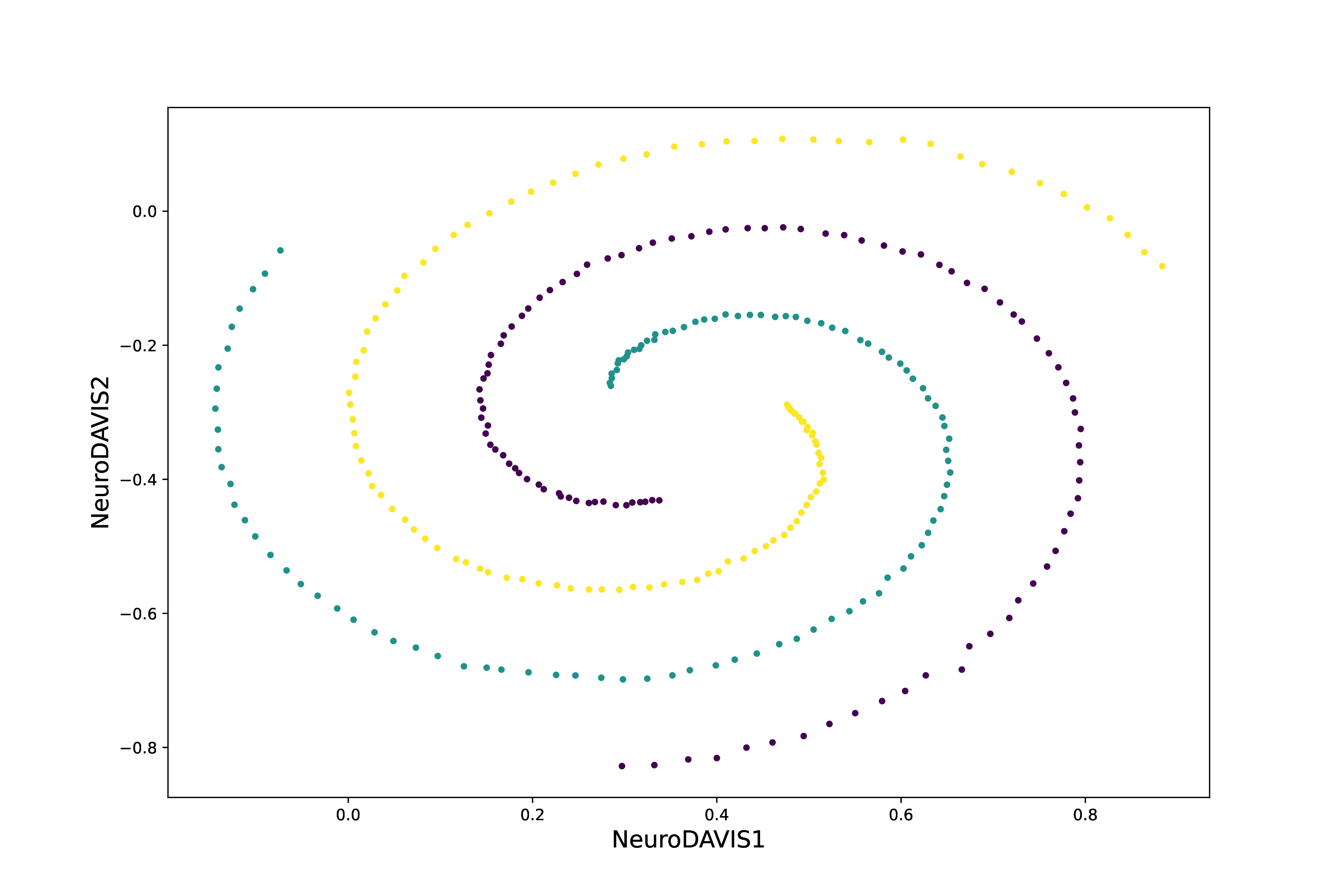}
		\label{fig:2D_Spiral_NeuroDAVIS}
	}
	\subfloat[t-SNE]{
		\centering
		\includegraphics[width=0.25\columnwidth,height=4cm]{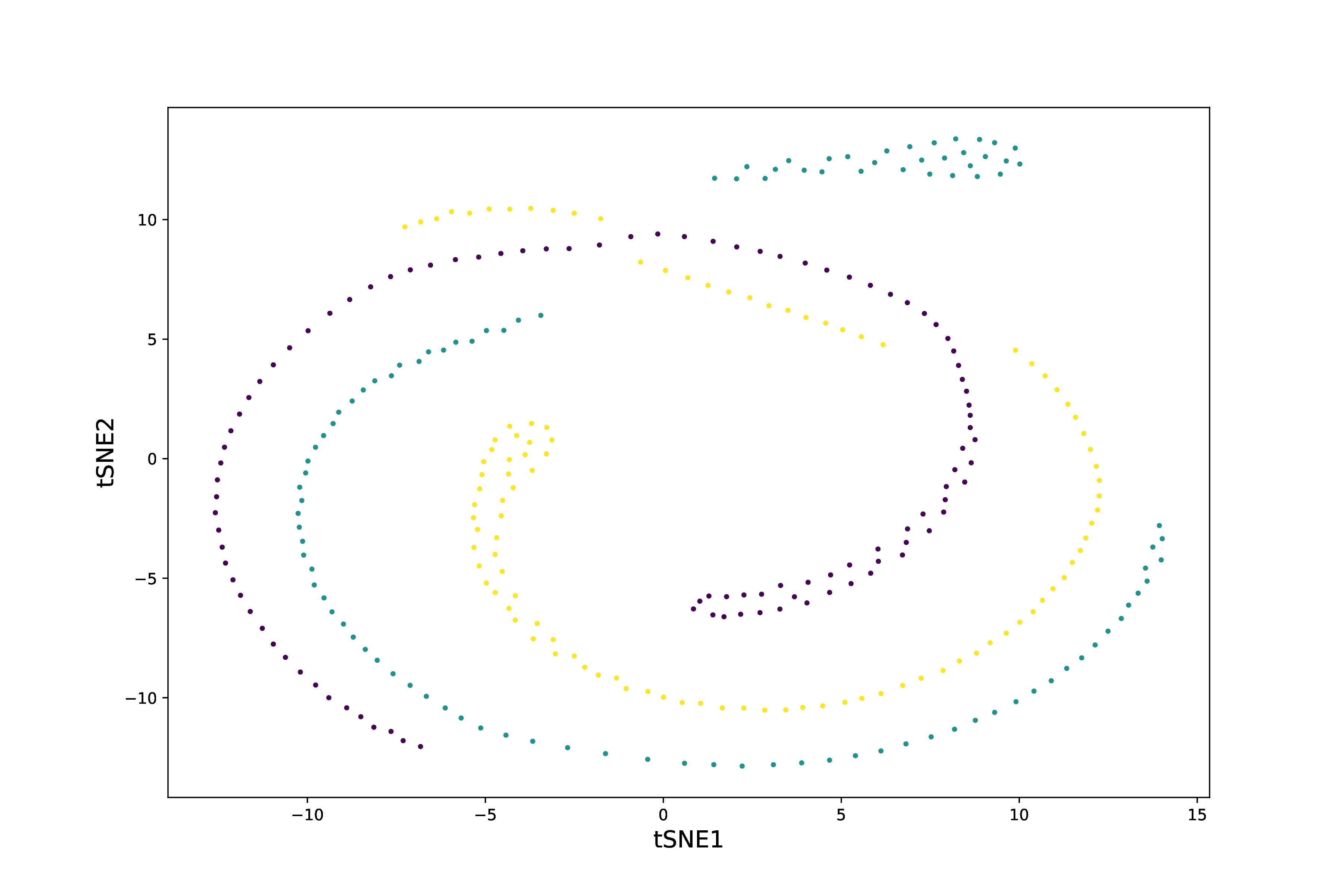}
		\label{fig:2D_Spiral_tSNE}
	}
	\subfloat[UMAP]{
		\centering
		\includegraphics[width=0.25\columnwidth,height=4cm]{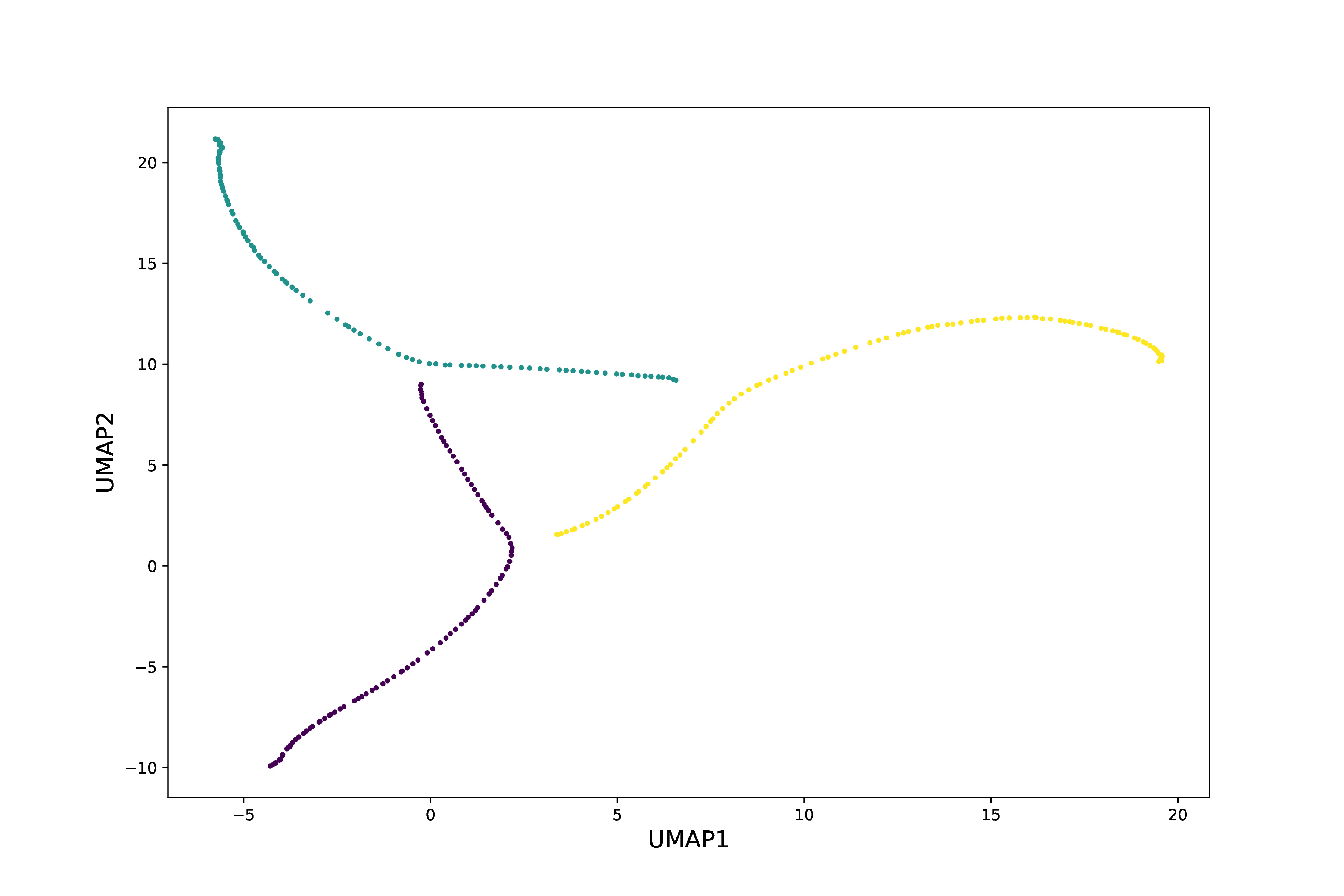}
		\label{fig:2D_Spiral_UMAP}
	}
	\\
	\subfloat[Original]{
		\centering
		\includegraphics[width=0.25\columnwidth,height=4cm]{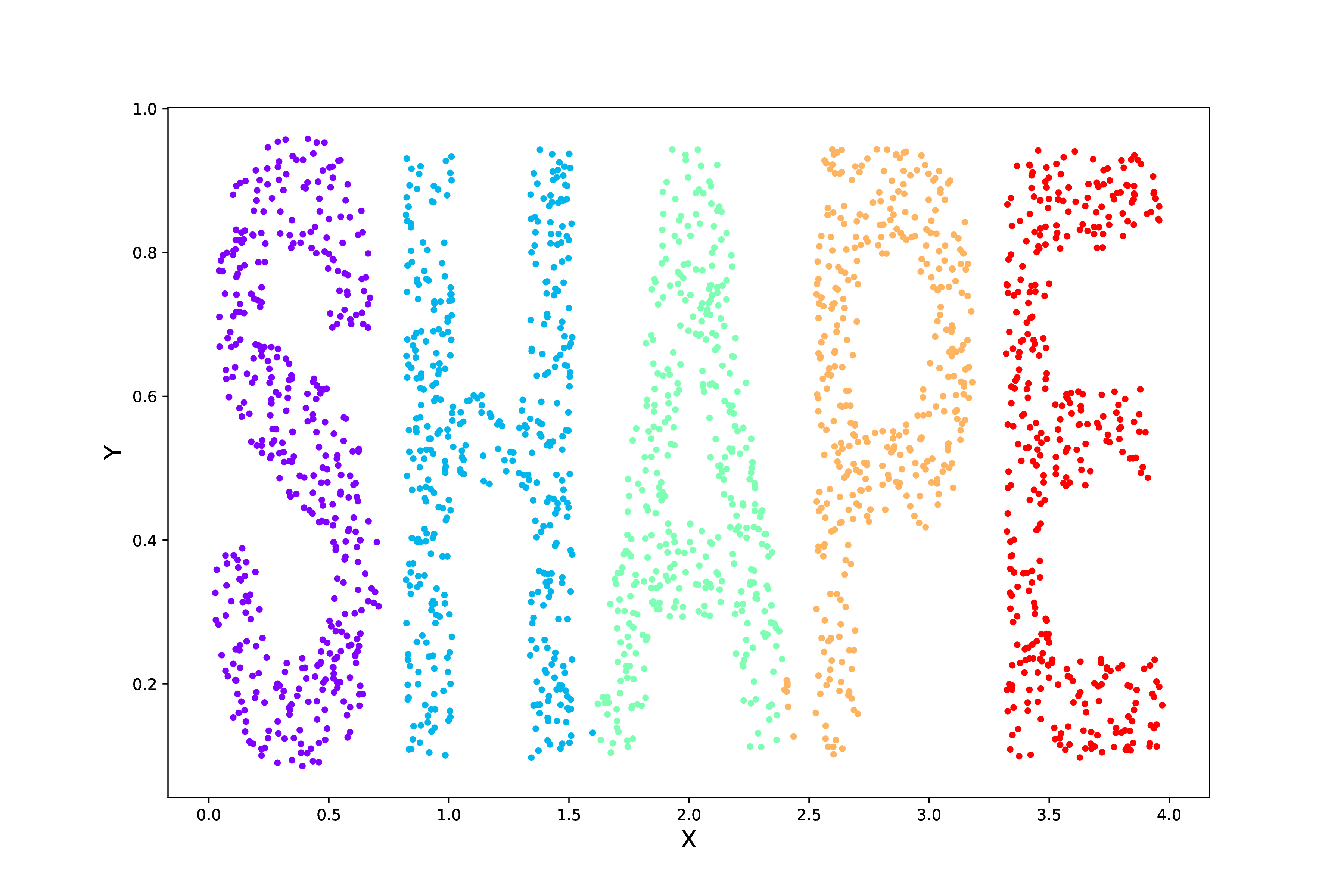}
		\label{fig:2D_Shape}
	}
	\subfloat[NeuroDAVIS]{
		\centering
		\includegraphics[width=0.25\columnwidth,height=4cm]{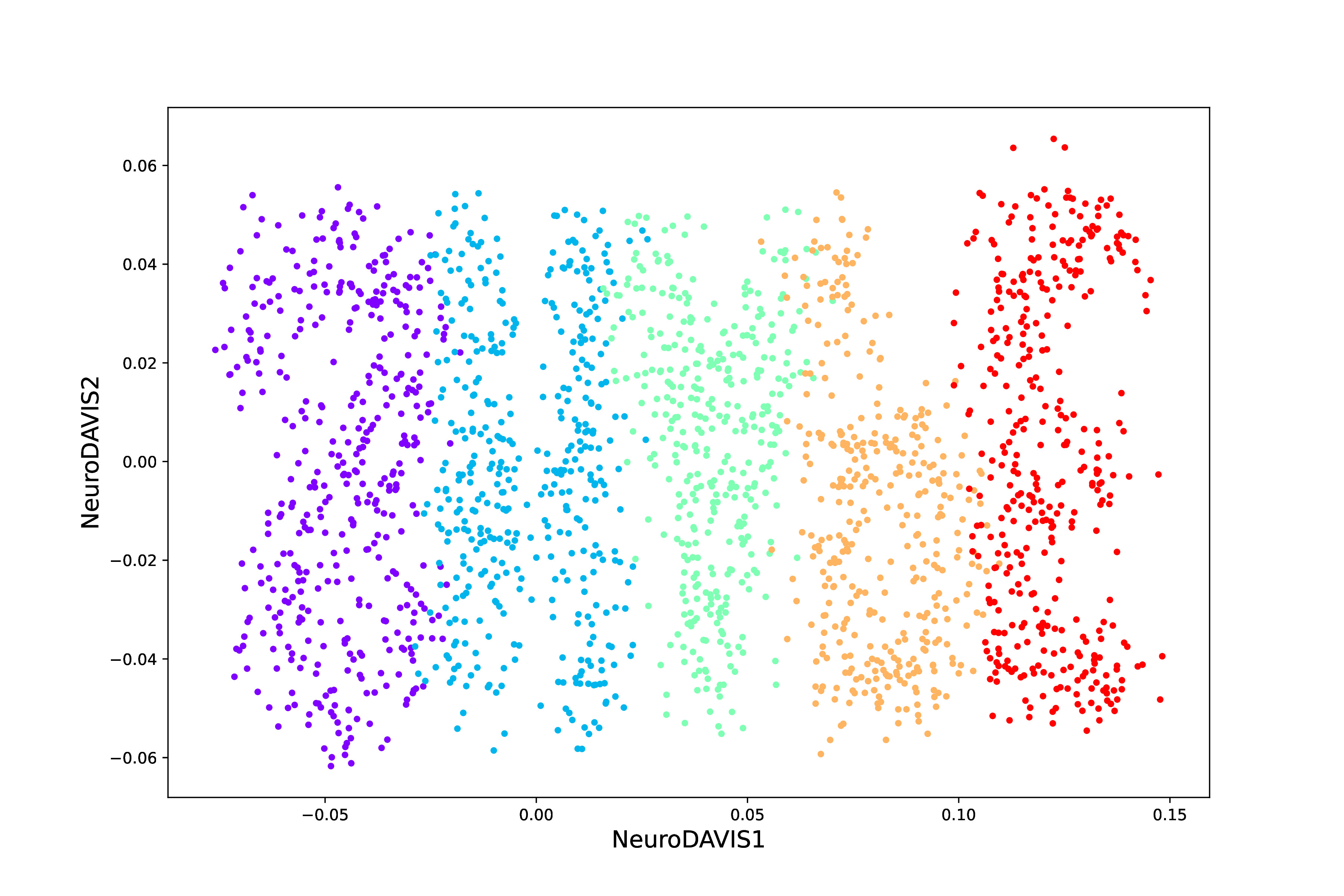}
		\label{fig:2D_Shape_NeuroDAVIS}
	}
	\subfloat[t-SNE]{
		\centering
		\includegraphics[width=0.25\columnwidth,height=4cm]{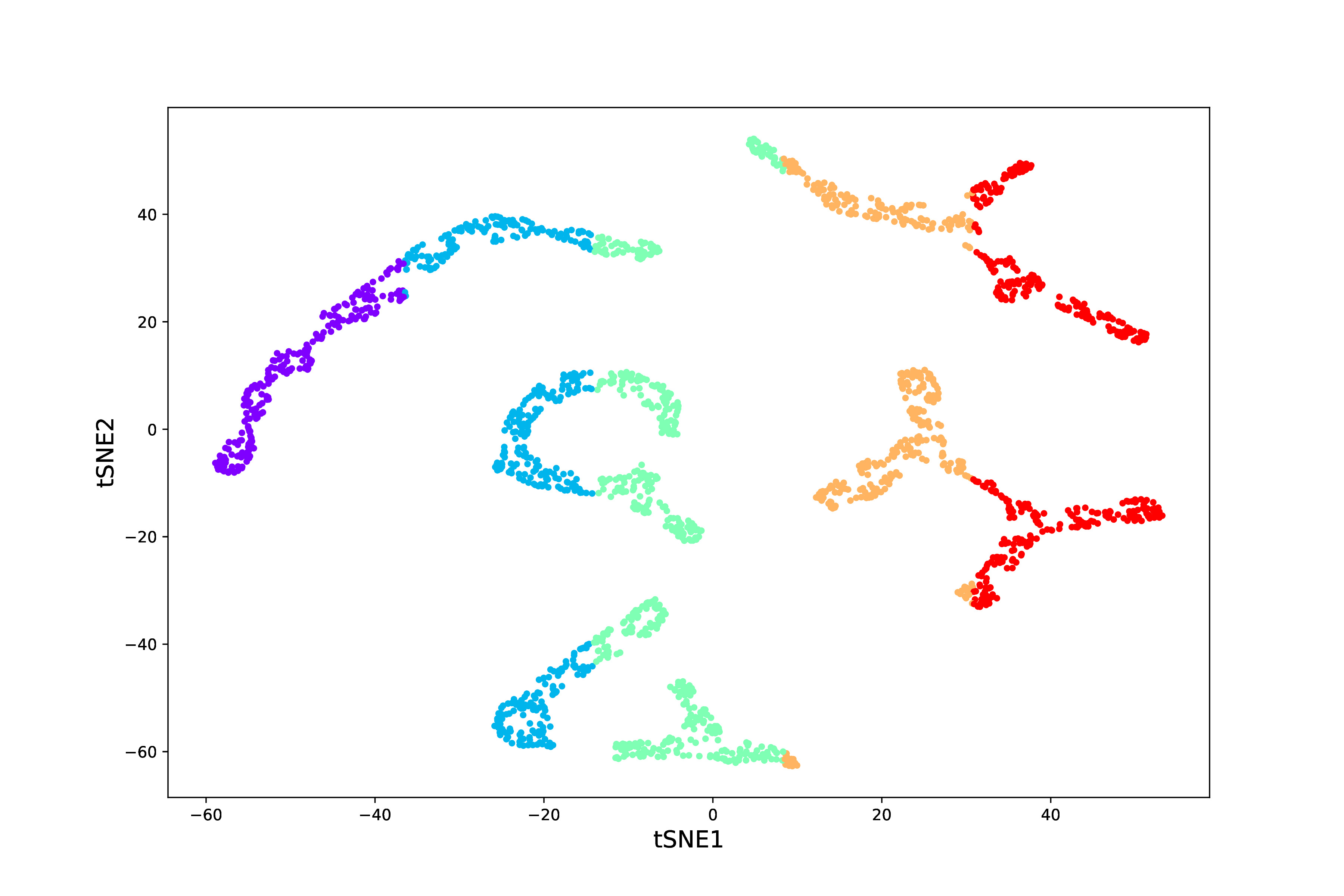}
		\label{fig:2D_Shape_tSNE}
	}
	\subfloat[UMAP]{
		\centering
		\includegraphics[width=0.25\columnwidth,height=4cm]{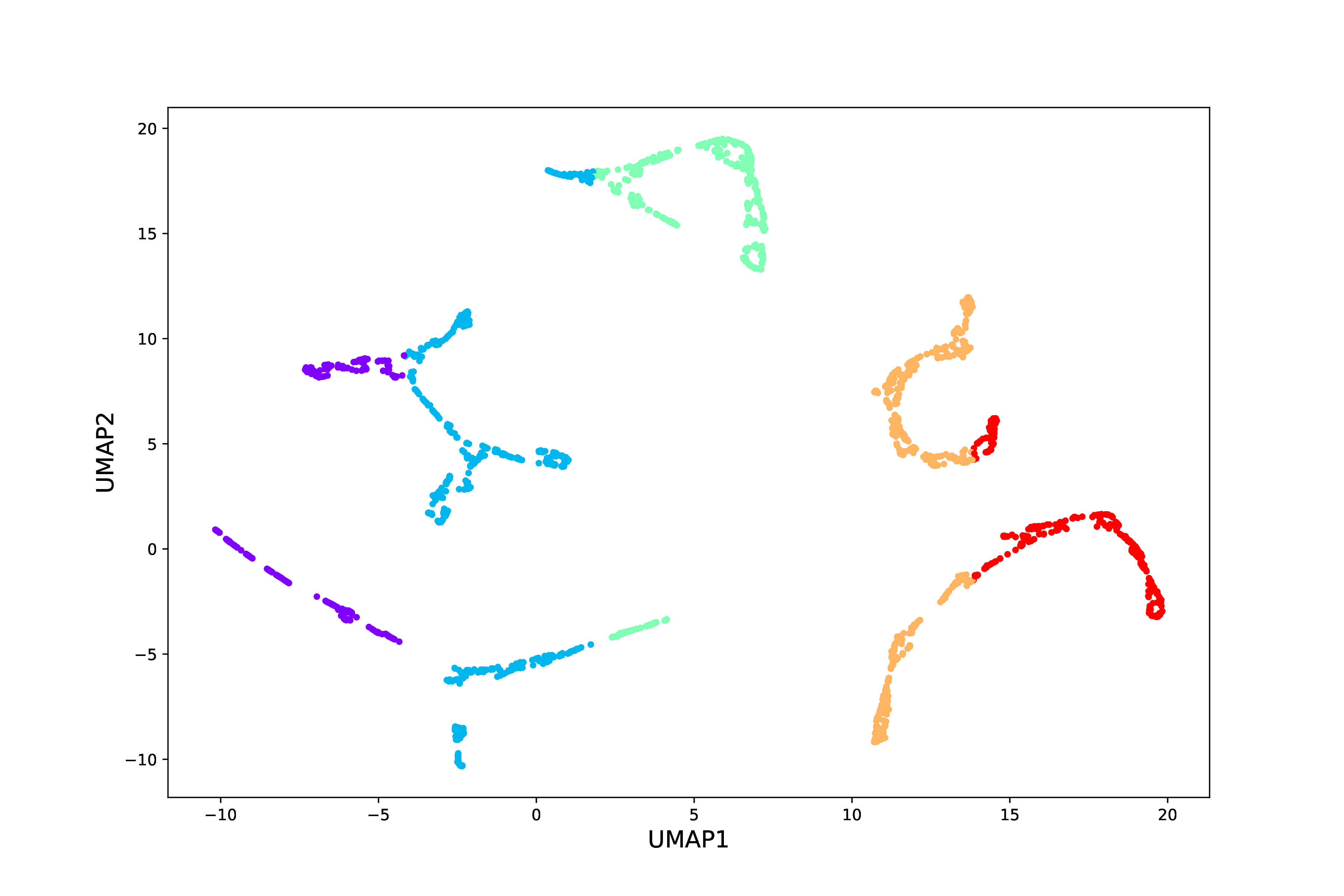}
		\label{fig:2D_Shape_UMAP}
	}
	\caption{The original distributions and the embeddings produced by NeuroDAVIS, t-SNE and UMAP for Elliptic Ring ((a)-(d)), Olympic ((e)-(h)), Spiral ((i)-(l)) and Shape ((m)-(p)) datasets.}
	\label{fig:2D}
\end{figure}

Elliptic Ring dataset (Figure \ref{fig:2D_EllipticRing}) consists of two small Gaussian balls within an outer Gaussian elliptic ring; Olympic dataset (Figure \ref{fig:2D_Olympic}) contains five circular rings representing the olympic logo; Spiral dataset (Figure \ref{fig:2D_Spiral}) contains three concentric spirals, while Shape dataset (Figure \ref{fig:2D_Shape}) consists of points representing the characters `S', `H', `A', `P' and `E'. 

We have observed that the NeuroDAVIS-generated embedding of Elliptic Ring dataset (Figure \ref{fig:2D_EllipticRing_NeuroDAVIS}) contains two Gaussian balls inside the outer Gaussian ring, similar to the original data. Likewise, the distances between the rings in Olympic dataset and that between the concentric spirals in Spiral dataset, as well as their shapes, have been preserved in their NeuroDAVIS-generated embeddings (Figures \ref{fig:2D_Olympic_NeuroDAVIS} and \ref{fig:2D_Spiral_NeuroDAVIS}). Furthermore for Shape dataset, the readability of the characters `S', `H', `A', `P' and `E' have not been compromised by NeuroDAVIS (Figure \ref{fig:2D_Shape_NeuroDAVIS}), which is totally lost in the cases of t-SNE and UMAP embeddings. Thus, we can say that NeuroDAVIS has been able to represent the clusters within the data similar to their original distributions, preserving both shape and size. Both t-SNE and UMAP, however, have failed to preserve original distributions, and produced compact clusters instead, disrupting both local and global structures within the data.

\begin{figure}
	\centering
	\includegraphics[width=\columnwidth,height=8cm]{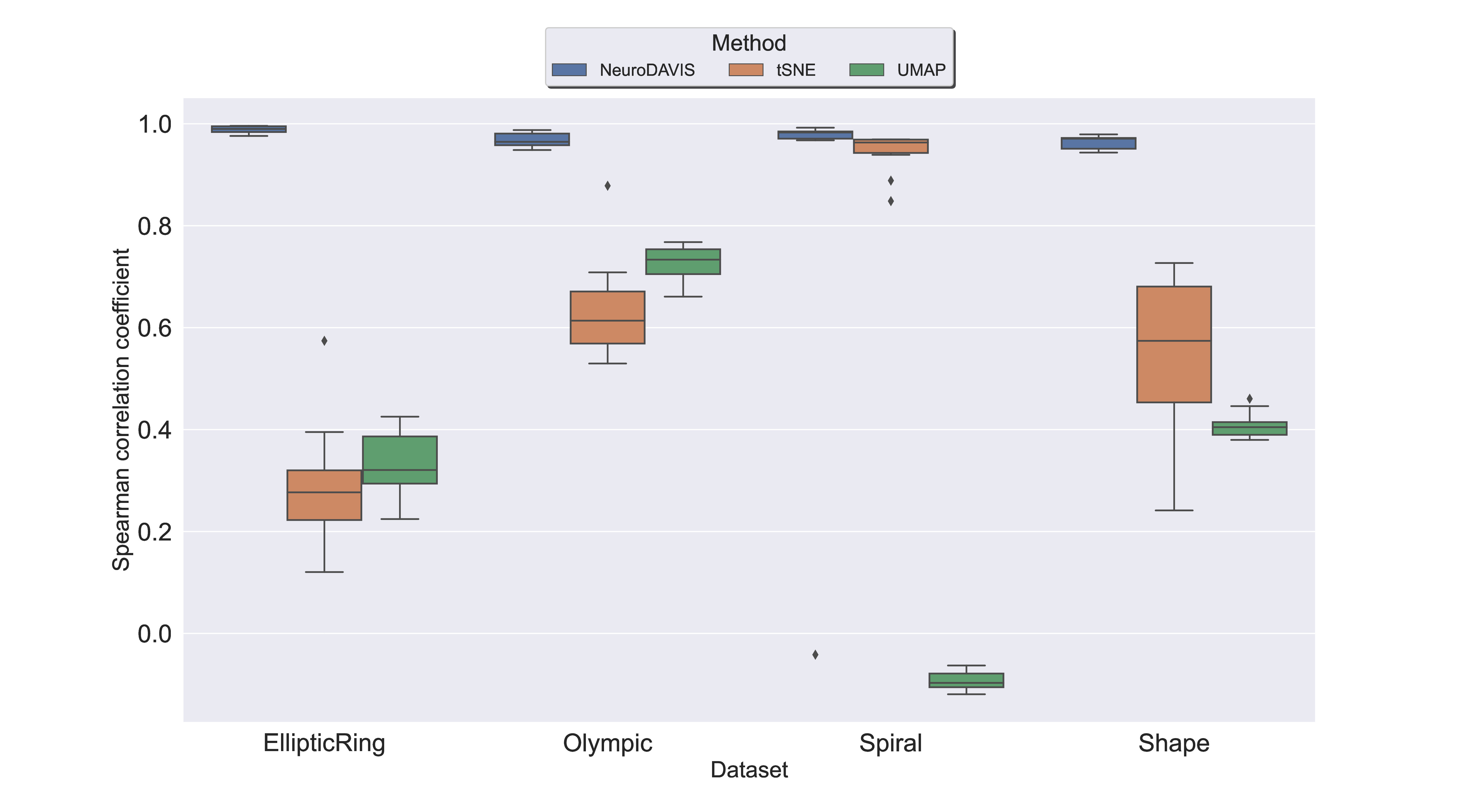}
	\caption{Spearman rank correlation between pairwise distances in the original distribution, and pairwise distances in t-SNE, UMAP and NeuroDAVIS-produced embeddings of the $2$D synthetic datasets Elliptic Ring, Olympic, Spiral and Shape. For Elliptic Ring dataset, median correlation coefficient values obtained are $0.98$ (NeuroDAVIS), $0.27$ (t-SNE) and $0.32$ (UMAP). For Olympic dataset, median correlation coefficient values obtained are $0.96$ (NeuroDAVIS), $0.61$ (t-SNE) and $0.73$ (UMAP). For Spiral dataset, median correlation coefficient values obtained are $0.98$ (NeuroDAVIS), $0.96$ (t-SNE) and $-0.09$ (UMAP). For Shape dataset, median correlation coefficient values obtained are $0.97$ (NeuroDAVIS), $0.57$ (t-SNE) and $0.40$ (UMAP). The corresponding p-values obtained using Mann-Whitney U test are $0.0001$ (both NeuroDAVIS-t-SNE and NeuroDAVIS-UMAP) for Elliptic Ring, Olympic and Shape datasets. For Spiral dataset, p-values obtained are $0.009$ (NeuroDAVIS-t-SNE) and $0.001$ (NeuroDAVIS-UMAP).}
	\label{fig:2D_Spearman}
\end{figure}

We have further compared the pairwise distances in the original distribution to that in the embedding produced by t-SNE, UMAP and NeuroDAVIS, using a Spearman rank correlation. Figure \ref{fig:2D_Spearman} shows the results for ten different executions of the same. We have observed that the correlation coefficient values obtained by NeuroDAVIS have been better than that produced by t-SNE and UMAP for all the datasets.

We have then performed another experiment on these synthetic datasets, projecting the two-dimensional space into nine-dimensional space using the transformation ($x+y$, $x-y$, $xy$, $x^2$, $y^2$, $x^2y$, $xy^2$, $x^3$, $y^3$) as perfomed in \cite{ding2018interpretable}, and applying NeuroDAVIS to project the data back into a two-dimensional space. Results have been compared to those obtained by t-SNE and UMAP for the same transformation. Figure S1 (in Supplementary Material) shows that NeuroDAVIS has once again produced better embeddings, preserving both cluster shapes and sizes, as compared to t-SNE and UMAP.

\subsubsection{Global structure preservation}
\label{sec:results_global}
\begin{figure}
	\subfloat[Original]{
		\centering
		\includegraphics[width=0.5\columnwidth,height=4cm]{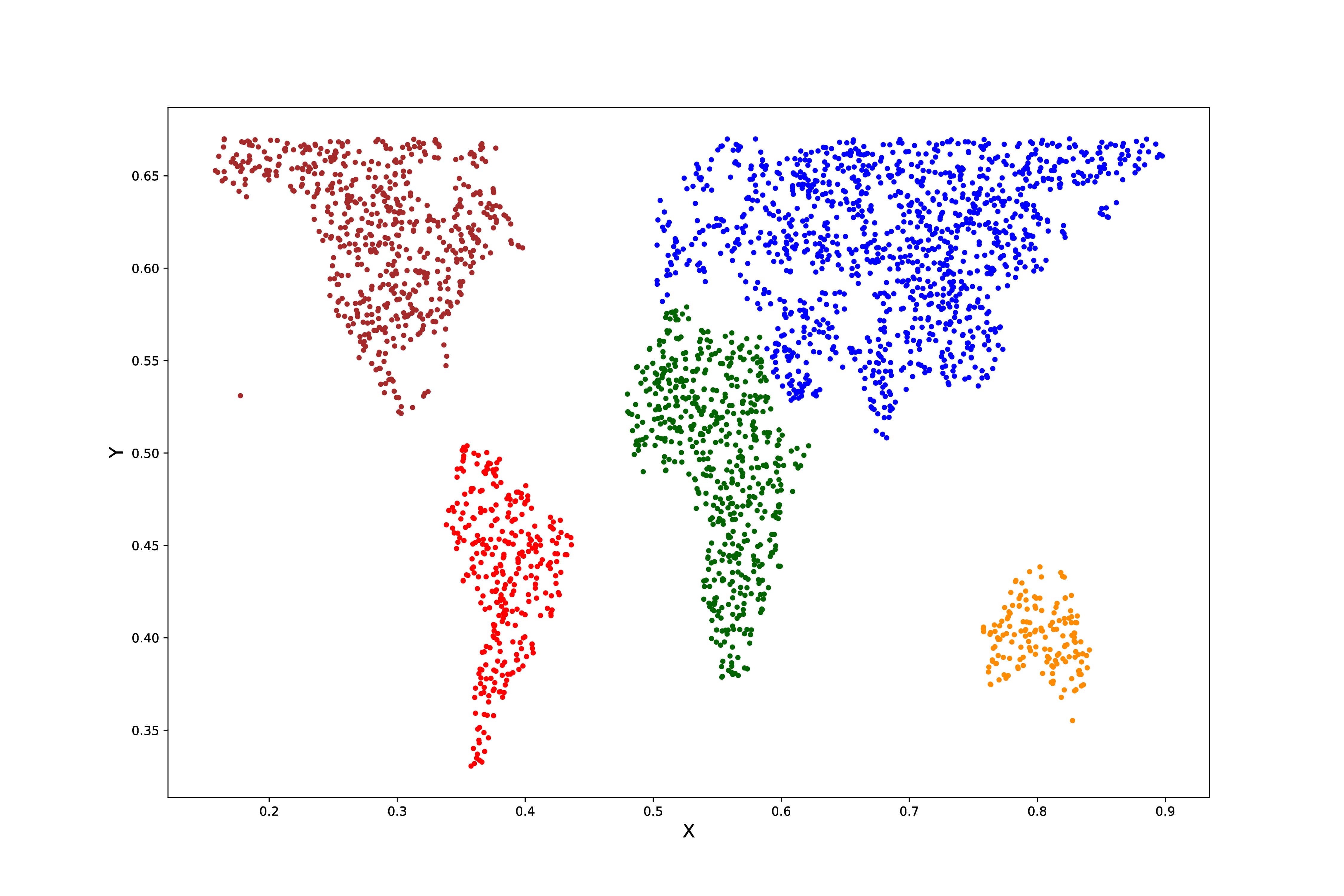}
		\label{fig:WorldMap_Orig}
	}
	\subfloat[NeuroDAVIS]{
		\centering
		\includegraphics[width=0.5\columnwidth,height=4cm]{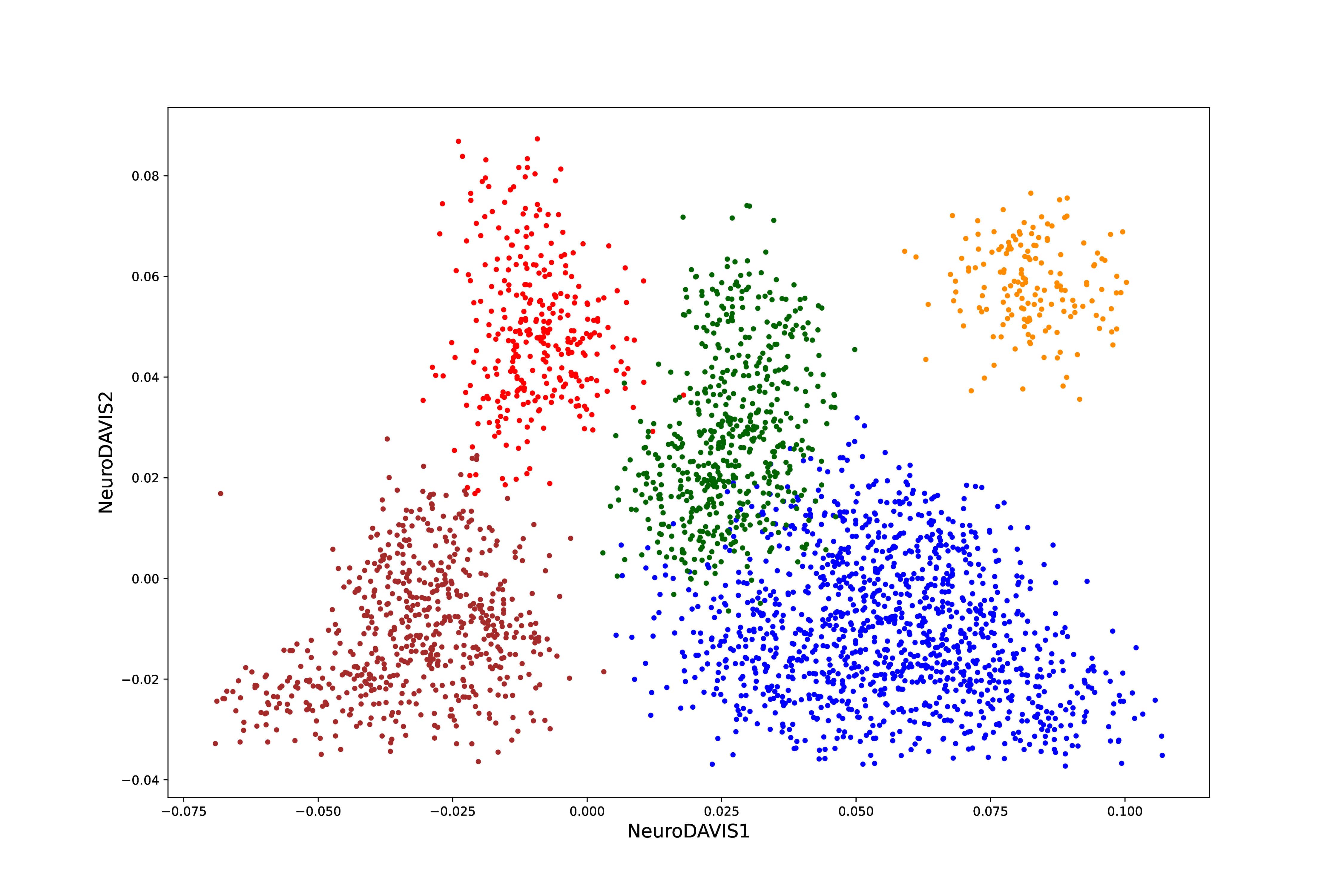}
		\label{fig:WorldMap_NeuroDAVIS}
	}\\
	\subfloat[t-SNE]{
		\centering
		\includegraphics[width=0.5\columnwidth,height=4cm]{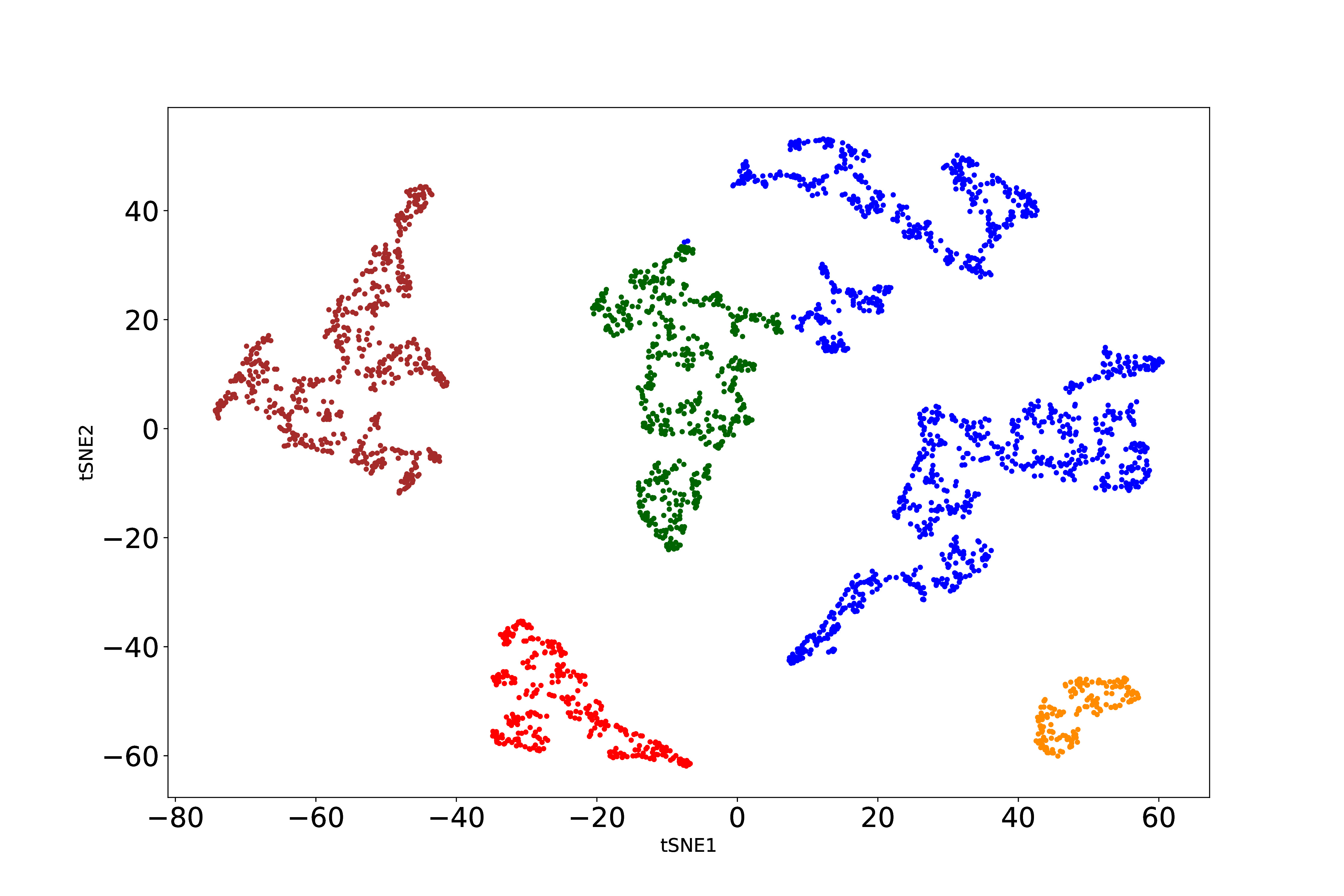}
		\label{fig:WorldMap_tSNE}
	}
	\subfloat[UMAP]{
		\centering
		\includegraphics[width=0.5\columnwidth,height=4cm]{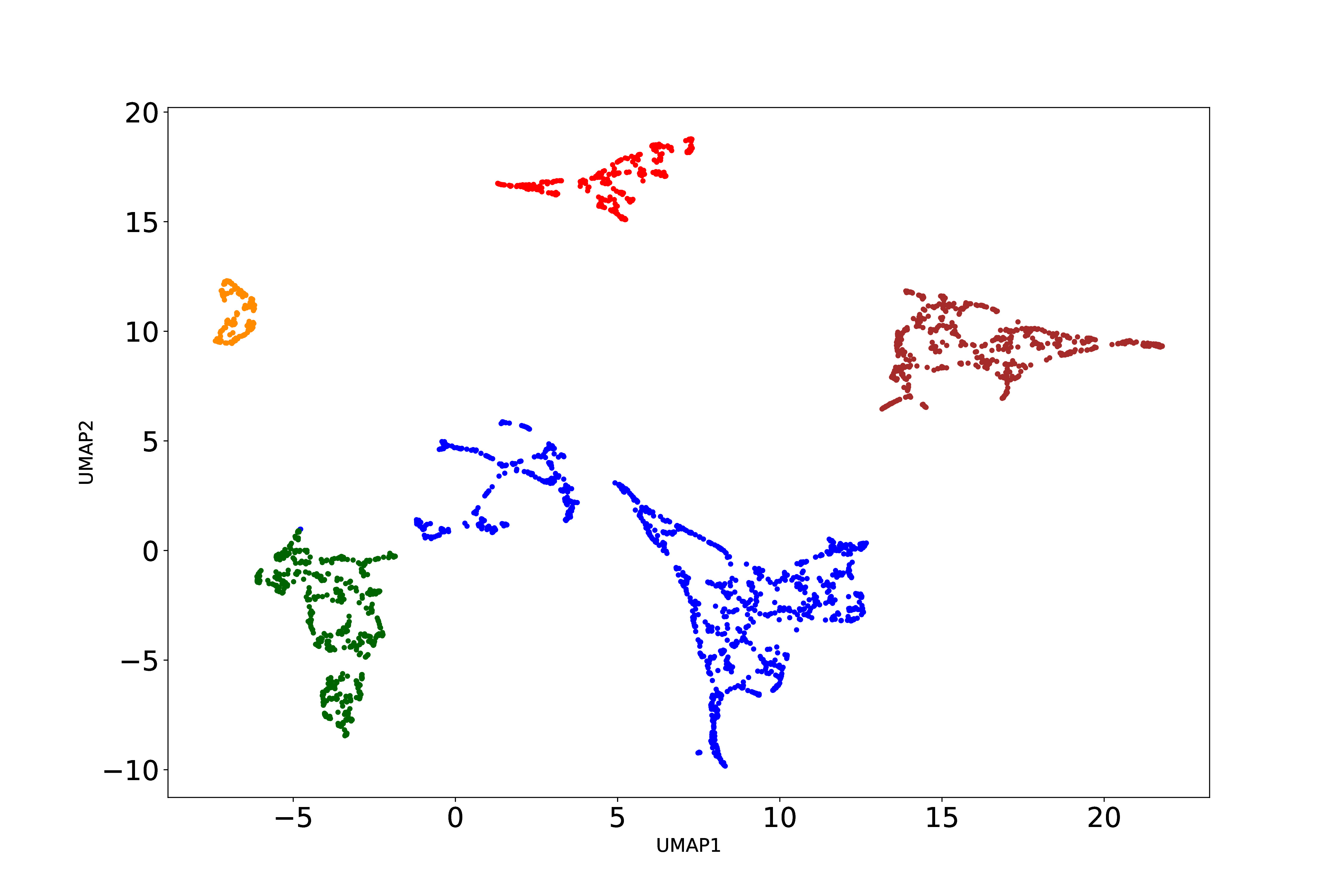}
		\label{fig:WorldMap_UMAP}
	}
	\caption{The original distribution and the embeddings produced by NeuroDAVIS, t-SNE and UMAP for World Map dataset.}
	\label{fig:WorldMap}
\end{figure}

In order to assess the performance of a dimension reduction/visualization method, we need to analyze the reduced dimensional embedding in terms of its capability to represent clusters in a way similar to their distribution in the high dimension. For this purpose, it is essential to study the inter-cluster separations within the data. Hence, we have performed a few more experiments to ascertain how well NeuroDAVIS preserves the inter-cluster distances in the low dimensional space. Information on the ground truth is important for this kind of analyses. For this reason, we have created a synthetic $2$D dataset representing the world map (a known structure) with five clusters / continents, viz., Eurasia, Australia, North America, South America and Africa (no Antarctica), following the tutorial (\url{https://towardsdatascience.com/tsne-vs-umap-global-structure-4d8045acba17}) (Figure \ref{fig:WorldMap_Orig}). The NeuroDAVIS-generated embedding for World Map dataset has been compared with its original data distribution, and the corresponding embeddings generated by t-SNE and UMAP (Figure \ref{fig:WorldMap}). As evident from Figure \ref{fig:WorldMap}, unlike t-SNE and UMAP, the shapes and sizes of the clusters in the NeuroDAVIS-generated embedding are quite similar to that in the original data.

\subsubsection{Inter-cluster distance preservation}
\label{sec:results_interClusterdist}
\begin{figure}
	\centering
	\includegraphics[width=\columnwidth,height=7cm]{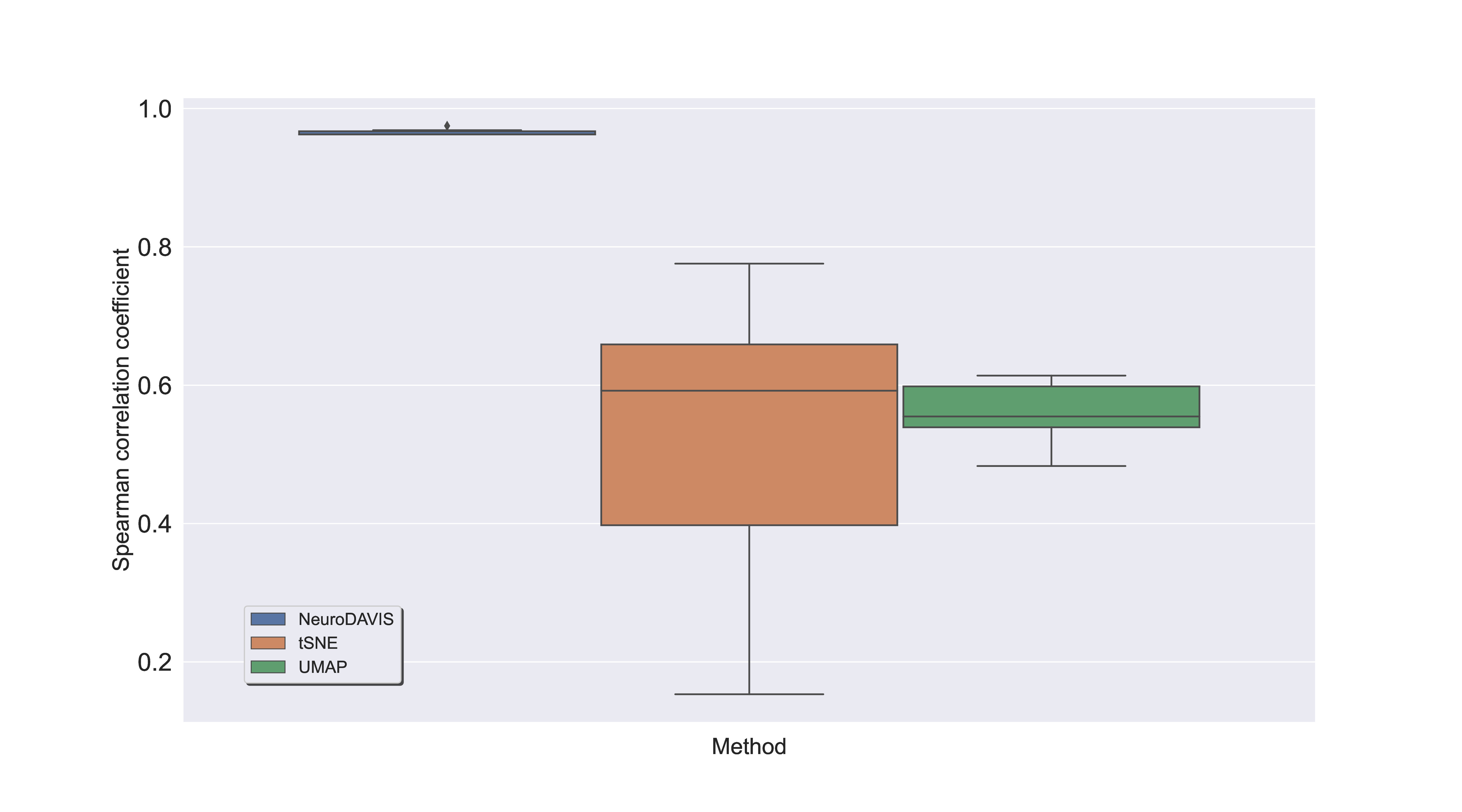}
	\caption{Spearman rank correlation between original distances among cluster centroids in the original dataset and distances between cluster centroids in the NeuroDAVIS, t-SNE and UMAP-generated embedding. The median correlation coefficient values are $0.96$ (NeuroDAVIS), $0.59$ (t-SNE) and $0.55$ (UMAP) for World Map dataset. The corresponding p-value obtained using Mann-Whitney U test is $0.0001$ (both NeuroDAVIS-t-SNE and NeuroDAVIS-UMAP)).}
	\label{fig:WorldMap_Cluster_Centroids}
\end{figure}

In order to check if the inter-cluster distances between the five continents have been preserved, we have calculated the Spearman rank correlation coefficients between the distances among the five cluster centroids obtained by NeuroDAVIS, and compared the result with those obtained using t-SNE and UMAP. As shown in Figure \ref{fig:WorldMap_Cluster_Centroids}, repeated runs of the same experiment have confirmed that NeuroDAVIS has performed much better than t-SNE and UMAP, with respect to distance preservation. 

\begin{figure}
		\subfloat[Distances between Eurasia and others]{
			\centering
			\includegraphics[width=0.5\columnwidth,height=5cm]{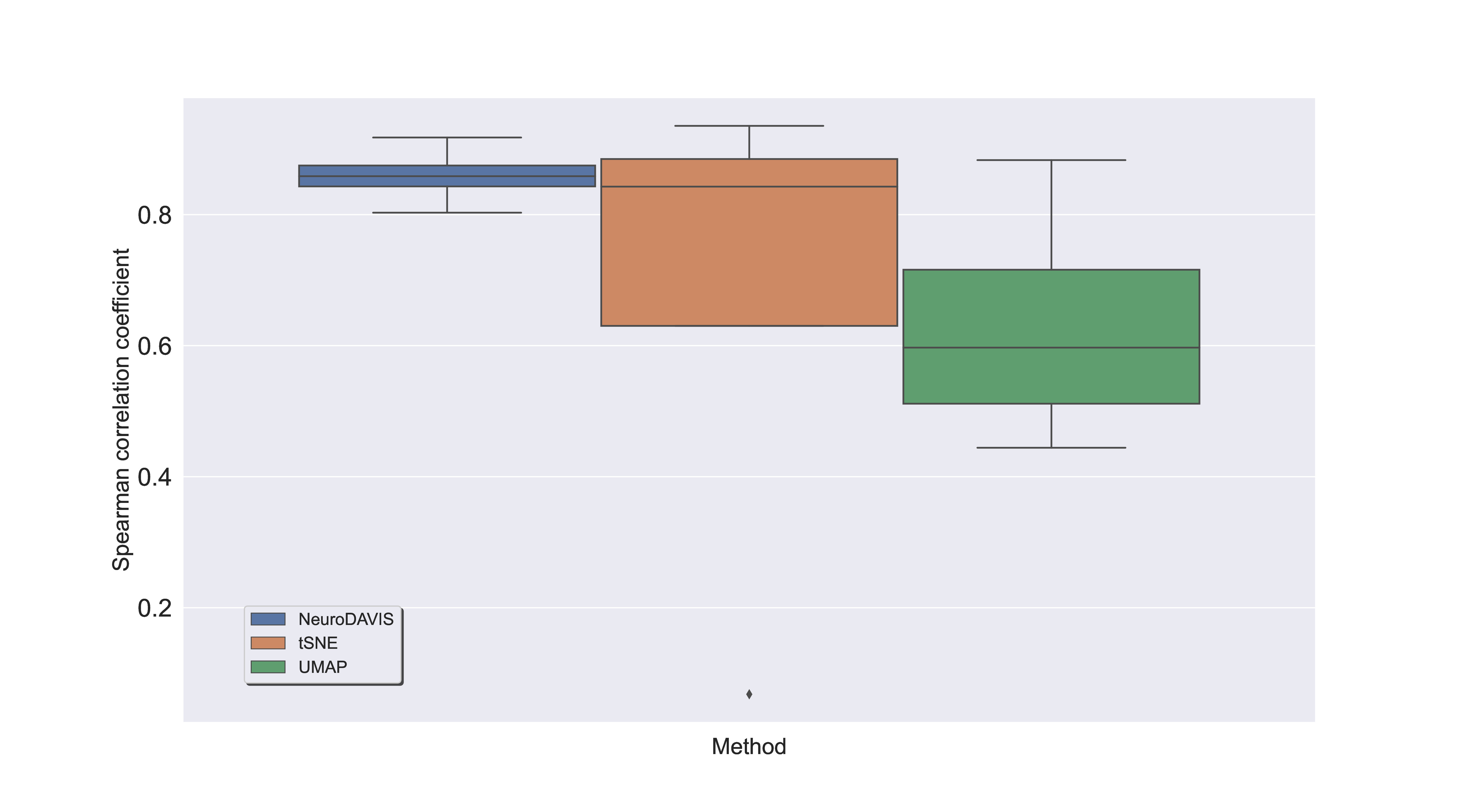}
			\label{fig:WorldMap_BP1}
		}
		\subfloat[Distances between Australia and others]{
			\centering
			\includegraphics[width=0.5\columnwidth,height=5cm]{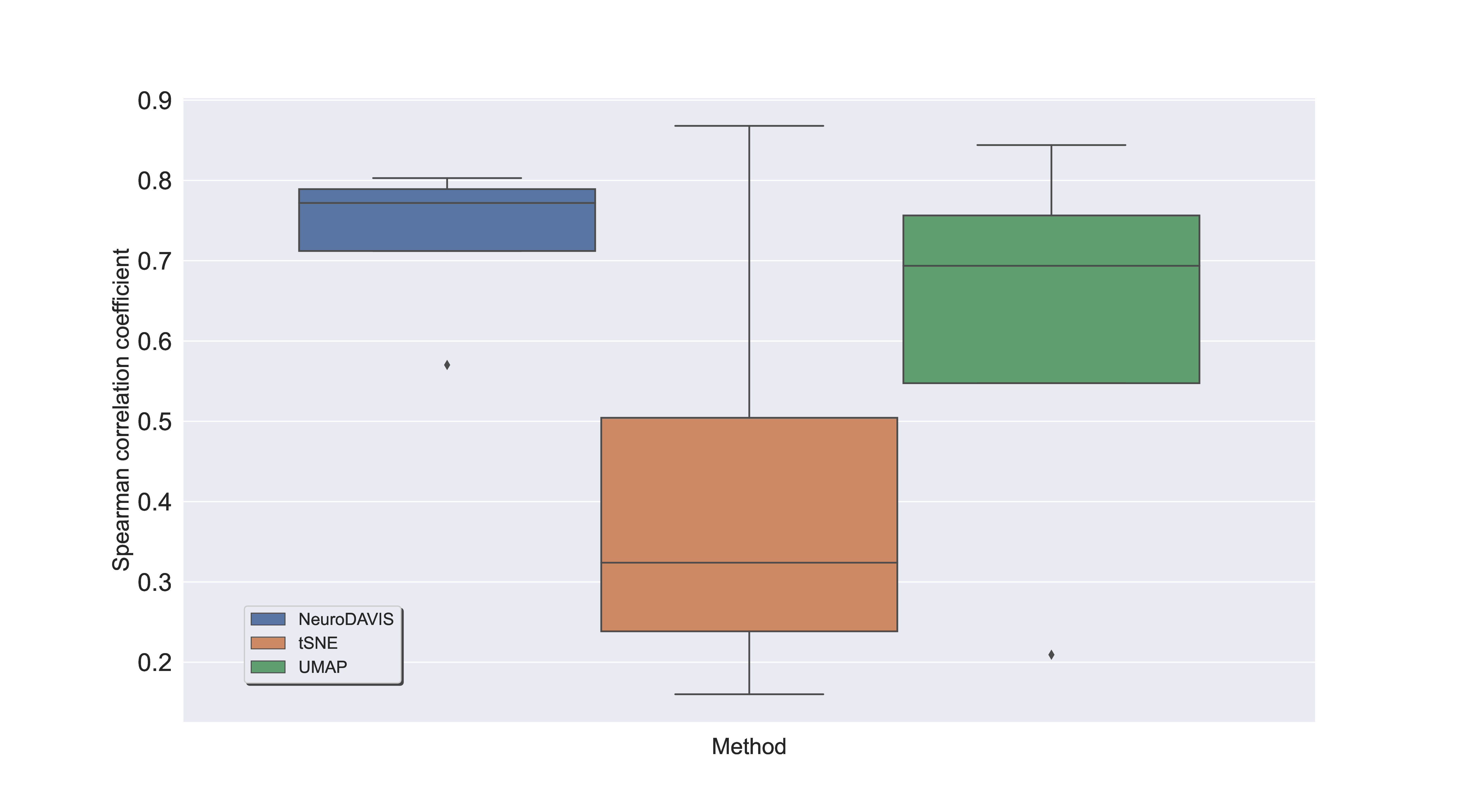}
			\label{fig:WorldMap_BP2}
		}
	\\
		\subfloat[Distances between North America and others]{
			\centering
			\includegraphics[width=0.5\columnwidth,height=5cm]{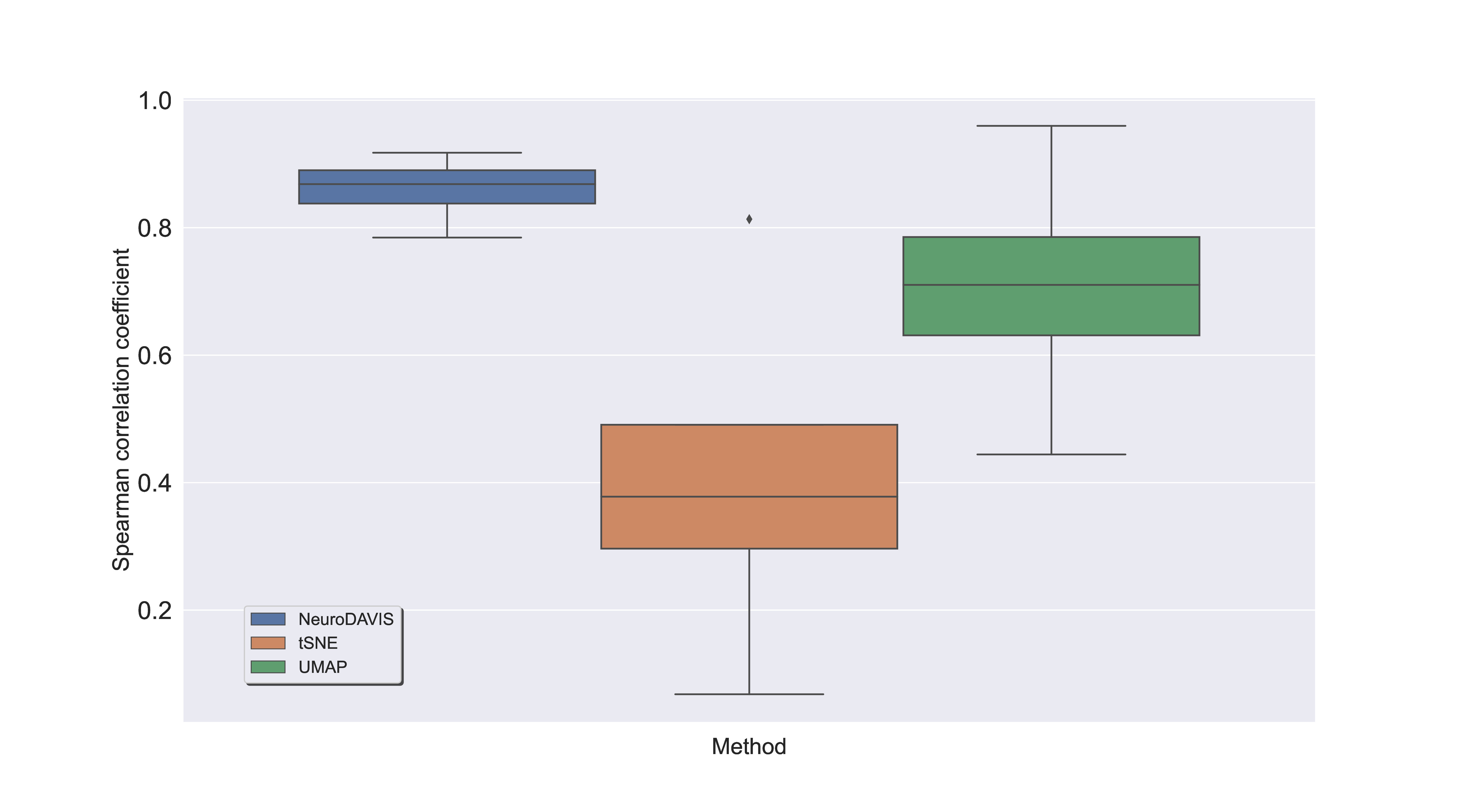}
			\label{fig:WorldMap_BP3}
		}
		\subfloat[Distances between South America and others]{
			\centering
			\includegraphics[width=0.5\columnwidth,height=5cm]{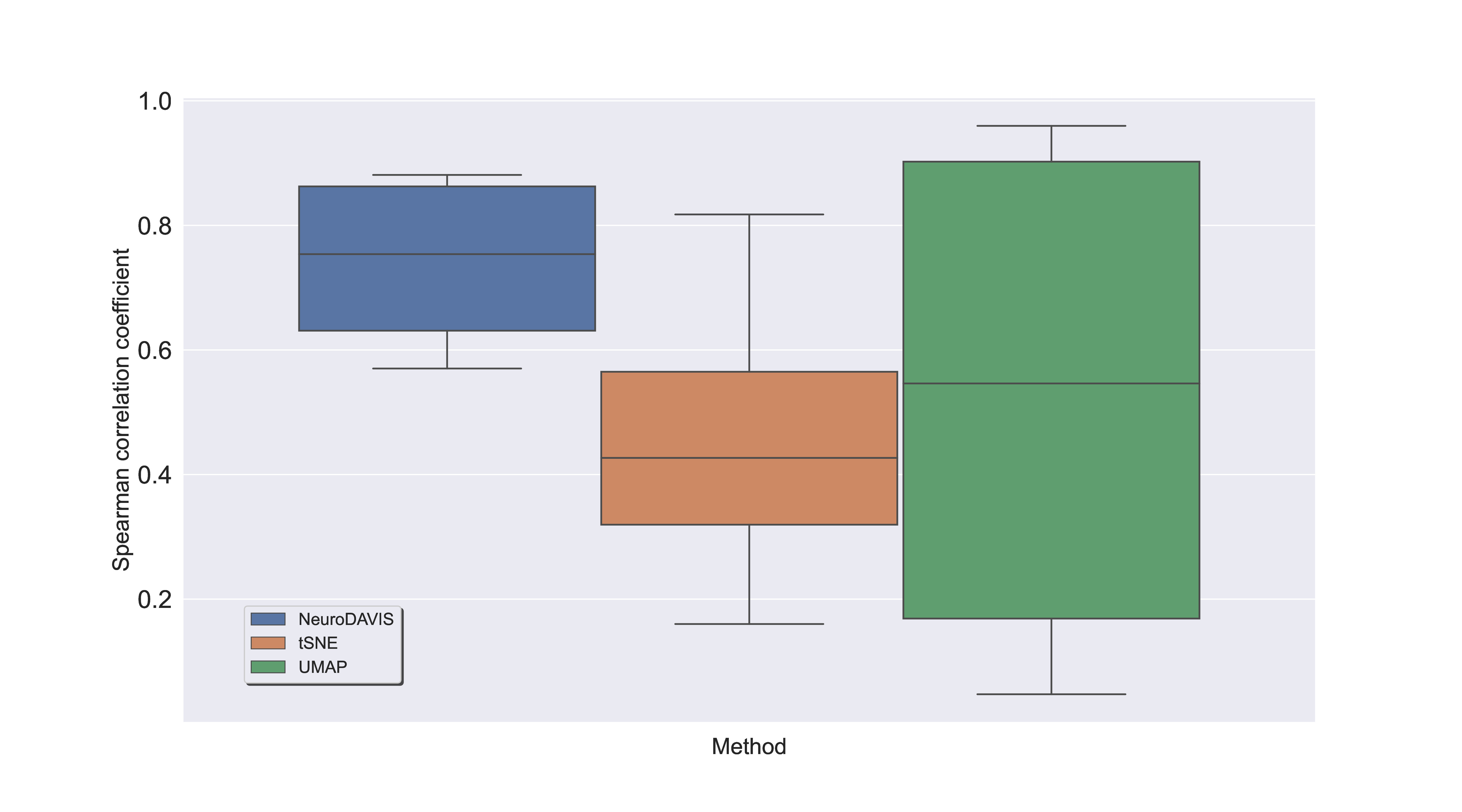}
			\label{fig:WorldMap_BP4}
		}
	\\
		\subfloat[Distances between Africa and others]{
			\centering
			\includegraphics[width=0.5\columnwidth,height=5cm]{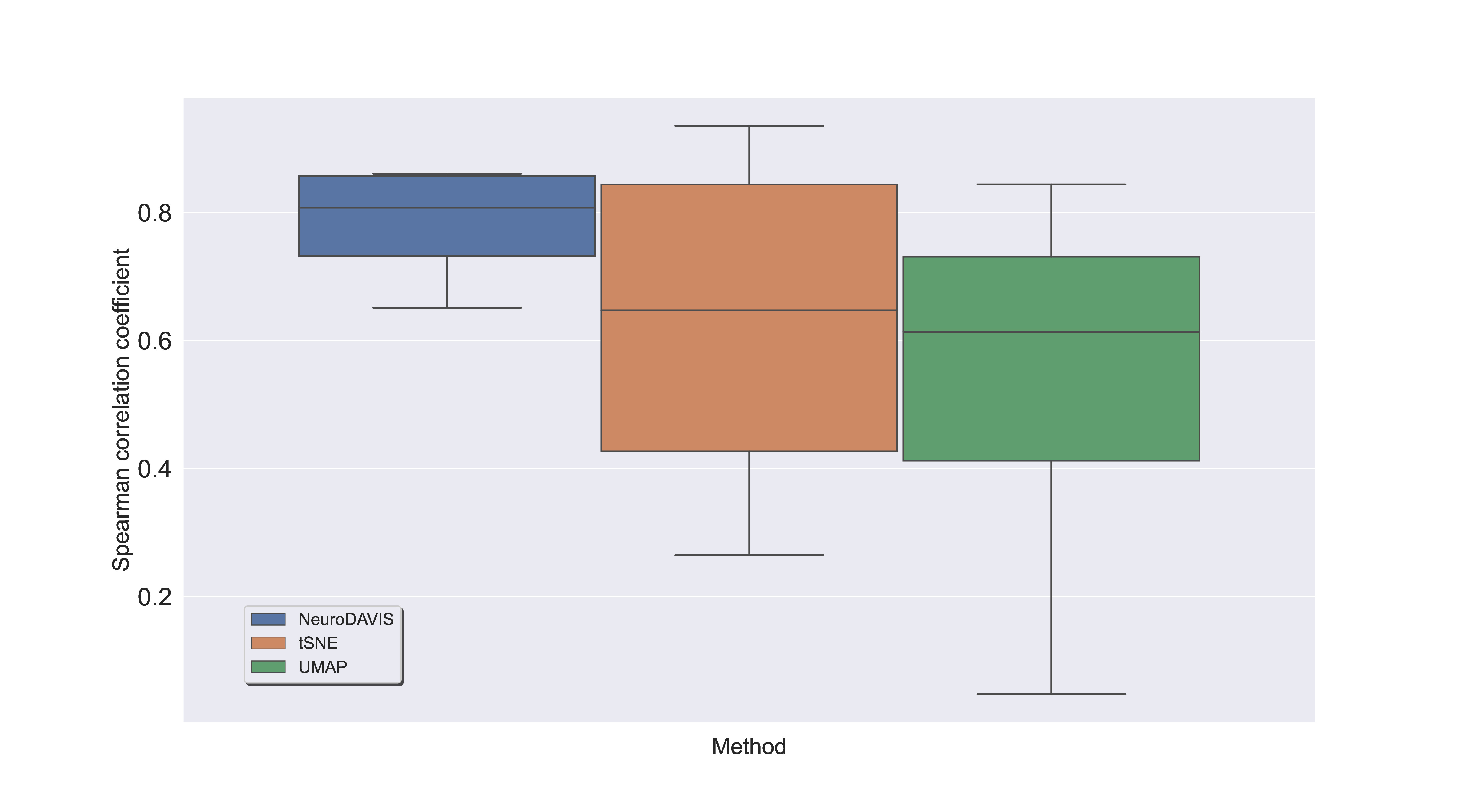}
			\label{fig:WorldMap_BP5}
		}
		\subfloat[Distances between all pairs of continents]{
			\centering
			\includegraphics[width=0.5\columnwidth,height=5cm]{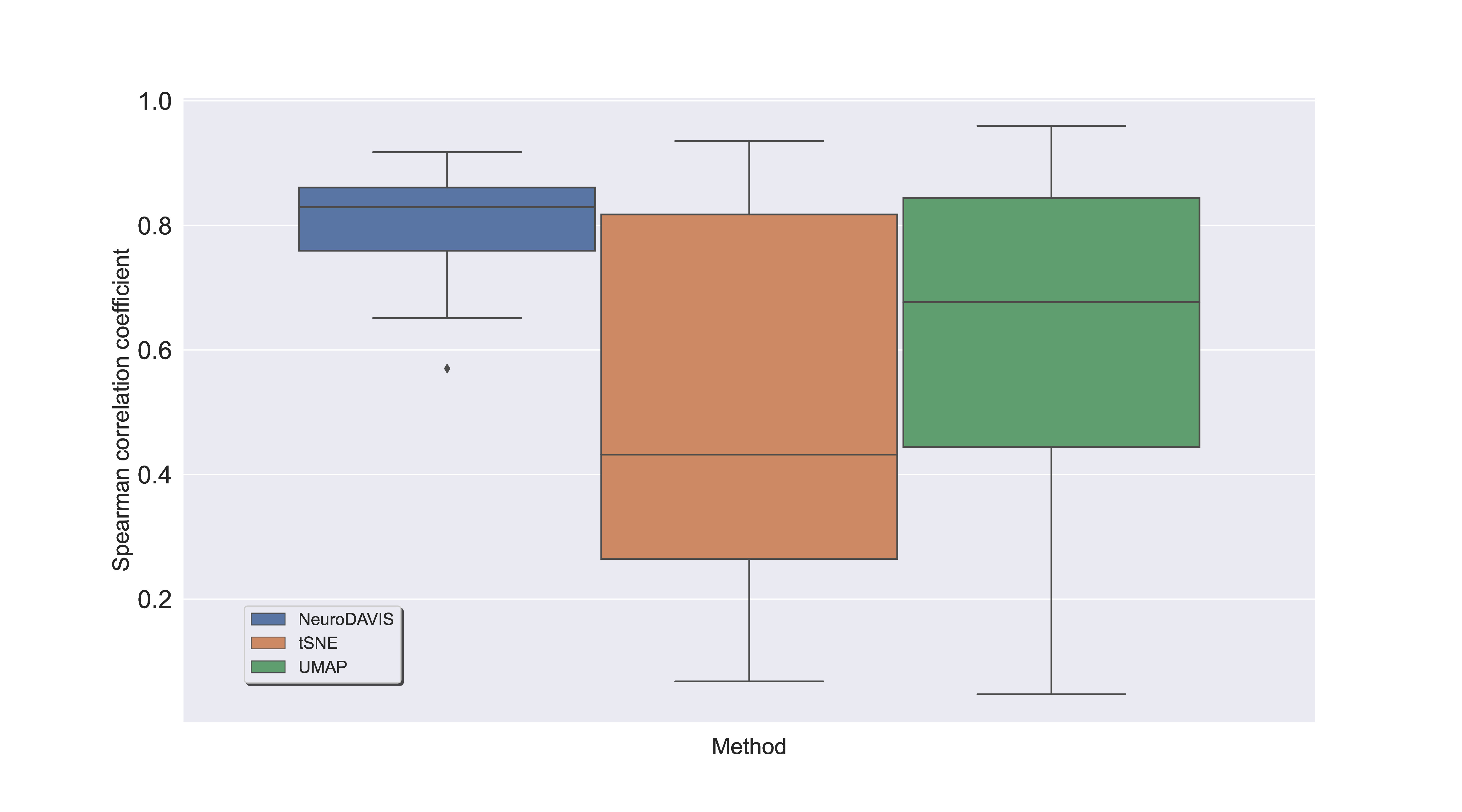}
			\label{fig:WorldMap_BP6}
		}
		\caption{Preservation of pairwise distances between clusters (continents) in World Map dataset by NeuroDAVIS, t-SNE and UMAP.}
		\label{fig:WorldMapBoxplots}
\end{figure}

We have further evaluated the NeuroDAVIS-embedding for World Map dataset for its ability to preserve pairwise euclidean distances between points in one cluster and points in the remaining clusters. Figure \ref{fig:WorldMapBoxplots} containing results from repeated runs of this experiment, shows that for each of the clusters Eurasia, Australia, North America, South America and Africa, NeuroDAVIS has been able to preserve the original pairwise distances better than t-SNE and UMAP.

\subsubsection{Cluster size preservation}
\label{sec:results_cluster_size}
We wondered whether the sizes of the clusters in the original data have been preserved in its NeuroDAVIS-generated embedding. Therefore, in order to ensure preservation of cluster sizes, we have carried out a few additional experiments on this World Map dataset. For each of the clusters in the dataset, we have estimated the area of the minimal rectangles bordering the clusters. This has been realized by considering the amount of spread of each cluster in either directions, as shown in Figure \ref{fig:WorldMap_Orig}. We have then computed the Pearson correlation coefficients between the original area of the clusters to their reconstructed counterparts. Repeating this experiment multiple times, we have arrived at the conclusion that NeuroDAVIS has been able to preserve original sizes of the clusters better than t-SNE and UMAP, as shown in Figure S2 (in Supplementary Material). All these experiments discussed above establish the superiority of NeuroDAVIS over t-SNE and UMAP in terms of global shape preservation.

\subsection{Performance evaluation on high-dimensional datasets}
\label{sec:results_hd}
Subsequently the effectiveness of NeuroDAVIS has been demonstrated on several categories of high-dimensional datasets including numeric, textual, image and biological data. The general steps used for evaluation are as follows. 

We have first reduced each dataset to two NeuroDAVIS dimensions. For each dataset, the quality of the NeuroDAVIS embedding has then been compared with that produced by each of t-SNE, UMAP and Fit-SNE, in two different aspects. First, we have calculated the Spearman correlation coefficient between the pairwise distances of the data points in original space and that in the latent space, produced by NeuroDAVIS, t-SNE, UMAP and Fit-SNE. Finally, for all the high dimensional datasets excluding the biological one, the embedding produced by NeuroDAVIS has been assessed for its classification performance. A train:test split in $80:20$ ratio has been used to generate training and test datasets from the NeuroDAVIS-generated embedding. Two classifiers, viz., k-nn and random forest (RF), have been trained and the held out test dataset has been used to evaluate the performance of the classifiers. The results have been compared with those obtained using the same classifiers on t-SNE, UMAP and Fit-SNE-embeddings. For the scRNA-seq (biological) dataset, we have performed cell-type clustering using k-means and hierarchical agglomerative clustering on the NeuroDAVIS-embedding, and compared the results with those obtained on t-SNE, UMAP and Fit-SNE embeddings. We have further compared the results obtained with two recent benchmarked methods, viz., PHATE \cite{moon2019visualizing} and IVIS \cite{szubert2019structure}, specially developed for scRNA-seq datasets. In order to measure the performance of NeuroDAVIS and the other methods, we have used some measures used in \cite{seal2022cassl}. Classification performance for the non-biological datasets has been measured using Accuracy and F1-scores, while two external indices Adjusted Rand Index (ARI) and Fowlkes Mallows Index (FMI) have been used to measure the quality of clusters produced in the case of scRNA-seq datasets.

\subsubsection{Numeric datasets}
\label{sec:results_numeric}
\begin{figure}
	\subfloat[NeuroDAVIS]{
		\centering
		\includegraphics[width=0.25\columnwidth,height=4cm]{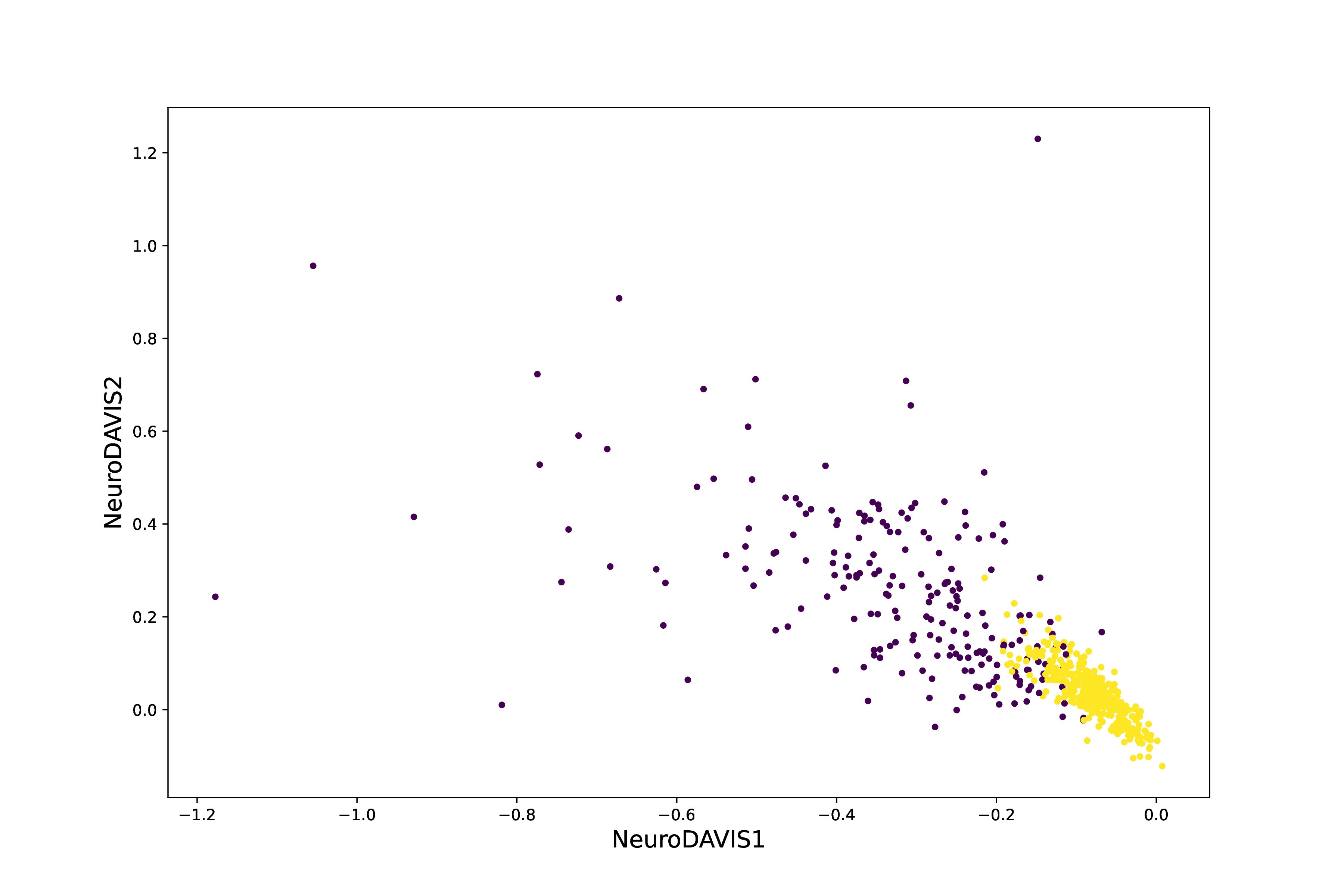}
		\label{fig:BreastCancer_NeuroDAVIS}
	} 
	\subfloat[t-SNE]{
		\centering
		\includegraphics[width=0.25\columnwidth,height=4cm]{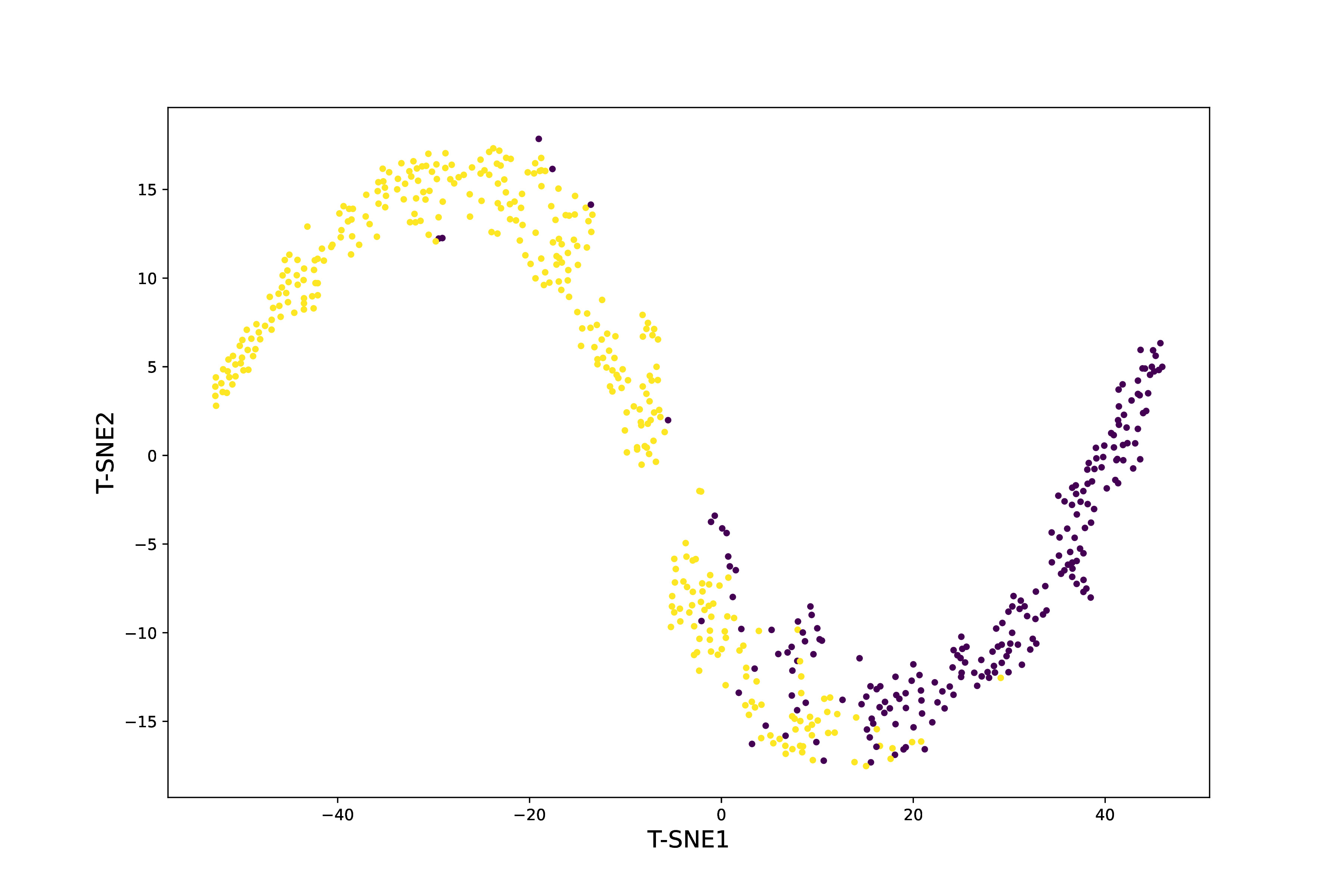}
		\label{fig:BreastCancer_tSNE}
	} 
	\subfloat[UMAP]{
		\centering
		\includegraphics[width=0.25\columnwidth,height=4cm]{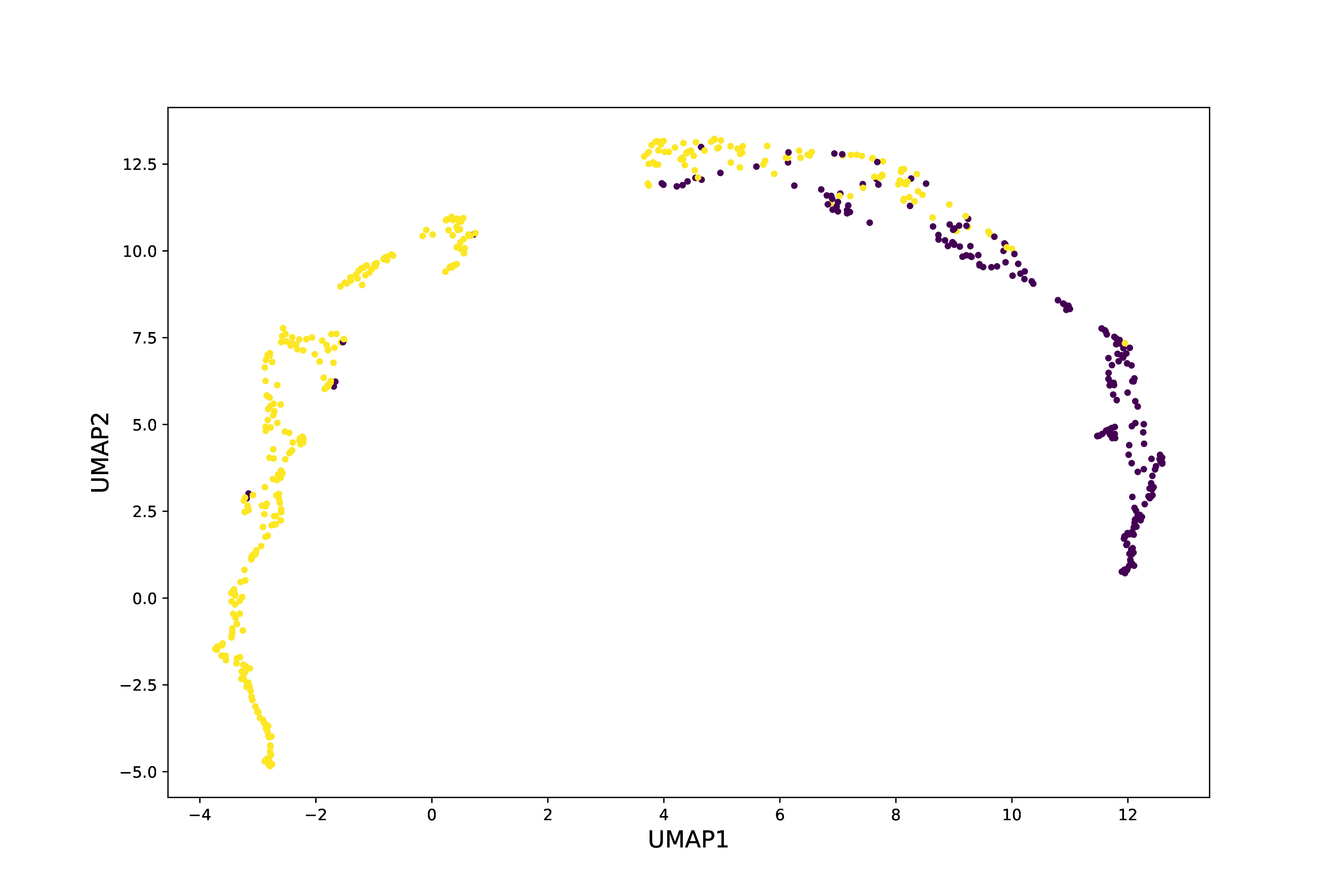}
		\label{fig:BreastCancer_UMAP}
	}
	\subfloat[Fit-SNE]{
		\centering
		\includegraphics[width=0.25\columnwidth,height=4cm]{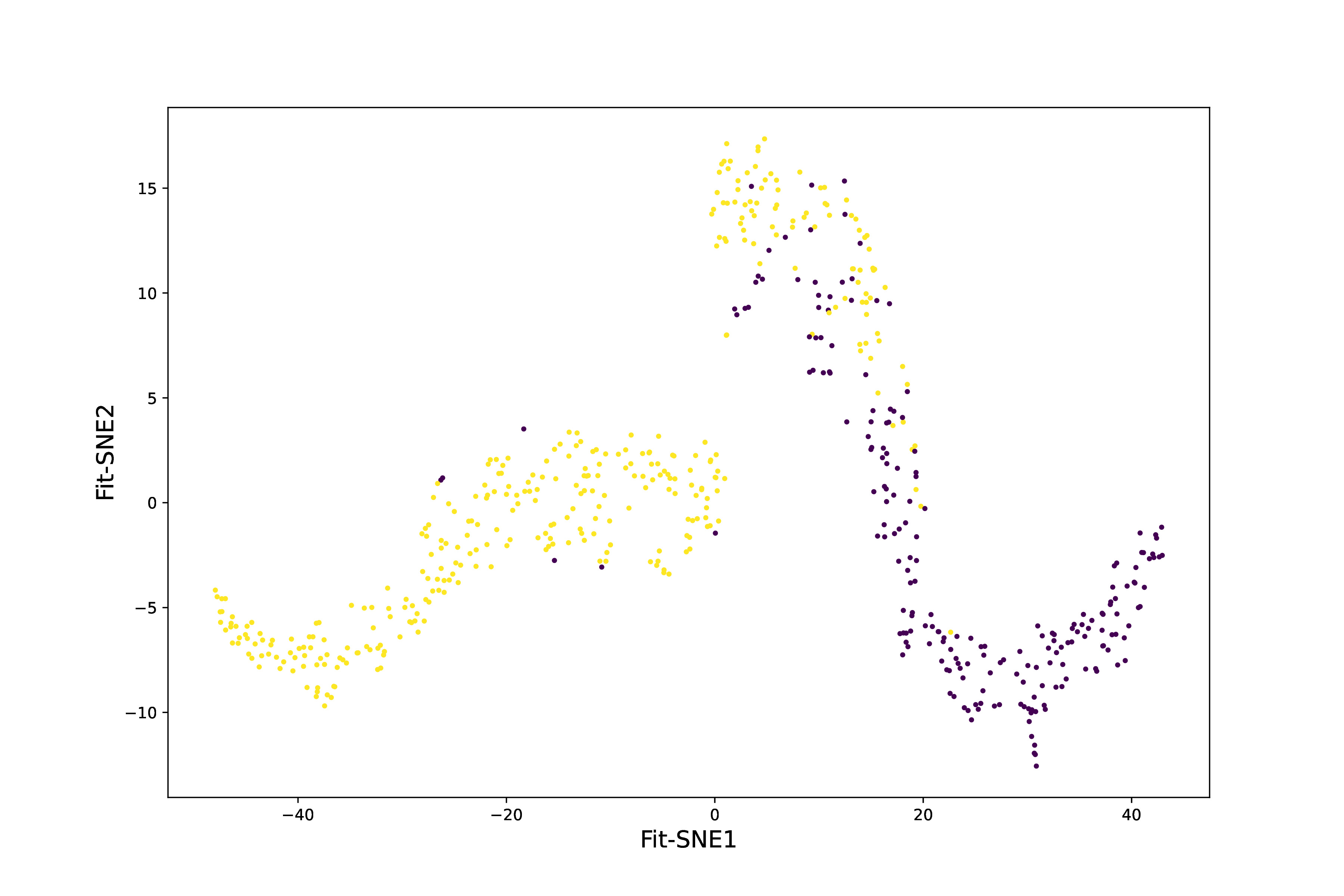}
		\label{fig:BreastCancer_fitsne}
	}
	\\
	\subfloat[NeuroDAVIS]{
		\centering
		\includegraphics[width=0.25\columnwidth,height=4cm]{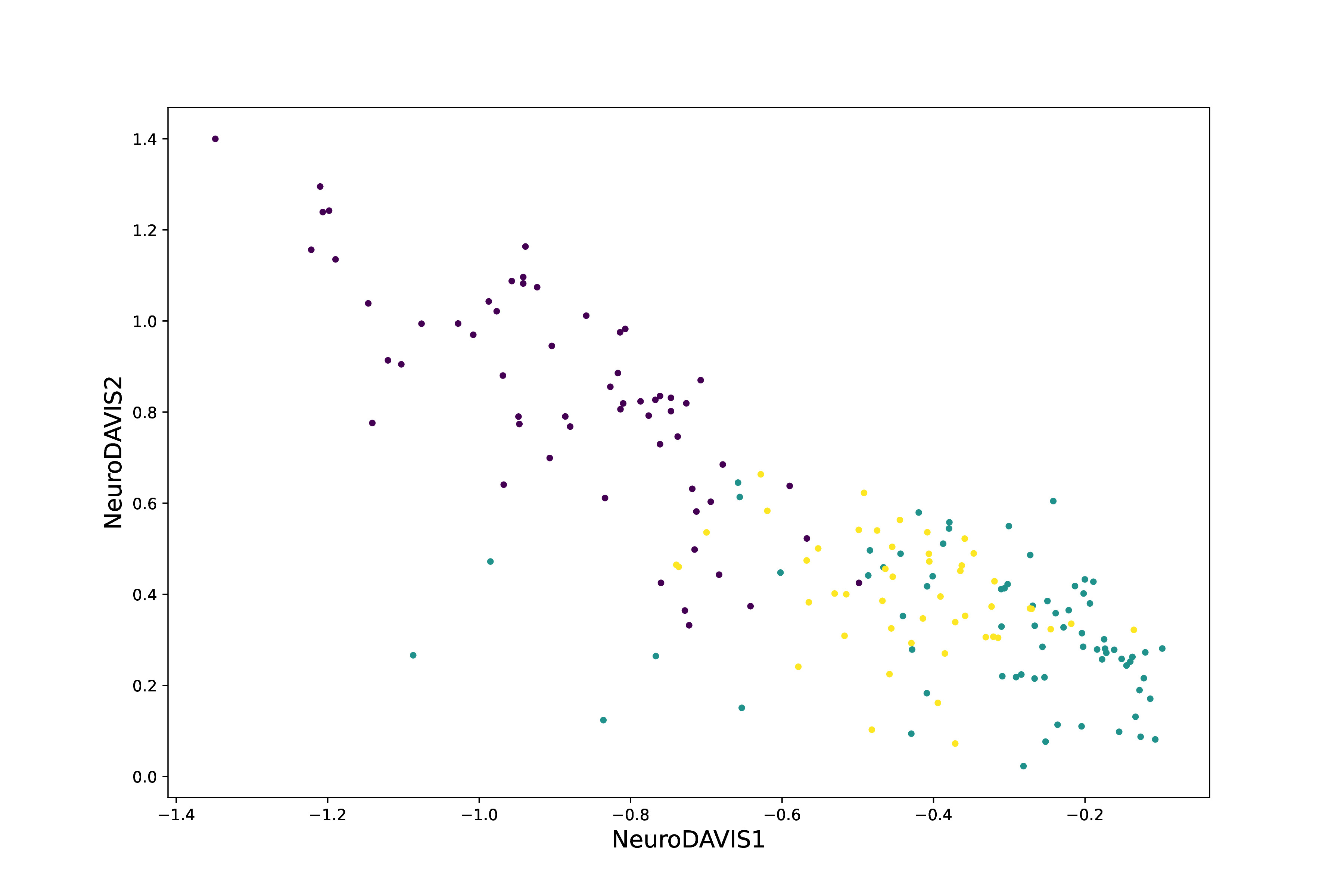}
		\label{fig:Wine_NeuroDAVIS}
	}
	\subfloat[t-SNE]{
		\centering
		\includegraphics[width=0.25\columnwidth,height=4cm]{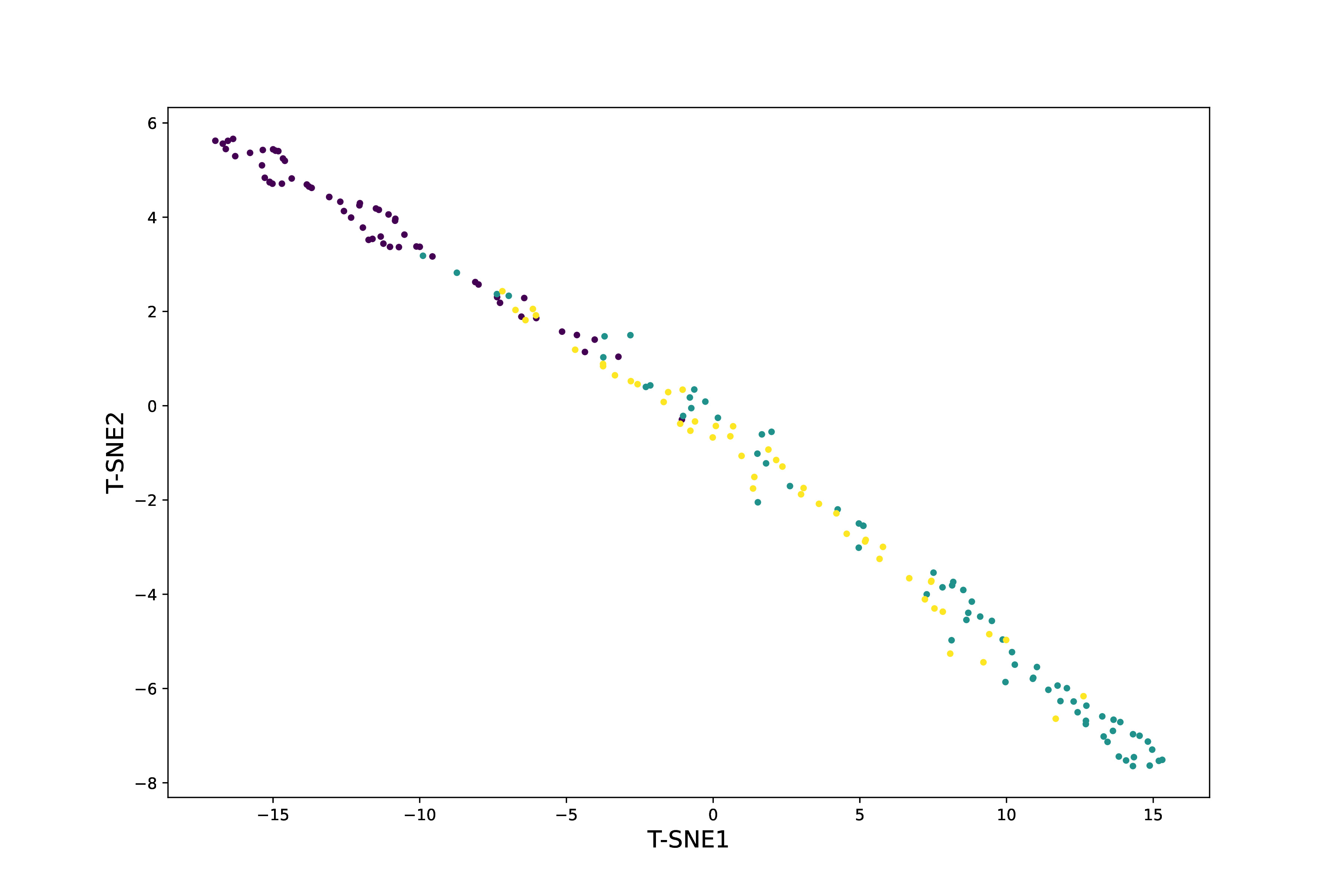}
		\label{fig:Wine_tSNE}
	}
	\subfloat[UMAP]{
		\centering
		\includegraphics[width=0.25\columnwidth,height=4cm]{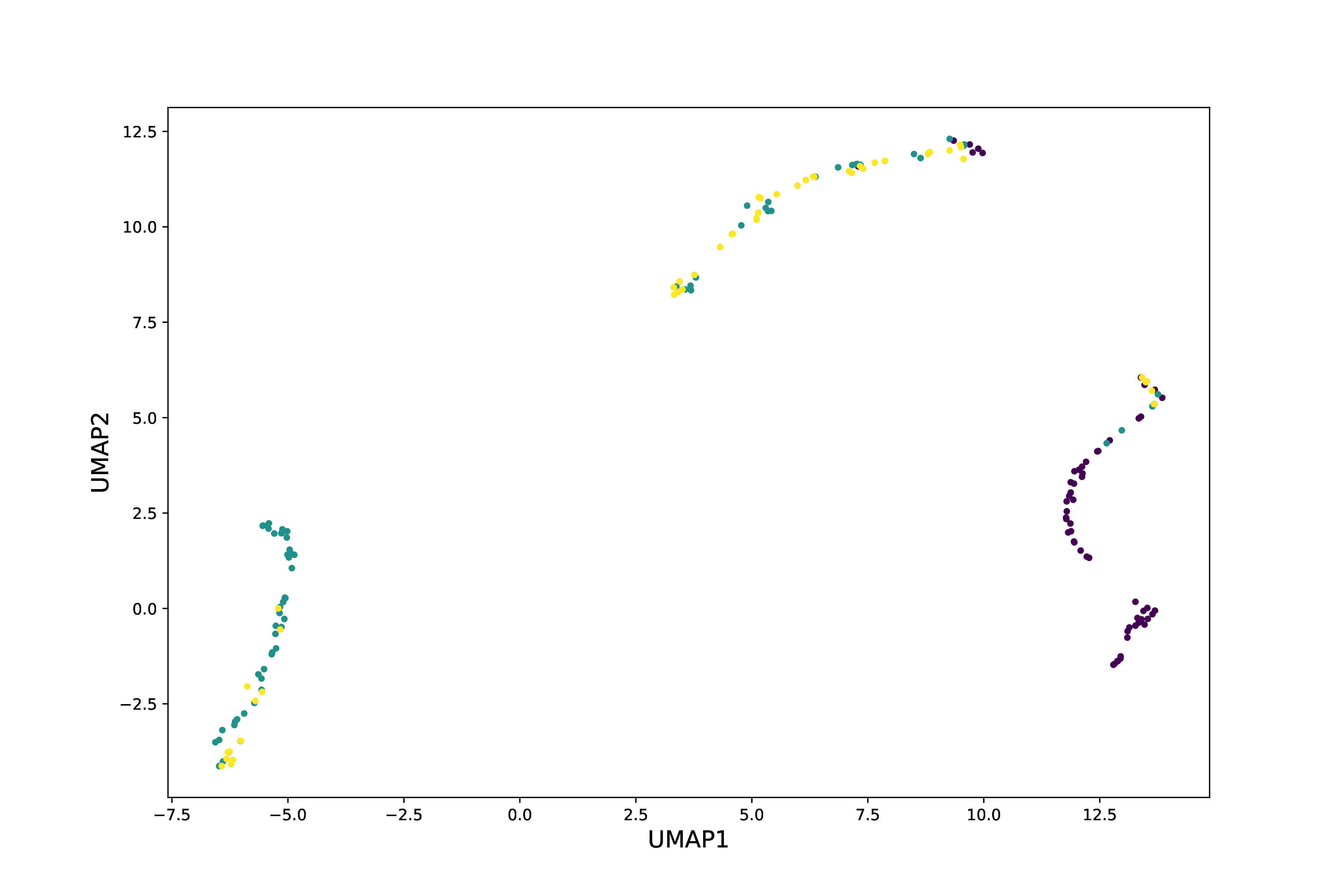}
		\label{fig:Wine_UMAP}
	}
	\subfloat[Fit-SNE]{
		\centering
		\includegraphics[width=0.25\columnwidth,height=4cm]{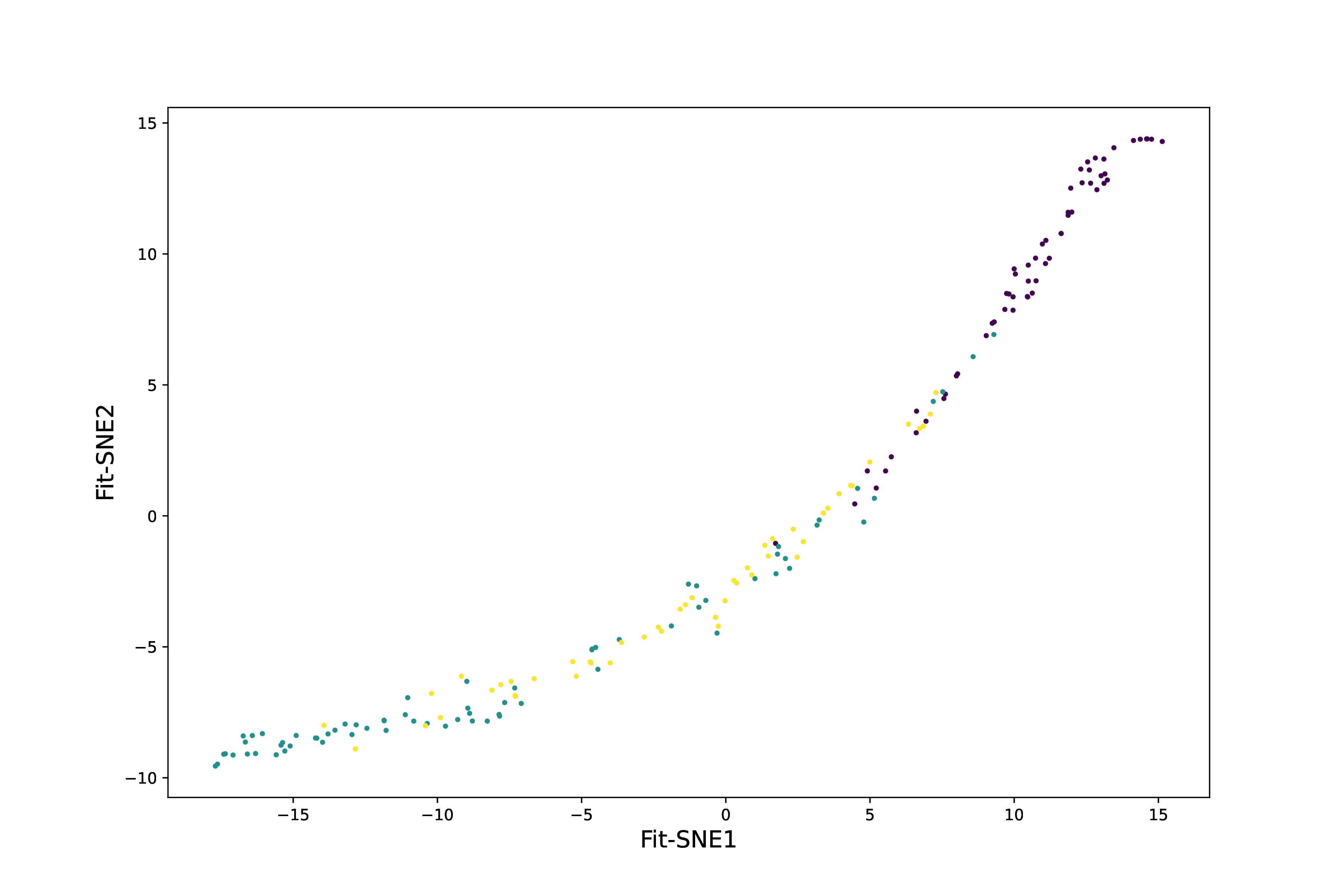}
		\label{fig:Wine_fitsne}
	}
	\\
	\subfloat[NeuroDAVIS]{
		\centering
		\includegraphics[width=0.25\columnwidth,height=4cm]{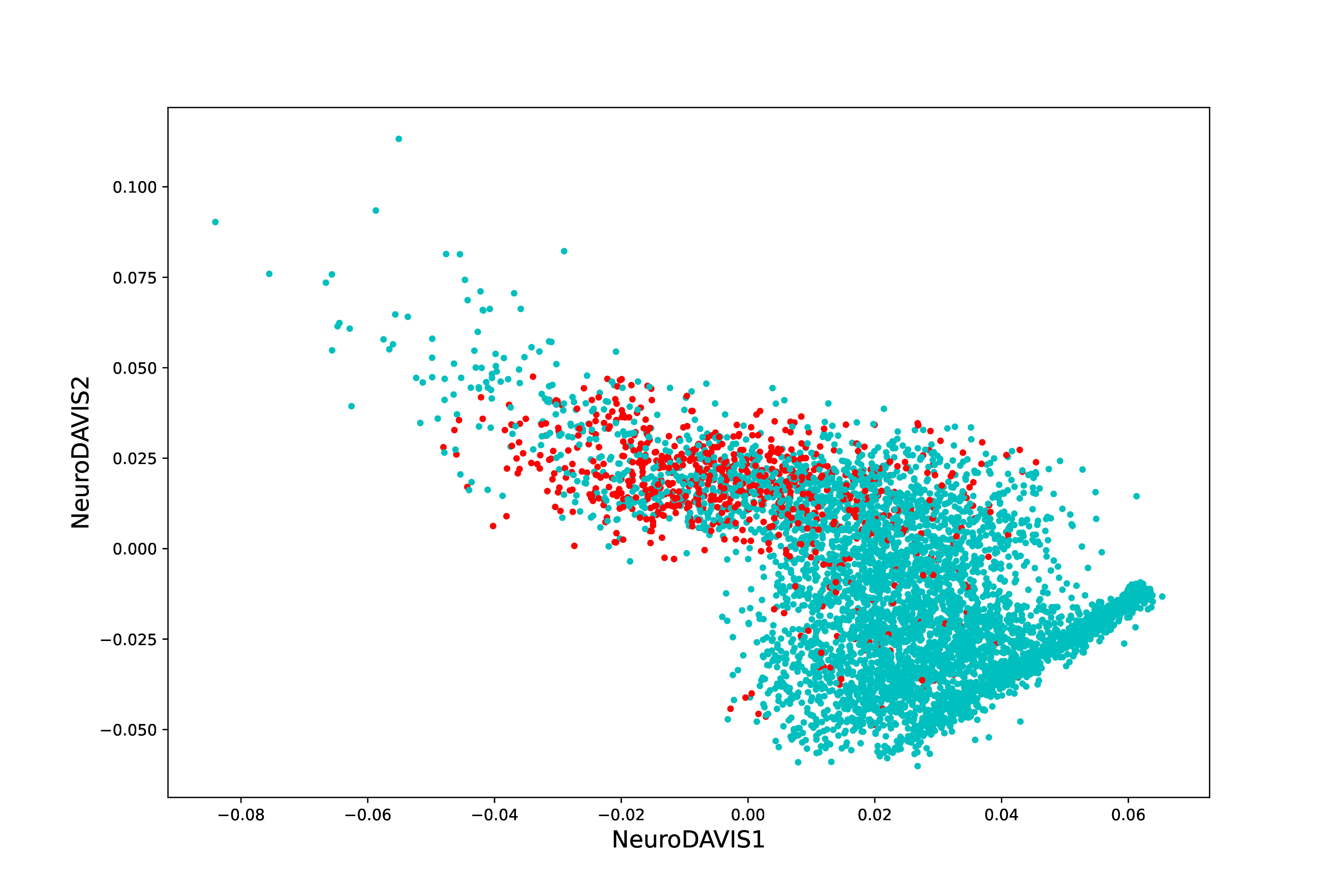}
		\label{fig:Spam_NeuroDAVIS}
	}
	\subfloat[t-SNE]{
		\centering
		\includegraphics[width=0.25\columnwidth,height=4cm]{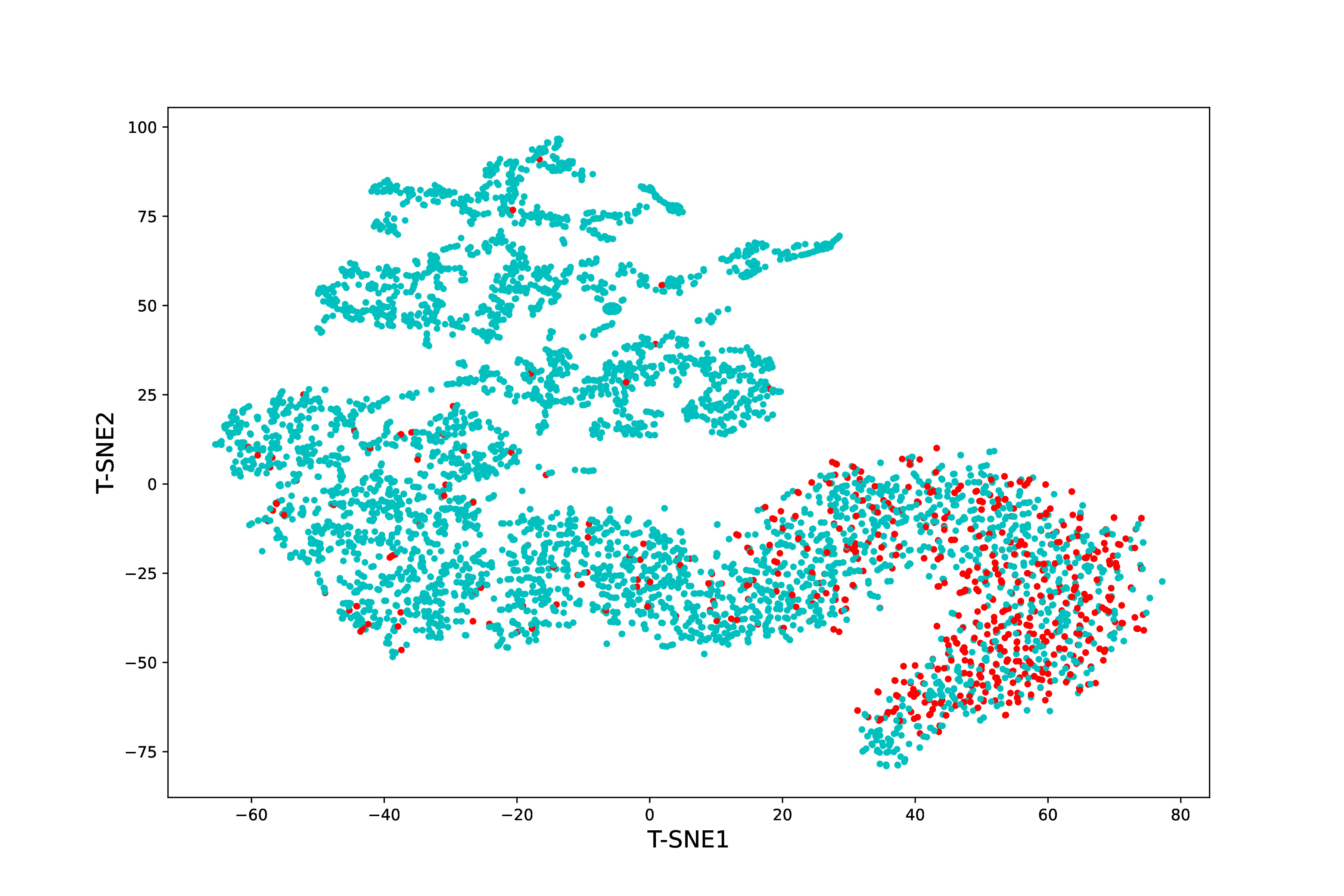}
		\label{fig:Spam_tSNE}
	}
	\subfloat[UMAP]{
		\centering
		\includegraphics[width=0.25\columnwidth,height=4cm]{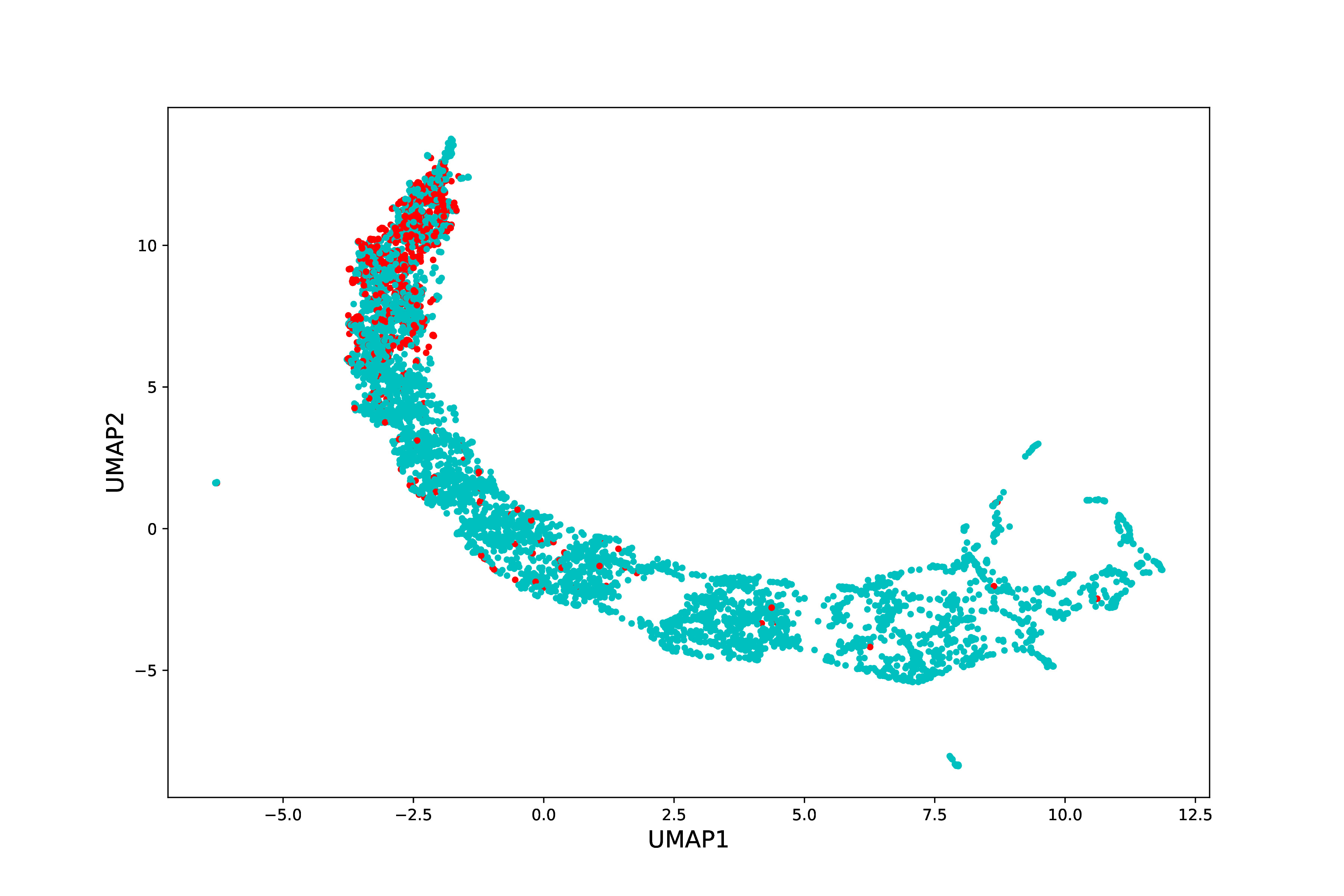}
		\label{fig:Spam_UMAP}
	}
	\subfloat[Fit-SNE]{
		\centering
		\includegraphics[width=0.25\columnwidth,height=4cm]{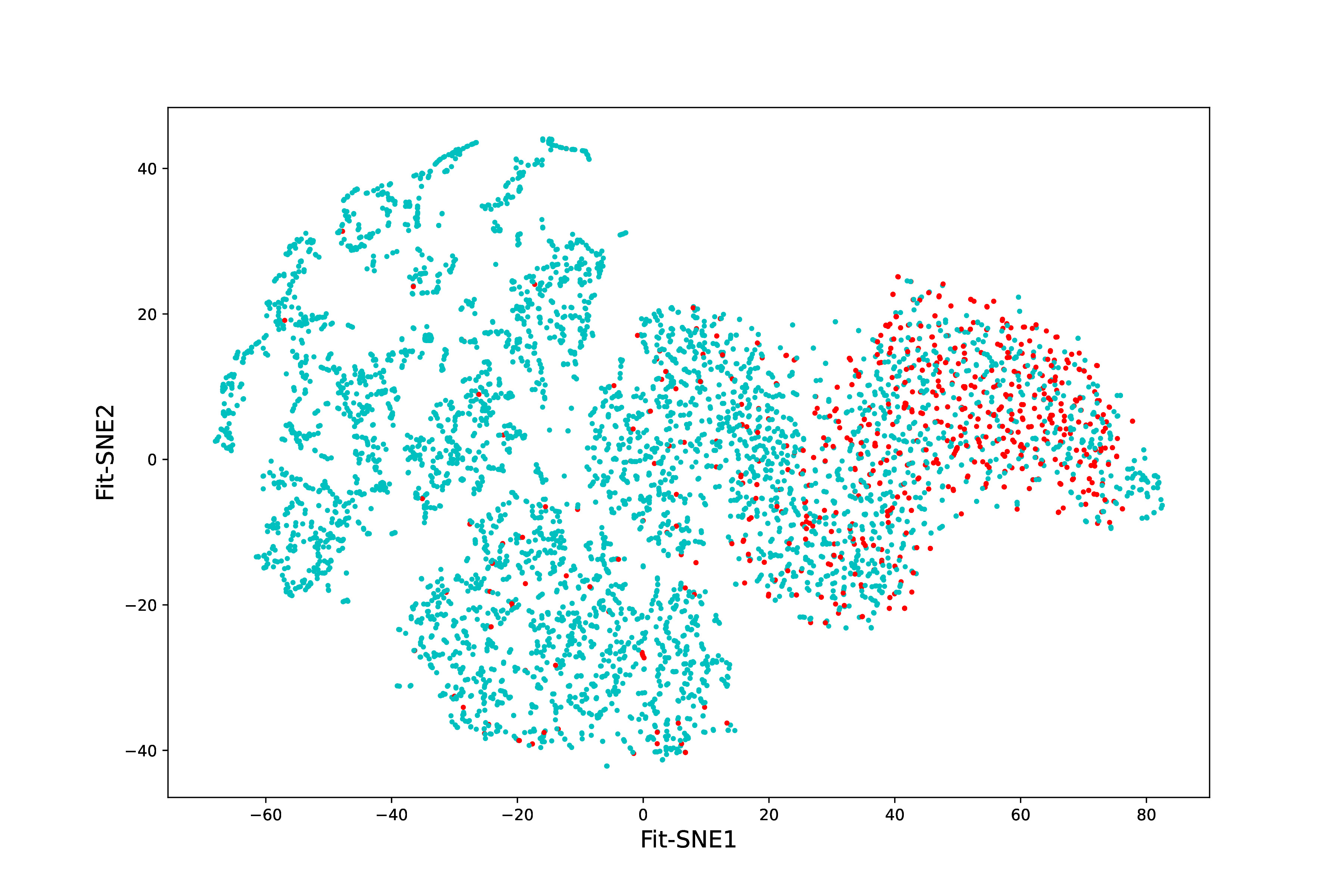}
		\label{fig:Spam_fitsne}
	}
	\\
	\subfloat[NeuroDAVIS]{
		\centering
		\includegraphics[width=0.25\columnwidth,height=4cm]{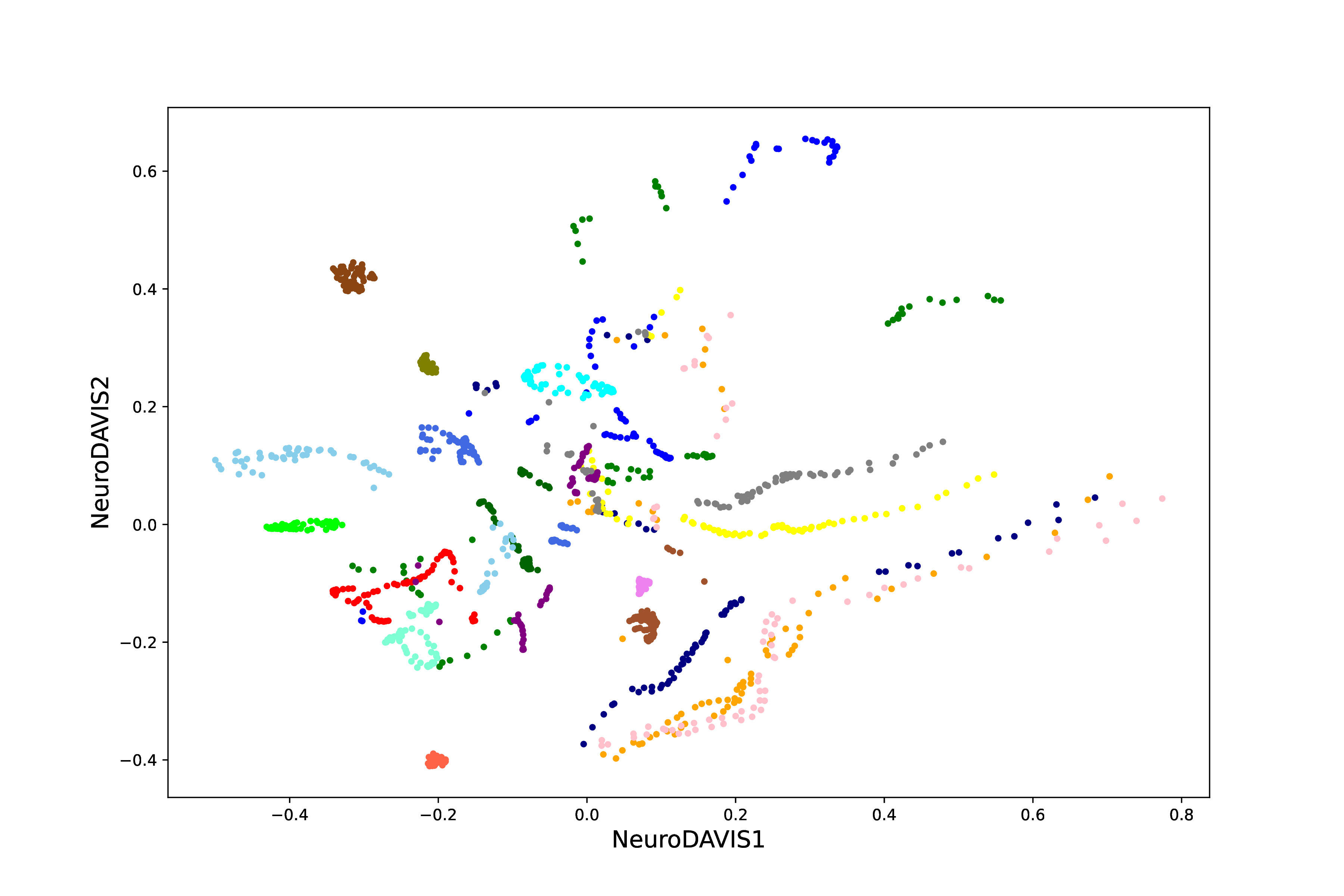}
		\label{fig:Coil20_NeuroDAVIS}
	}
	\subfloat[t-SNE]{
		\centering
		\includegraphics[width=0.25\columnwidth,height=4cm]{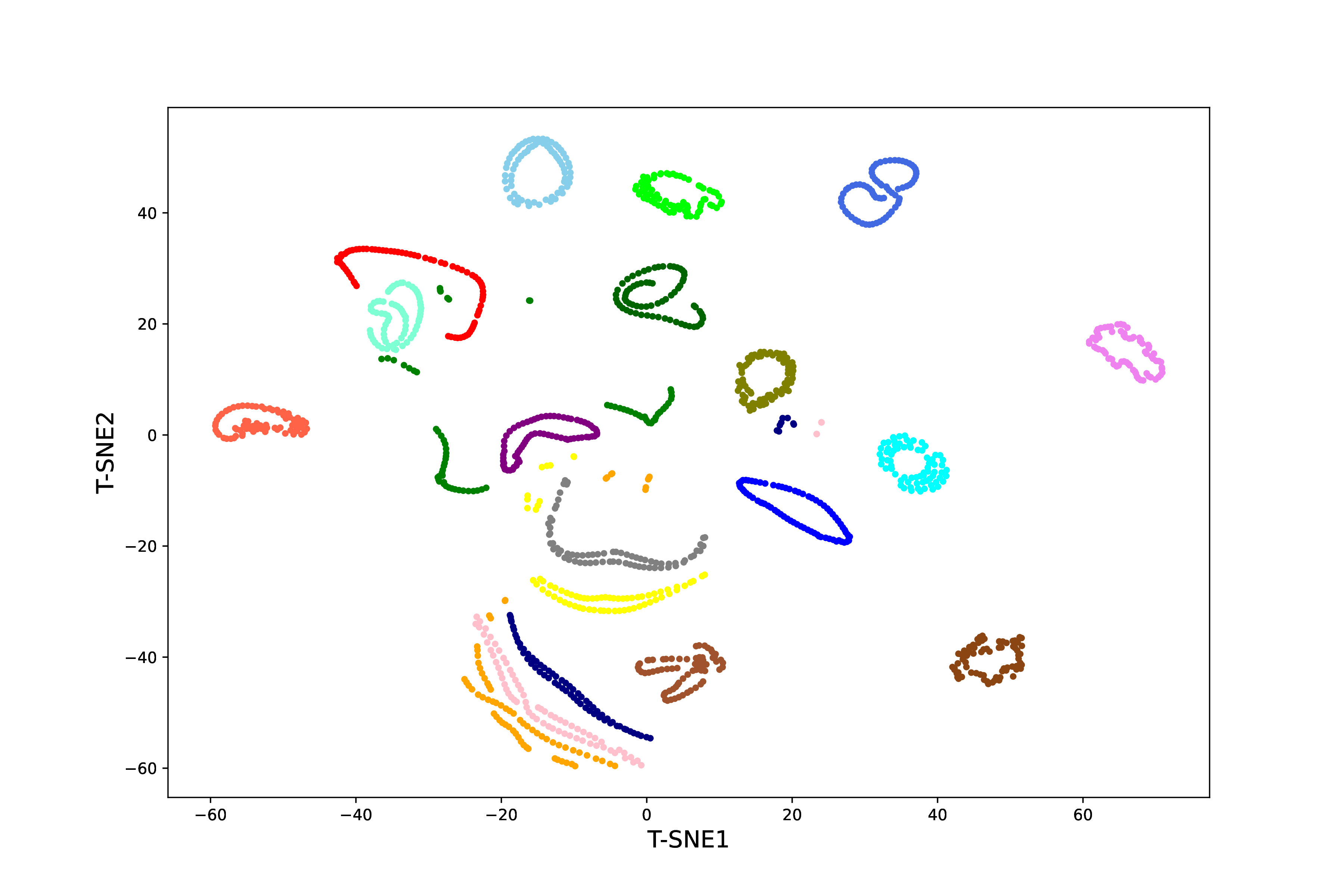}
		\label{fig:Coil20_tSNE}
	}
	\subfloat[UMAP]{
		\centering
		\includegraphics[width=0.25\columnwidth,height=4cm]{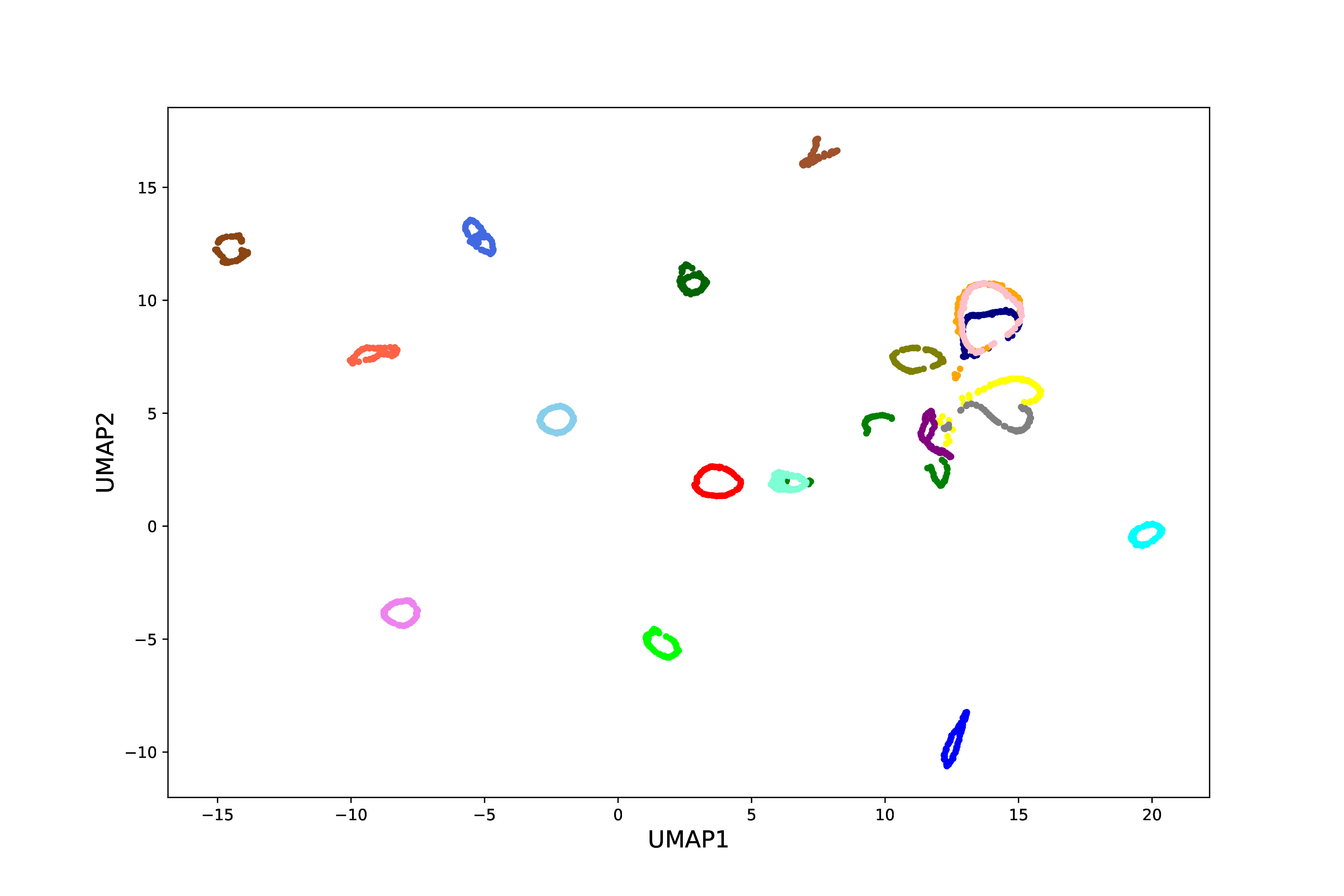}
		\label{fig:Coil20_UMAP}
	}
	\subfloat[Fit-SNE]{
		\centering
		\includegraphics[width=0.25\columnwidth,height=4cm]{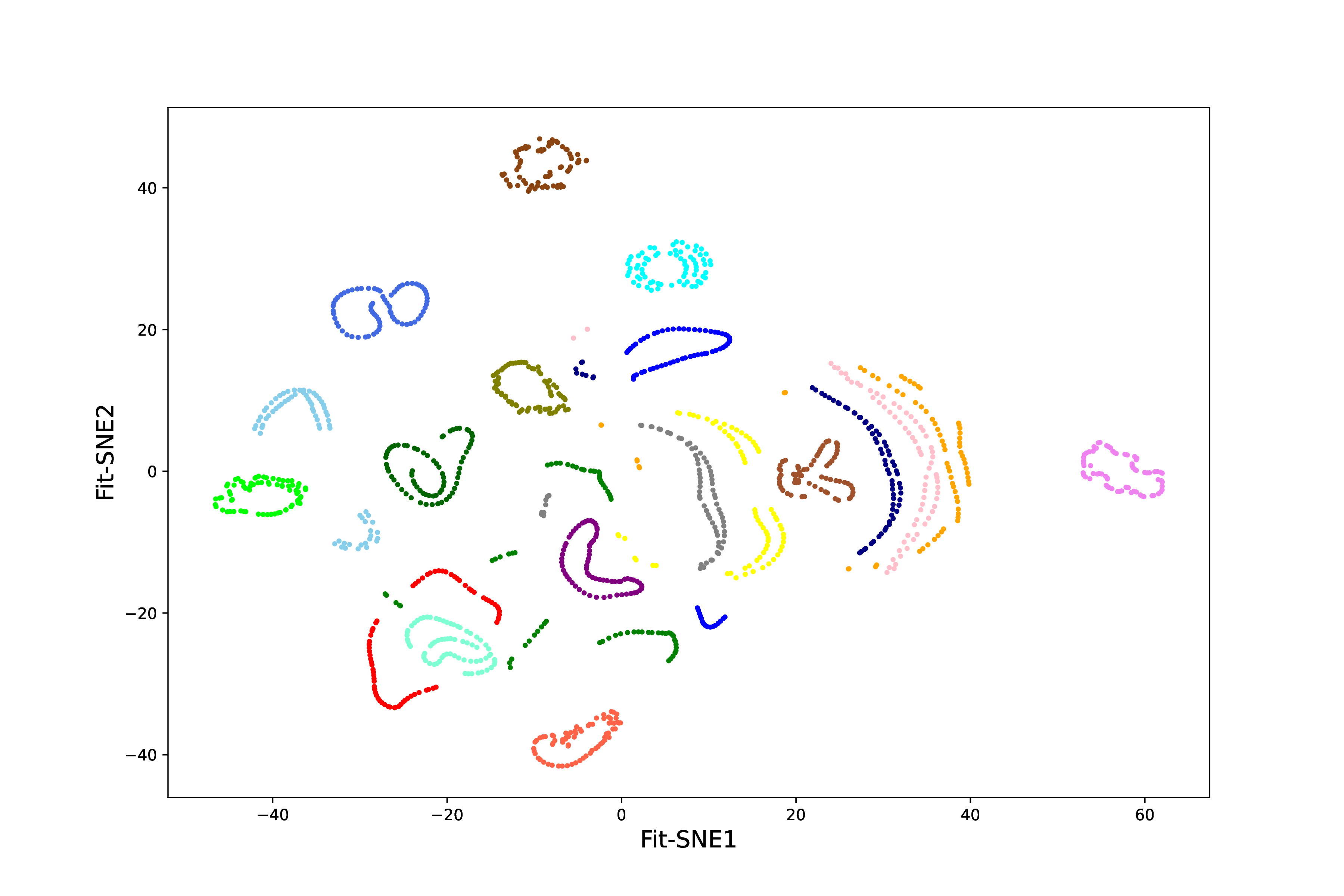}
		\label{fig:Coil20_fitsne}
	}
	\caption{Embeddings produced by NeuroDAVIS, t-SNE, UMAP and Fit-SNE for Breast cancer ((a)-(d)), Wine ((e)-(h)), Spam ((i)-(l)) and Coil20 ((m)-(p)) datasets}
	\label{fig:HD}
\end{figure}

\begin{figure}
	\centering
	\includegraphics[width=\columnwidth,height=8cm]{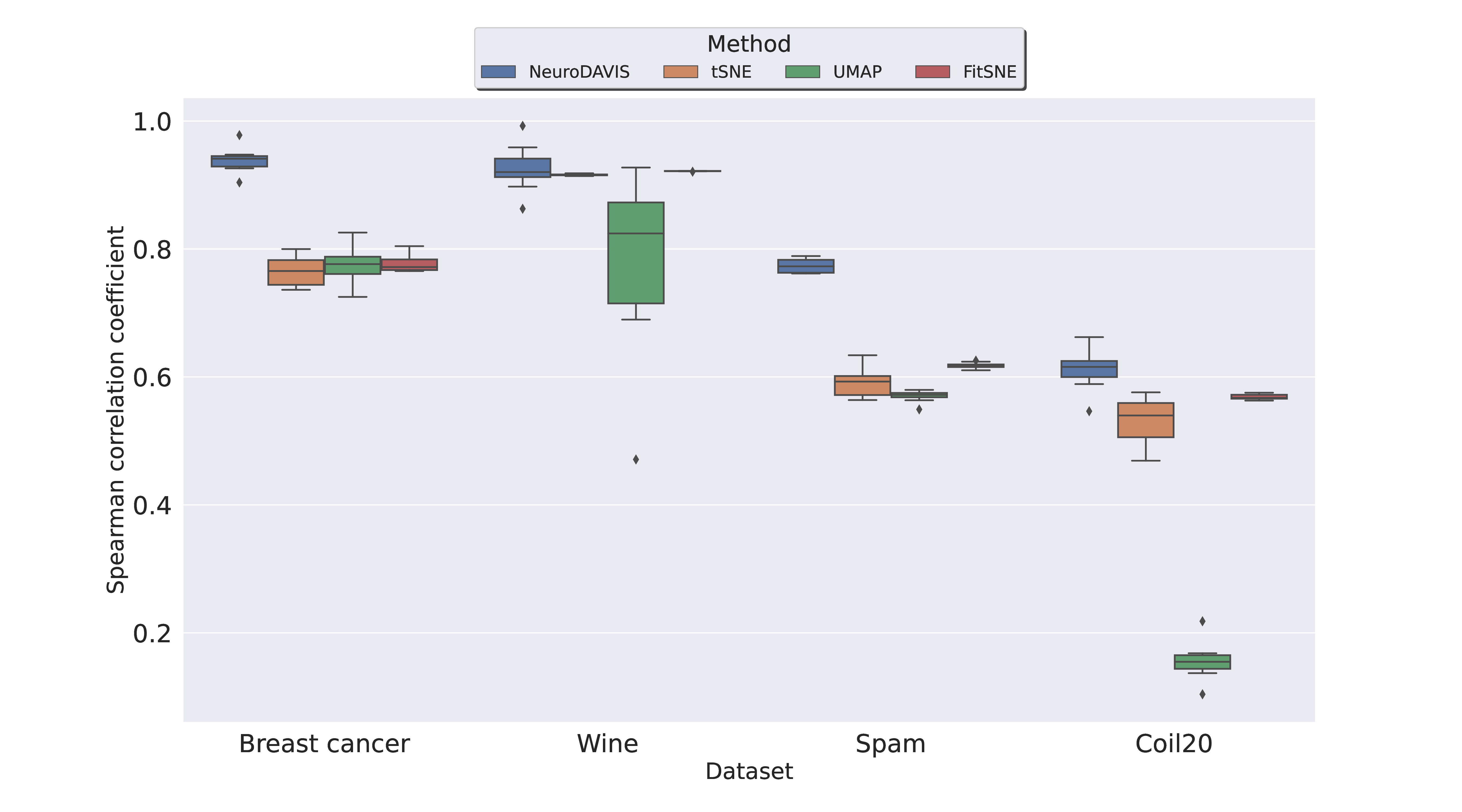}
	\caption{Spearman rank correlation coefficient between pairwise distances in original distribution, and pairwise distances in NeuroDAVIS, t-SNE, UMAP and Fit-SNE embeddings of Breast Cancer, Wine, Spam and Coil20 datasets. For Breast Cancer dataset, median correlation coefficient values obtained are $0.94$ (NeuroDAVIS), $0.76$ (t-SNE), $0.77$ (UMAP) and $0.77$ (Fit-SNE). For Wine dataset, median correlation coefficient values obtained are $0.92$ (NeuroDAVIS), $0.91$ (t-SNE), $0.82$ (UMAP) and $0.92$ (Fit-SNE). For Spam dataset, median correlation coefficient values obtained are $0.77$ (NeuroDAVIS), $0.59$ (t-SNE), $0.57$ (UMAP) and $0.61$ (Fit-SNE), while for Coil20 dataset, these values are $0.61$ (NeuroDAVIS), $0.53$ (t-SNE), $0.15$ (UMAP) and $0.56$ (Fit-SNE). The corresponding p-values  obtained using Mann-Whitney U test are $0.0001$ (both NeuroDAVIS-t-SNE and NeuroDAVIS-UMAP) for both Breast Cancer and Spam datasets. For Wine dataset, p-values obtained using Mann-Whitney U test are $0.73$ (NeuroDAVIS-t-SNE), $0.002$ (NeuroDAVIS-UMAP) and $1.00$ (NeuroDAVIS-Fit-SNE), while for Coil20 dataset, p-values obtained using Mann-Whitney U test are $0.0007$ (NeuroDAVIS-t-SNE), $0.0001$ (NeuroDAVIS-UMAP) and $0.002$ (NeuroDAVIS-Fit-SNE).}
	\label{fig:HD_Spearman}
\end{figure}

Two numeric datasets Breast cancer and Wine, originally available at \url{https://archive.ics.uci.edu/ml/datasets.php}, have been downloaded from \cite{scikit-learn} in order to evaluate the performance of NeuroDAVIS. The NeuroDAVIS-generated embedding as compared to those obtained using t-SNE, UMAP and Fit-SNE for Breast cancer dataset are shown in Figures \ref{fig:BreastCancer_NeuroDAVIS},  \ref{fig:BreastCancer_tSNE}, \ref{fig:BreastCancer_UMAP} and \ref{fig:BreastCancer_fitsne}, while Figures \ref{fig:Wine_NeuroDAVIS}, \ref{fig:Wine_tSNE}, \ref{fig:Wine_UMAP} and \ref{fig:Wine_fitsne} show similar embeddings for Wine dataset. The correlation coefficient values, as shown in Figure \ref{fig:HD_Spearman}, reveal that pairwise distances in high dimensions are more correlated with those in NeuroDAVIS embedding (median correlation coefficient = $0.94$ (Breast cancer), $0.92$ (Wine)) than those in t-SNE (median correlation coefficient = $0.76$ (Breast cancer), $0.91$ (Wine)), UMAP (median correlation coefficient = $0.77$ (Breast cancer), $0.82$ (wine)) and Fit-SNE embedding (median correlation coefficient = $0.77$ (Breast cancer), $0.92$ (wine)). The corresponding p-values obtained using Mann-Whitney U test are $0.0001$ (for all pairs, viz., NeuroDAVIS-t-SNE, NeuroDAVIS-UMAP and NeuroDAVIS-Fit-SNE) for Breast cancer dataset, and $0.0001$ (NeuroDAVIS-UMAP), $0.733$ (NeuroDAVIS-t-SNE) and $1.000$ (NeuroDAVIS-Fit-SNE) for Wine dataset.

The results for classification (accuracy and F1-score) using k-nn and RF, on the NeuroDAVIS-embedding, as shown in Figure S3 (in Supplementary Material), also reflect superior performance of NeuroDAVIS over t-SNE,  UMAP and Fit-SNE for Wine dataset, while for Breast cancer dataset, NeuroDAVIS-embedding has displayed comparable classification performance to both t-SNE, UMAP and Fit-SNE embeddings. Results have been recorded by repeating each experiment multiple times.

\subsubsection{Textual dataset}
\label{sec:results_textual}
Next, we have used a textual dataset to demonstrate the effectiveness of NeuroDAVIS. Spam dataset, originally available at \url{https://archive.ics.uci.edu/ml/datasets.php}, has been downloaded from \url{https://www.kaggle.com/datasets/team-ai/spam-text-message-classification}. Out of $5572$ messages in this dataset, $747$ messages are classified as spam messages. This dataset has undergone preprocessing using standard pipelines used for textual datasets, which include the following major steps:
\begin{enumerate*}[series = tobecont, itemjoin = \quad]
	\item Removal of URLs, \item Conversion into lower case, \item Removal of punctuations, \item Removal of extra whitespaces, \item Removal of stopwords, \item Lemmatization, \item Tokenization, and \item min-max scaling.
\end{enumerate*}

The two-dimensional embedding obtained by NeuroDAVIS on the preprocessed Spam dataset has been shown in Figure \ref{fig:Spam_NeuroDAVIS}. The corresponding t-SNE, UMAP and Fit-SNE embeddings on the same dataset have been shown in Figures \ref{fig:Spam_tSNE}, \ref{fig:Spam_UMAP} and \ref{fig:Spam_fitsne} respectively. Although the embeddings look quite similar to each other, on close observation, it can be seen that quite contrary to NeuroDAVIS and UMAP, the ham (not spam) cluster in both the t-SNE and Fit-SNE embedding portray multiple sub-clusters, which is unrealistic. Figure \ref{fig:HD_Spearman} also reveals that the correlation coefficient values for the NeuroDAVIS embedding (median correlation coefficient = $0.77$) is far better than those obtained for the t-SNE (median correlation coefficient = $0.59$), UMAP (median correlation coefficient = $0.57$) and Fit-SNE (median correlation coefficient = $0.61$) embeddings. The correponding p-value obtained using Mann–Whitney U test is $0.0001$ (for NeuroDAVIS-t-SNE, NeuroDAVIS-UMAP and NeuroDAVIS-Fit-SNE). The classification performance on NeuroDAVIS embedding of Spam dataset is, however, not as good as that on the t-SNE, UMAP or Fit-SNE embeddings, as shown in Figure S3 (in Supplementary Material).

\subsubsection{Image datasets}

\begin{figure}
	\subfloat[NeuroDAVIS embedding of original data]{
		\centering
		\includegraphics[width=0.5\columnwidth,height=4cm]{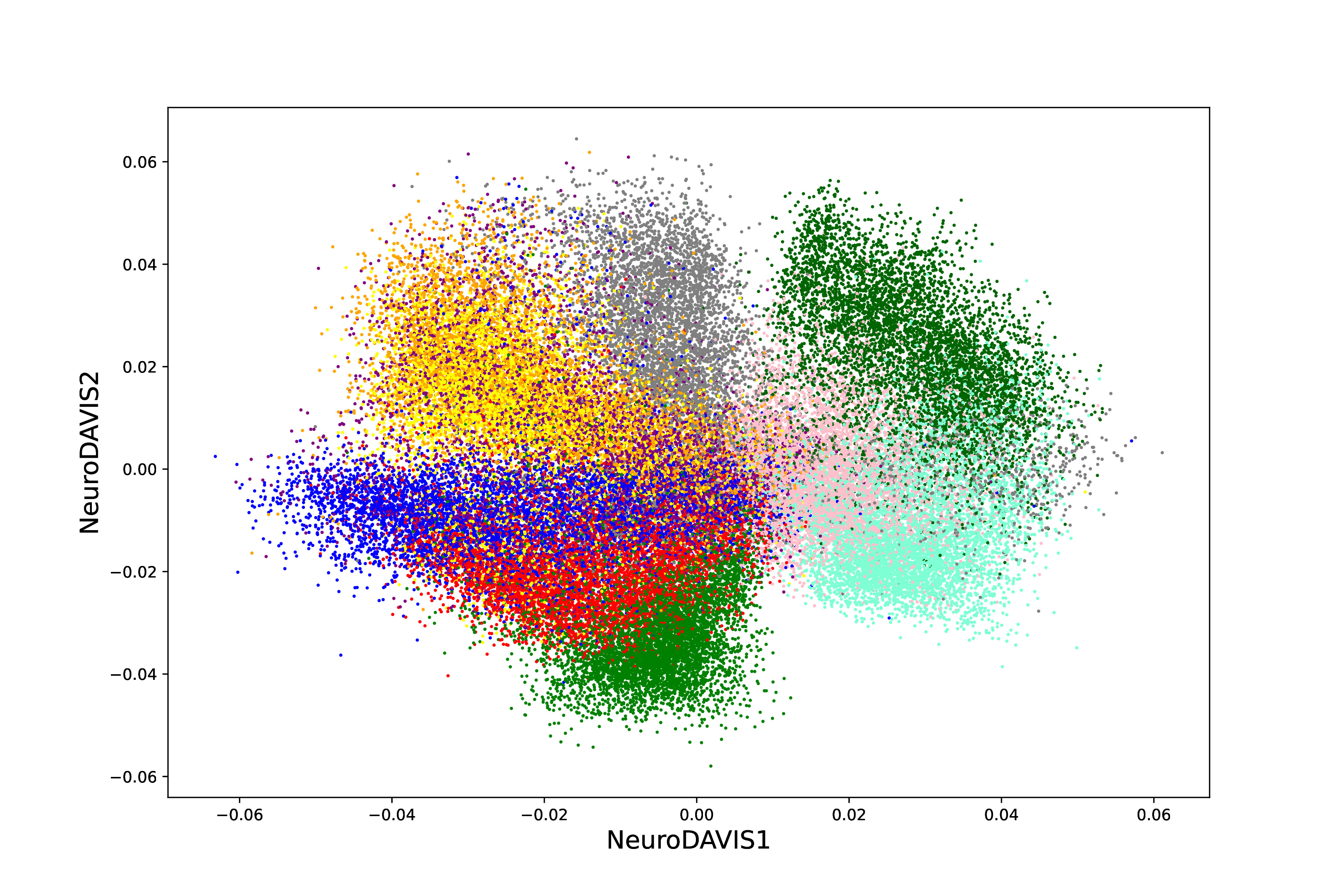}
		\label{fig:FMnist_NeuroDAVIS_60k}
	}
	\subfloat[PCA prior to NeuroDAVIS]{
		\centering
		\includegraphics[width=0.5\columnwidth,height=4cm]{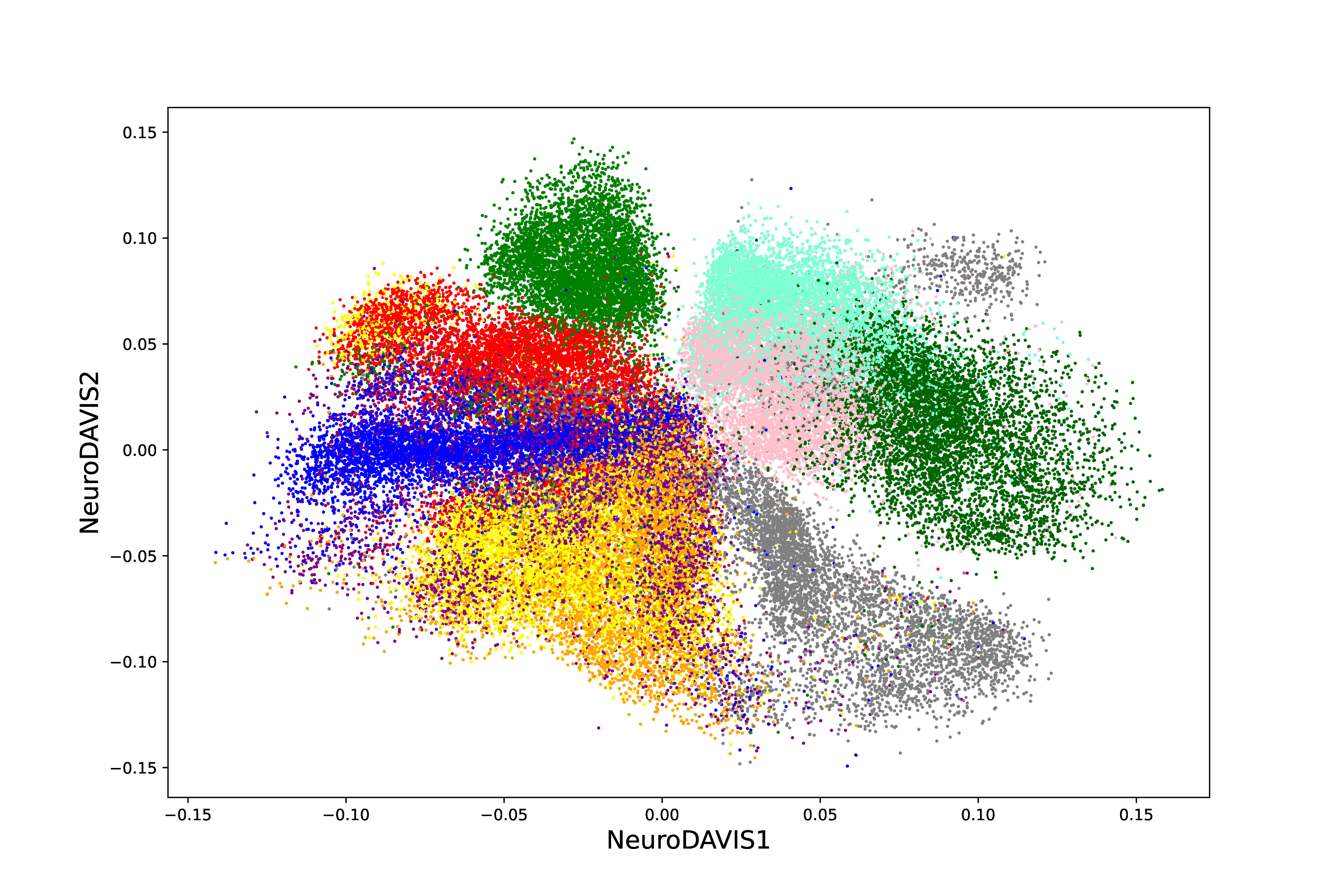}
		\label{fig:FMnist_pca_NeuroDAVIS_60k}
	}\\
	\subfloat[t-SNE embedding of original data]{
		\centering
		\includegraphics[width=0.5\columnwidth,height=4cm]{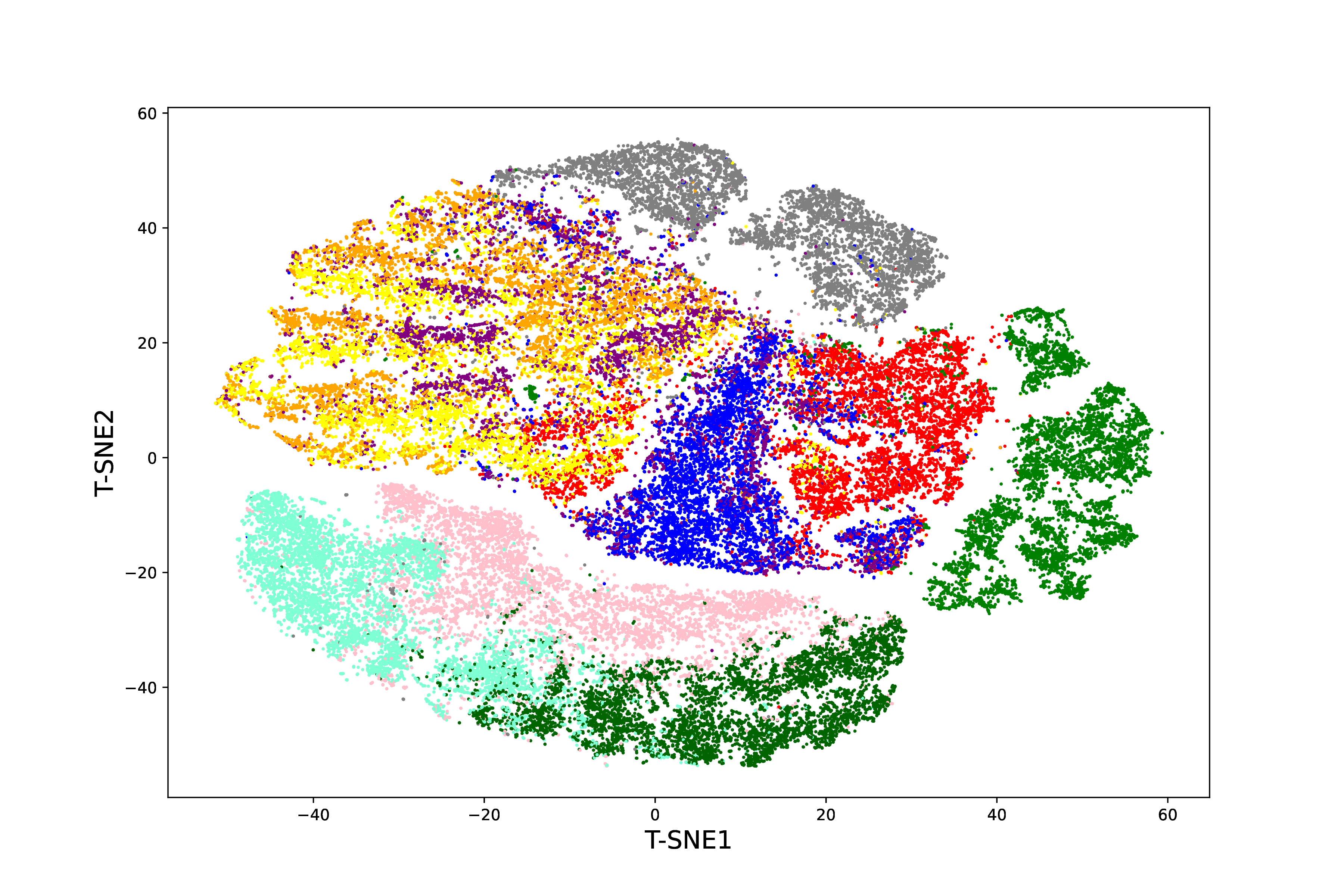}
		\label{fig:FMnist_tSNE_60k}
	}
	\subfloat[PCA prior to t-SNE ]{
		\centering
		\includegraphics[width=0.5\columnwidth,height=4cm]{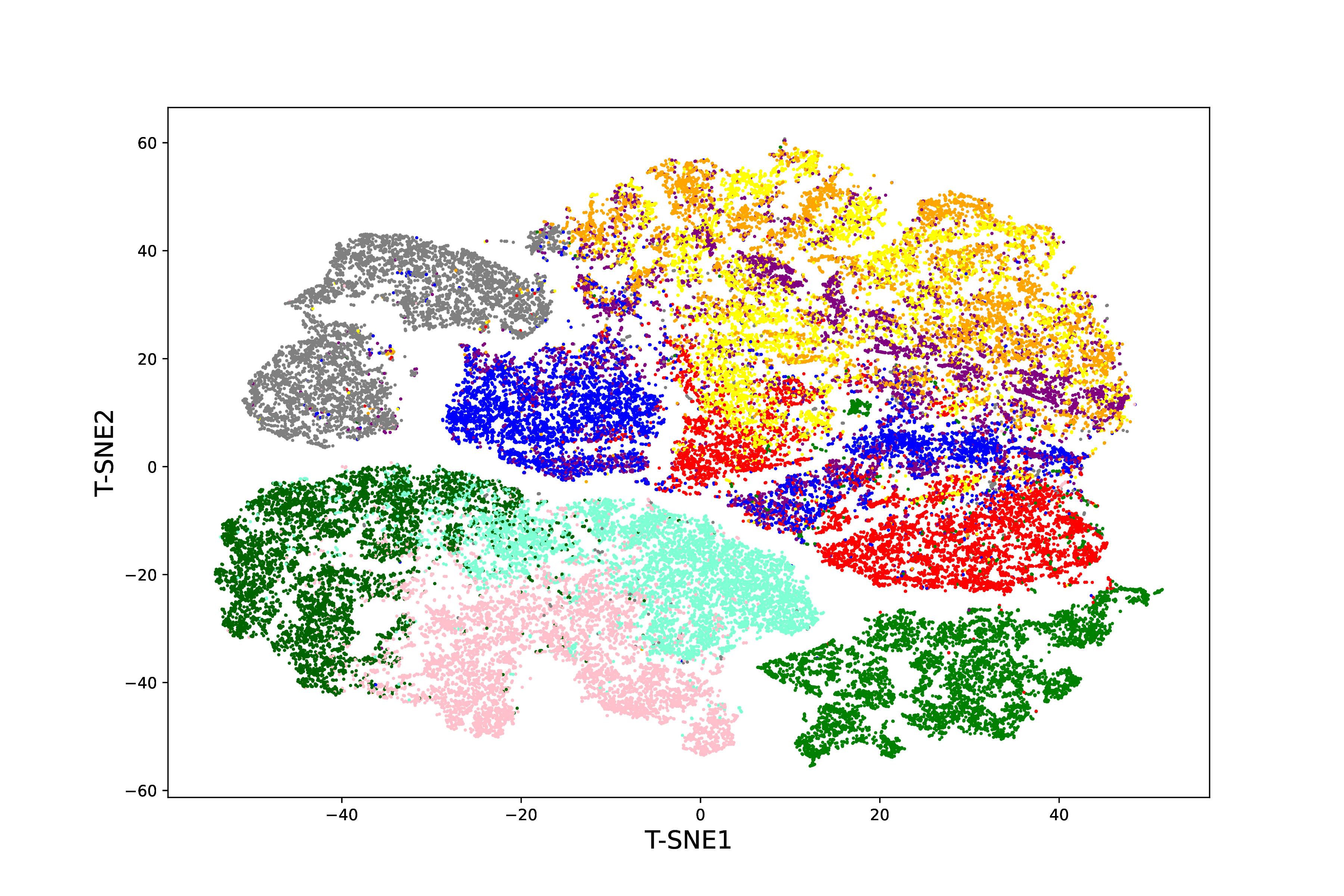}
		\label{fig:FMnist_pca_tSNE_60k}
	}\\
	\subfloat[UMAP embedding of original data]{
		\centering
		\includegraphics[width=0.5\columnwidth,height=4cm]{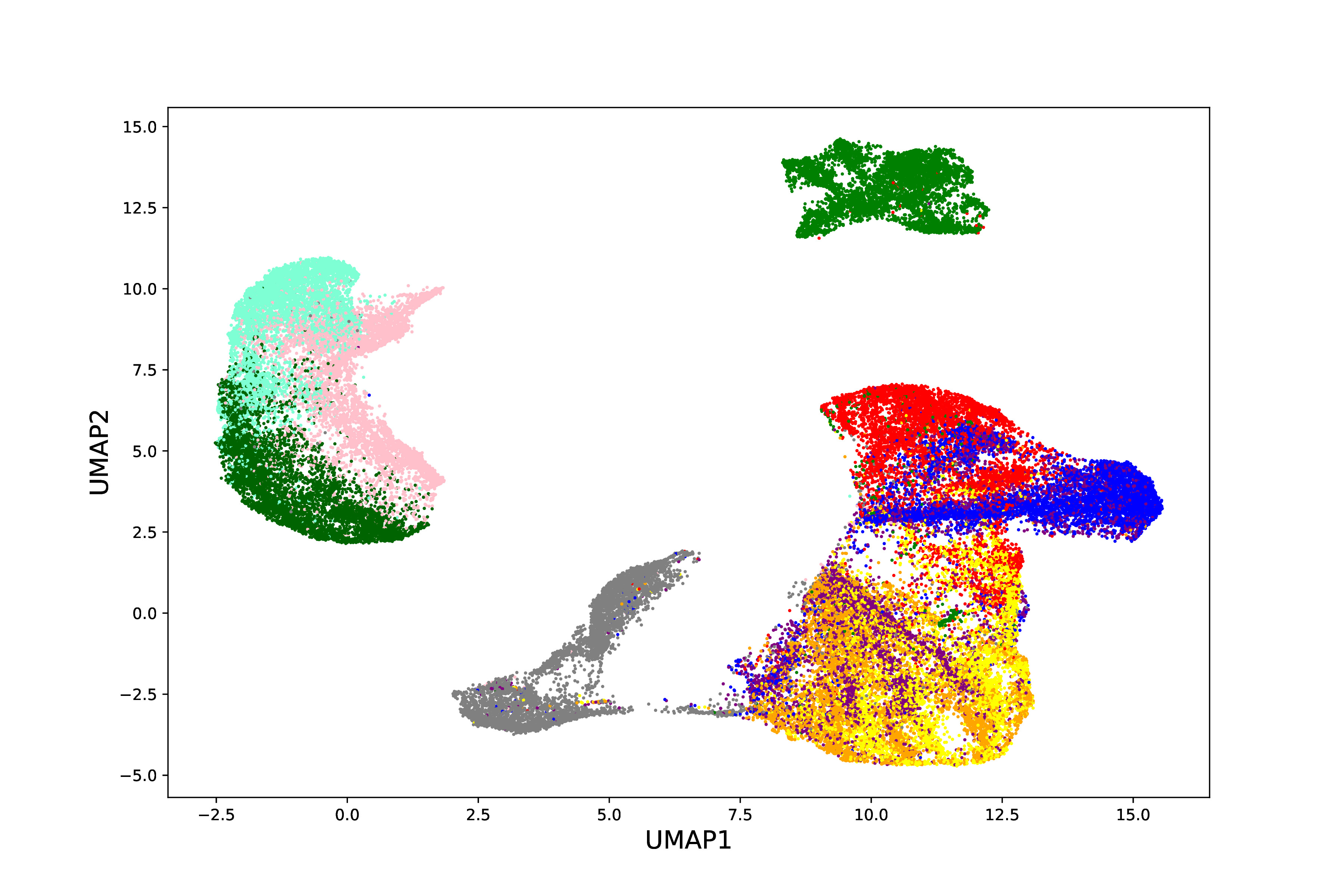}
		\label{fig:FMnist_UMAP_60k}
	}
	\subfloat[PCA prior to UMAP]{
		\centering
		\includegraphics[width=0.5\columnwidth,height=4cm]{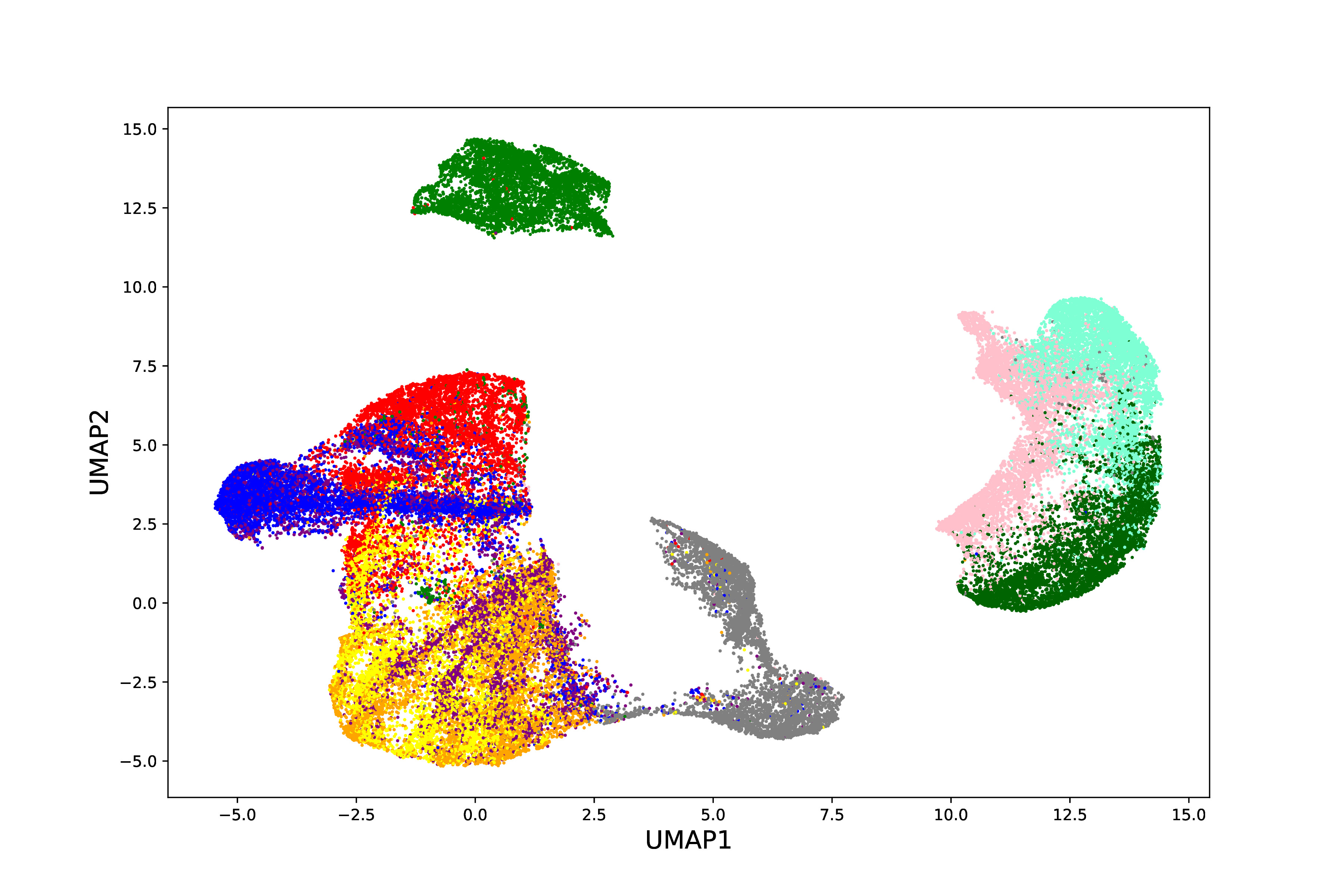}
		\label{fig:FMnist_pca_UMAP_60k}
	}\\
	\subfloat[Fit-SNE embedding of original data]{
		\centering
		\includegraphics[width=0.5\columnwidth,height=4cm]{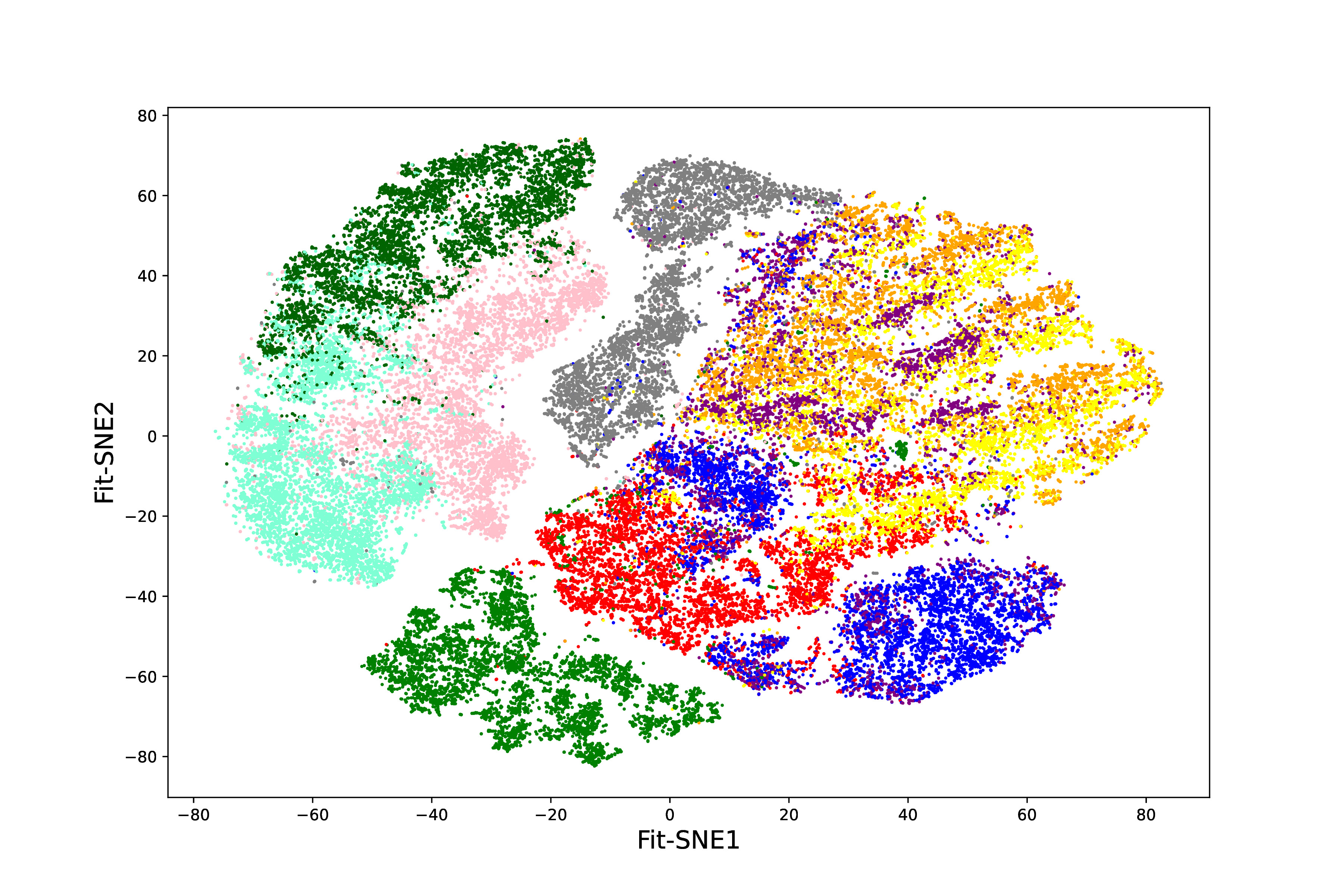}
		\label{fig:FMnist_FitSNE_60k}
	}
	\subfloat[PCA prior to Fit-SNE]{
		\centering
		\includegraphics[width=0.5\columnwidth,height=4cm]{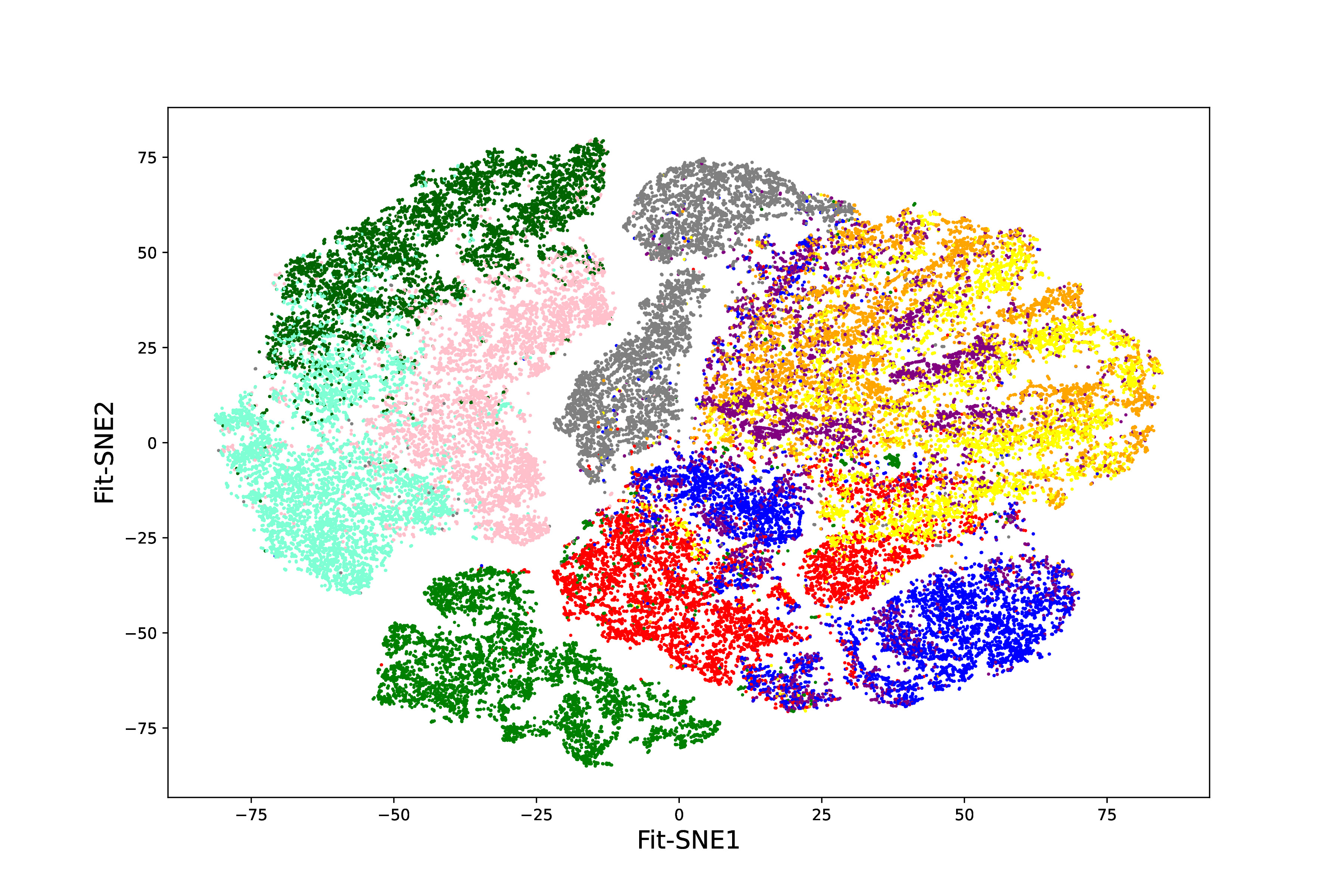}
		\label{fig:FMnist_pca_FitSNE_60k}
	}
	\caption{(a)NeuroDAVIS (c) t-SNE (e) UMAP (g) Fit-SNE embeddings obtained on Fashion MNIST dataset, (b) NeuroDAVIS (d) t-SNE (f) UMAP (h) Fit-SNE embeddings obtained on the first 50 PCA components of the original Fashion MNIST dataset.}
	\label{fig:FMnist}
\end{figure}

\label{sec:results_image}
Subsequently, we have evaluated NeuroDAVIS on an image dataset Coil20. Coil20 dataset \cite{coil20} contains images of $20$ different objects with backgrounds discarded. Figures \ref{fig:Coil20_NeuroDAVIS}, \ref{fig:Coil20_tSNE}, \ref{fig:Coil20_UMAP} and \ref{fig:Coil20_fitsne} show the NeuroDAVIS, t-SNE, UMAP and Fit-SNE embeddings obtained on this dataset respectively. We have observed that similar to the numeric and textual datasets, NeuroDAVIS has been able to preserve the distances between objects in high dimension, better than t-SNE and UMAP. The median correlation coefficient value reported by NeuroDAVIS has been $0.61$, while that for the t-SNE, UMAP and Fit-SNE embeddings have been $0.53$, $0.15$ and $0.56$ respectively. Figure \ref{fig:HD_Spearman} shows the distribution of correlation coefficient values obtained from multiple executions of this experiment. The corresponding p-values obtained using Mann–Whitney U test are $0.0007$ (NeuroDAVIS-t-SNE), $0.0001$ (NeuroDAVIS-UMAP) and $0.002$(NeuroDAVIS-Fit-SNE).

The classification accuracy and F1-scores achieved by k-nn and RF classifiers on the NeuroDAVIS embedding for Coil20 image dataset have been better than that achieved by UMAP. However, t-SNE and Fit-SNE have produced better classification results on this dataset, as depicted in Figure S3 (in Supplementary Material).

In order to demonstrate the effectiveness of NeuroDAVIS on large datasets, we have additionally evaluated NeuroDAVIS on another image dataset, called Fashion MNIST \cite{fmnist}, containing $60$k training images of clothings of $28 \times 28$ pixels each. Figures \ref{fig:FMnist_NeuroDAVIS_60k}, \ref{fig:FMnist_tSNE_60k}, \ref{fig:FMnist_UMAP_60k} and \ref{fig:FMnist_FitSNE_60k} show NeuroDAVIS, t-SNE, UMAP and Fit-SNE embeddings of Fashion MNIST datasets respectively. To obtain high-quality embeddings from high dimensional datasets, researchers often use PCA as a preprocessing step \cite{szubert2019structure}. For this reason, we have performed a futher investigation applying NeuroDAVIS, t-SNE, UMAP and Fit-SNE on the first $50$ principal components obtained from Fashion MNIST data, and comparing the results with that produced by the sole usage of NeuroDAVIS, t-SNE, UMAP and Fit-SNE. Figures  \ref{fig:FMnist_pca_NeuroDAVIS_60k}, \ref{fig:FMnist_pca_tSNE_60k}, \ref{fig:FMnist_pca_UMAP_60k} and \ref{fig:FMnist_pca_FitSNE_60k} show the effect of using PCA as a preprocessing step before applying NeuroDAVIS, t-SNE, UMAP or Fit-SNE. 

\begin{figure}
	\centering
	\includegraphics[width=\columnwidth,height=7cm]{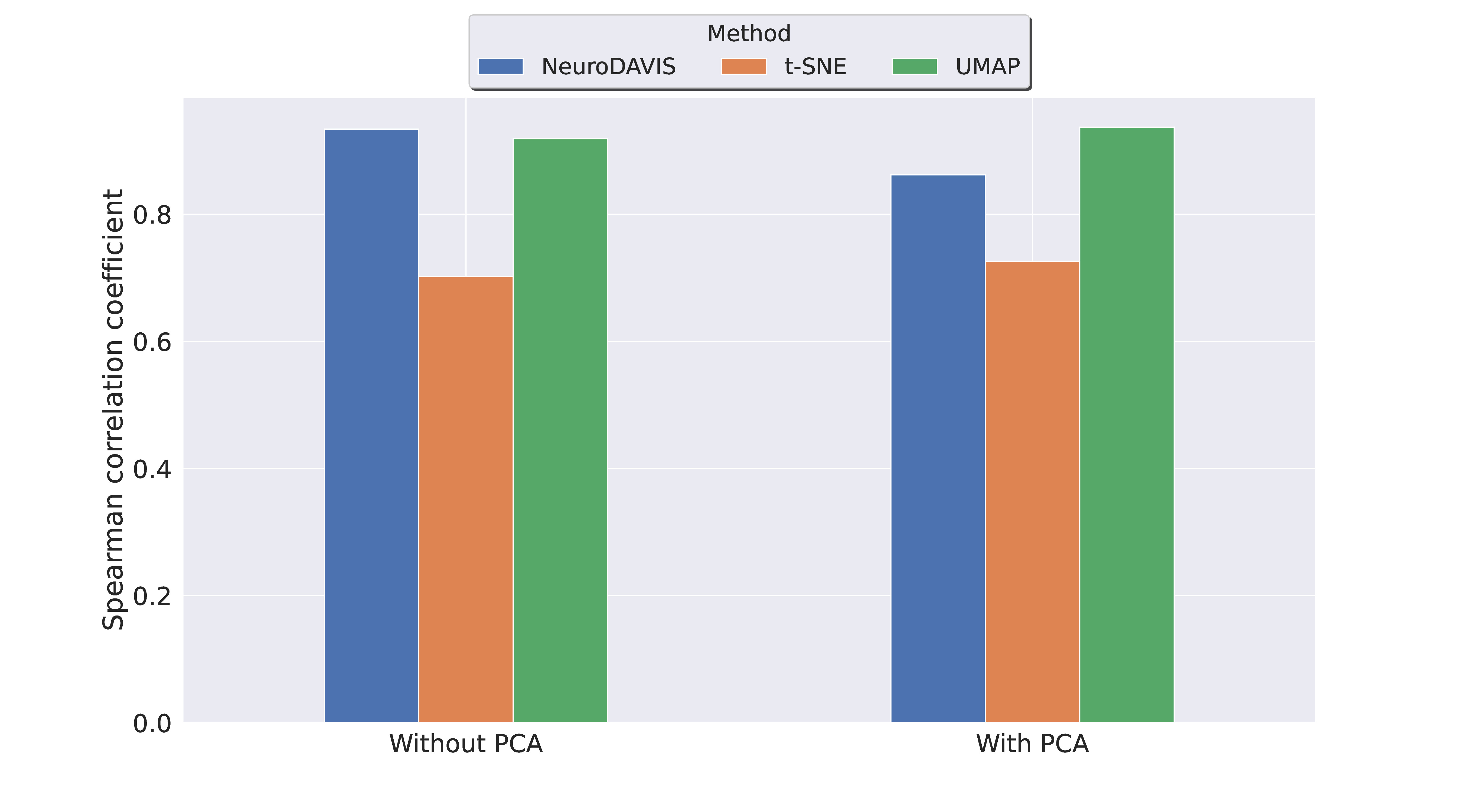}
	\caption{Spearman rank correlation coefficients between pairwise distances of the cluster centroids in high-dimension and the pairwise distances of the cluster centroids in the low-dimensional embedding of Fashion MNIST dataset produced by NeuroDAVIS, t-SNE, UMAP and Fit-SNE with and without PCA-based preprocessing.}
	\label{fig:FMnist_Spearman}
\end{figure}

It is often difficult to assess the quality of clusters visually. Hence, in order to quantify the quality of clusters, we have obtained the Spearman rank correlation coefficients between the pairwise distances of the cluster centroids in high-dimension and the pairwise distances of the cluster centroids in the low-dimensional embedding. A high correlation coefficient signifies better preservation of inter-cluster distances. Here, we have observed that NeuroDAVIS has produced a higher correlation coefficient of $0.93$, compared to that produced by t-SNE (correlation coefficient = $0.70$), UMAP (correlation coefficient  = $0.91$) and Fit-SNE (correlation coefficient  = $0.89$), as shown in Figure \ref{fig:FMnist_Spearman}. Interestingly, applying PCA for preprocessing has produced lower correlation coefficient for NeuroDAVIS (correlation coefficient = $0.86$). It has improved the result for t-SNE (correlation coefficient = $0.72$) and UMAP (correlation coefficient = $0.93$), while FitSNE has not been affected by the usage of PCA as a preprocessing step (correlation coefficient  = $0.89$), which are shown in Figure \ref{fig:FMnist_Spearman}. These results have led us to infer that t-SNE or UMAP perform better when PCA is used as a preprocessing tool. NeuroDAVIS, on the other hand, is able to produce high quality embeddings independently. PCA may have hampered the embedding quality due to loss of information during PCA-based preprocessing prior to applying NeuroDAVIS. 

\begin{figure}[h]
	\centering
	\subfloat[NeuroDAVIS]{
		\includegraphics[width=.3\columnwidth,height=5cm]{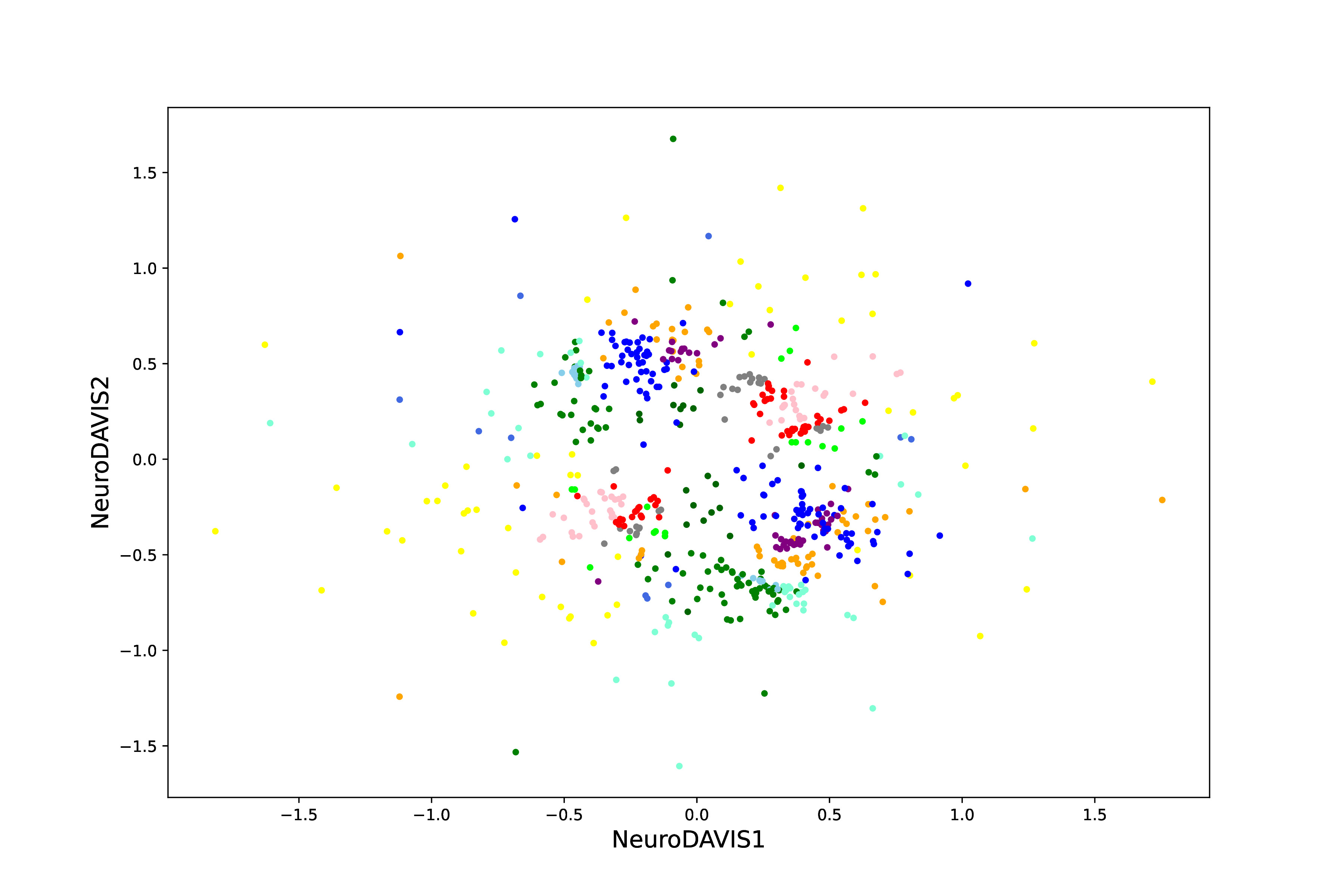}}\quad
	\subfloat[t-SNE]{
		\includegraphics[width=.3\columnwidth,height=5cm]{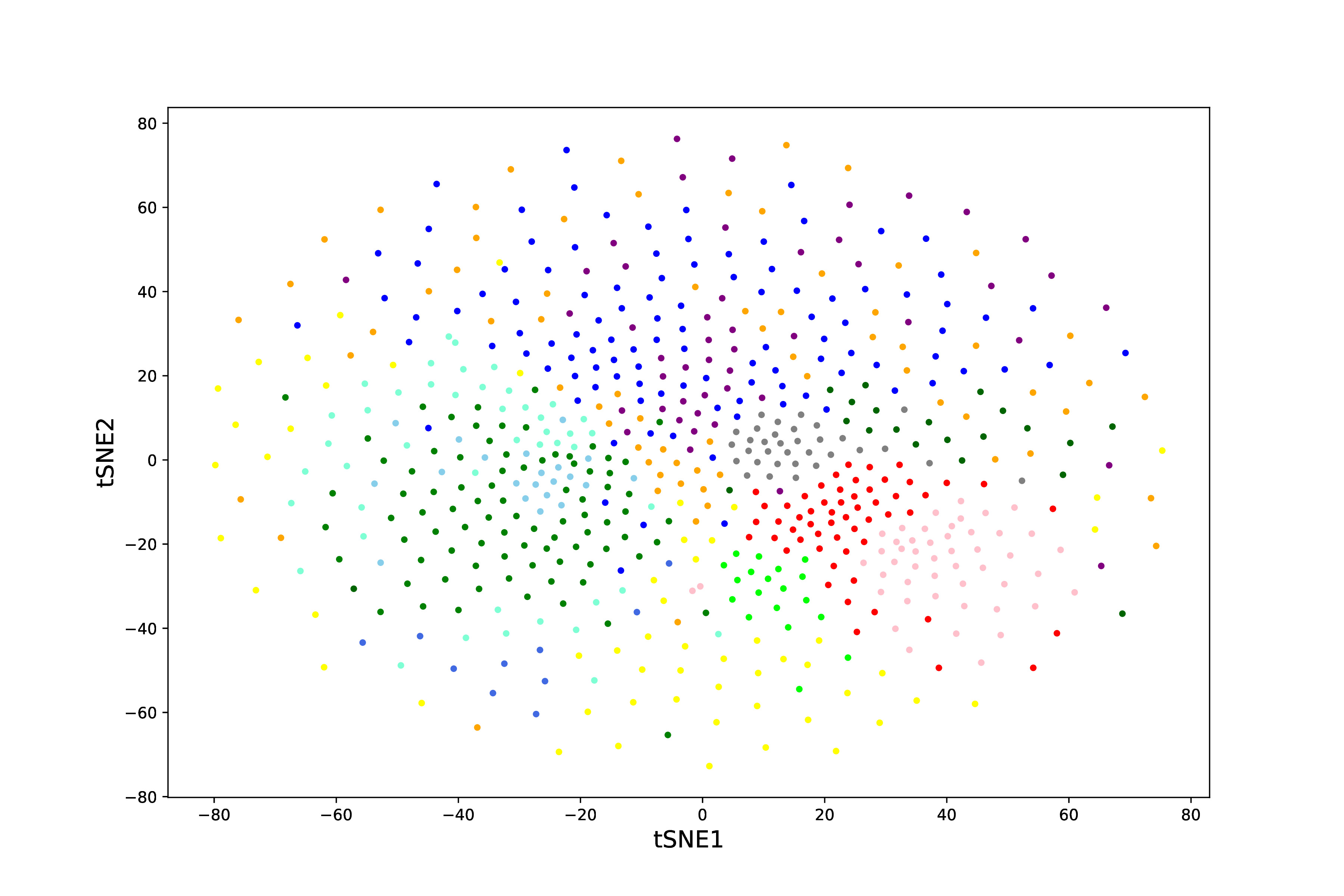}}\quad
	\subfloat[UMAP]{
		\includegraphics[width=.3\columnwidth,height=5cm]{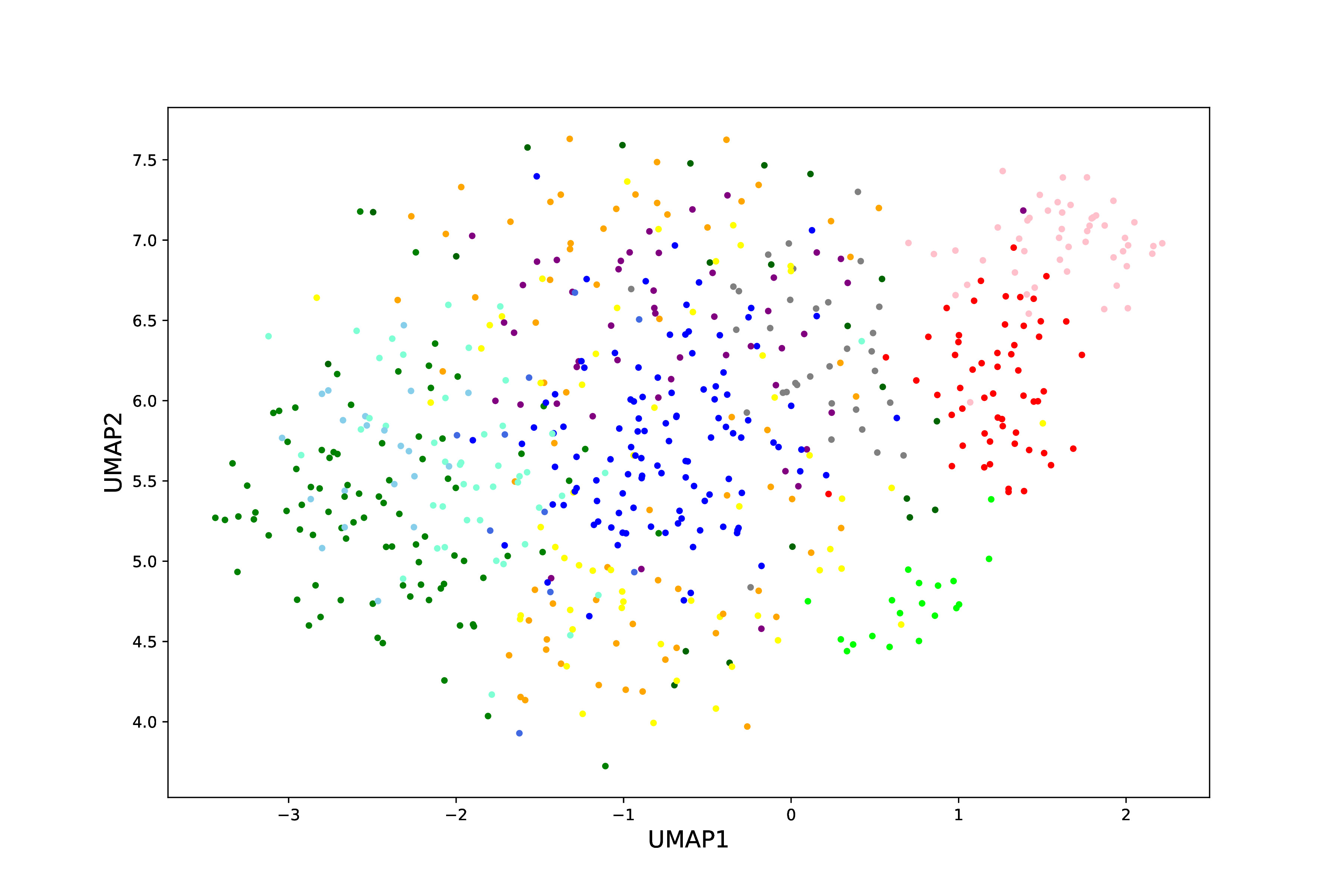}}
	
	\subfloat[Fit-SNE]{
		\includegraphics[width=.3\columnwidth,height=5cm]{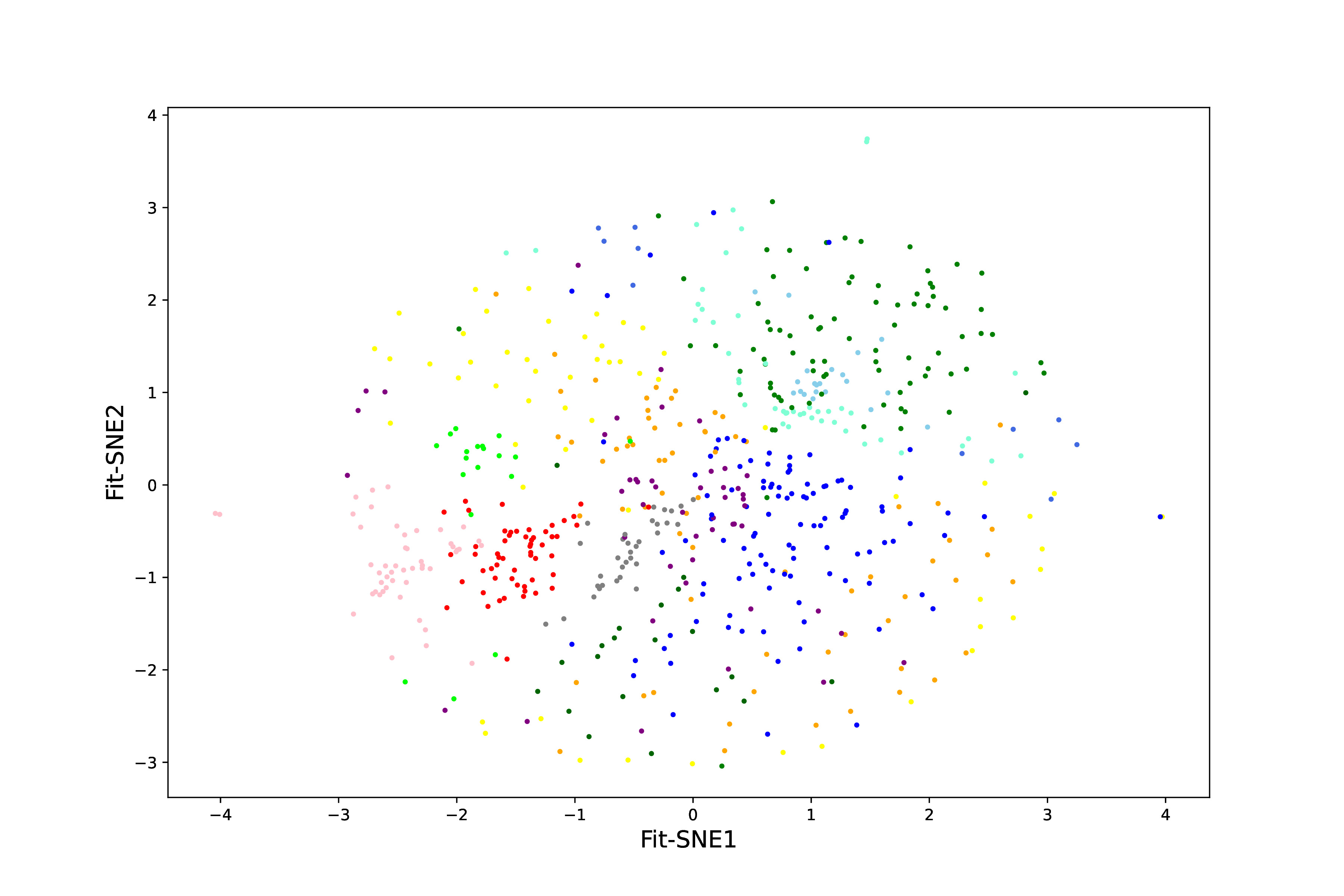}}\quad
	\subfloat[PHATE]{
		\includegraphics[width=.3\columnwidth,height=5cm]{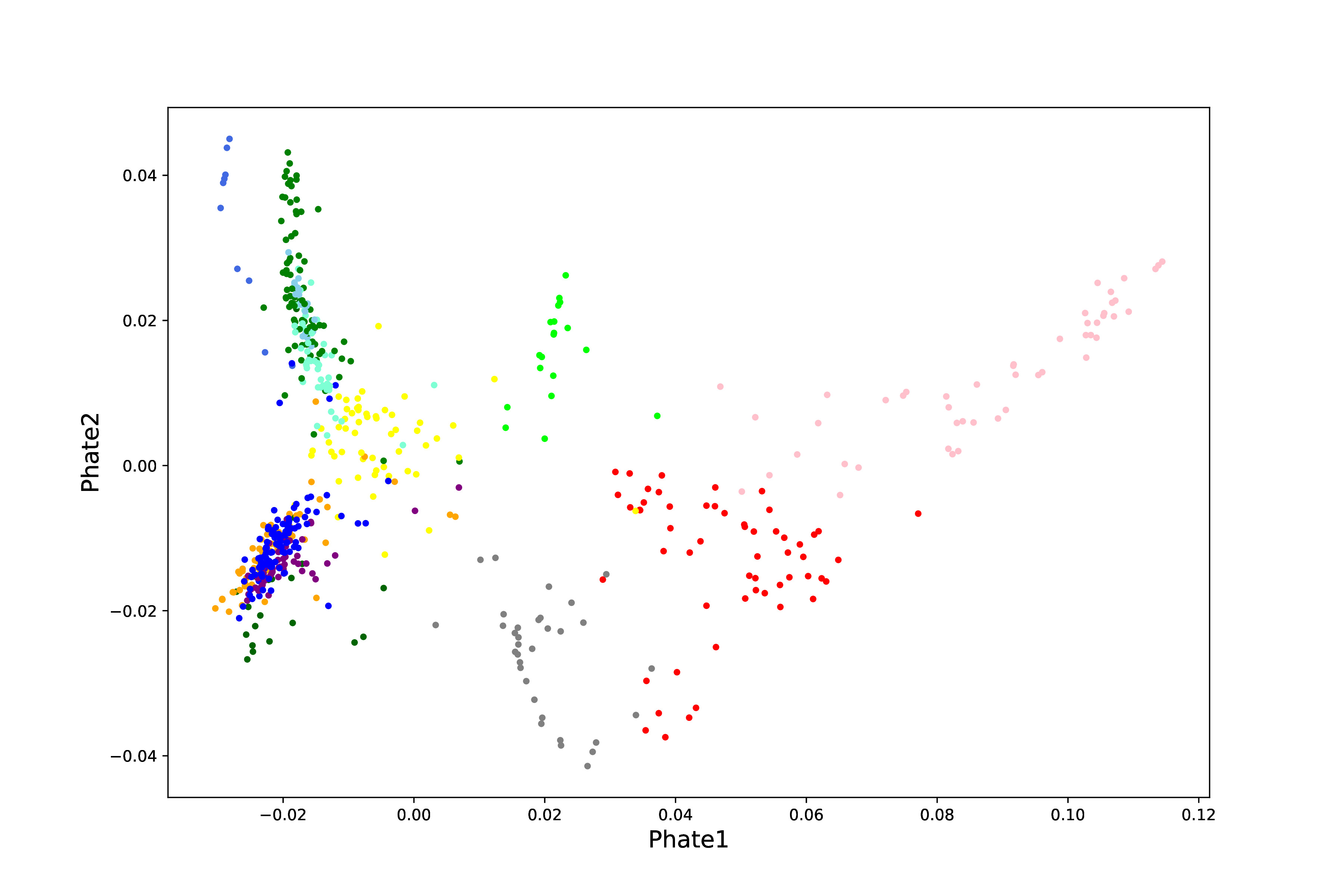}}\quad
	\subfloat[IVIS]{
		\includegraphics[width=.3\columnwidth,height=5cm]{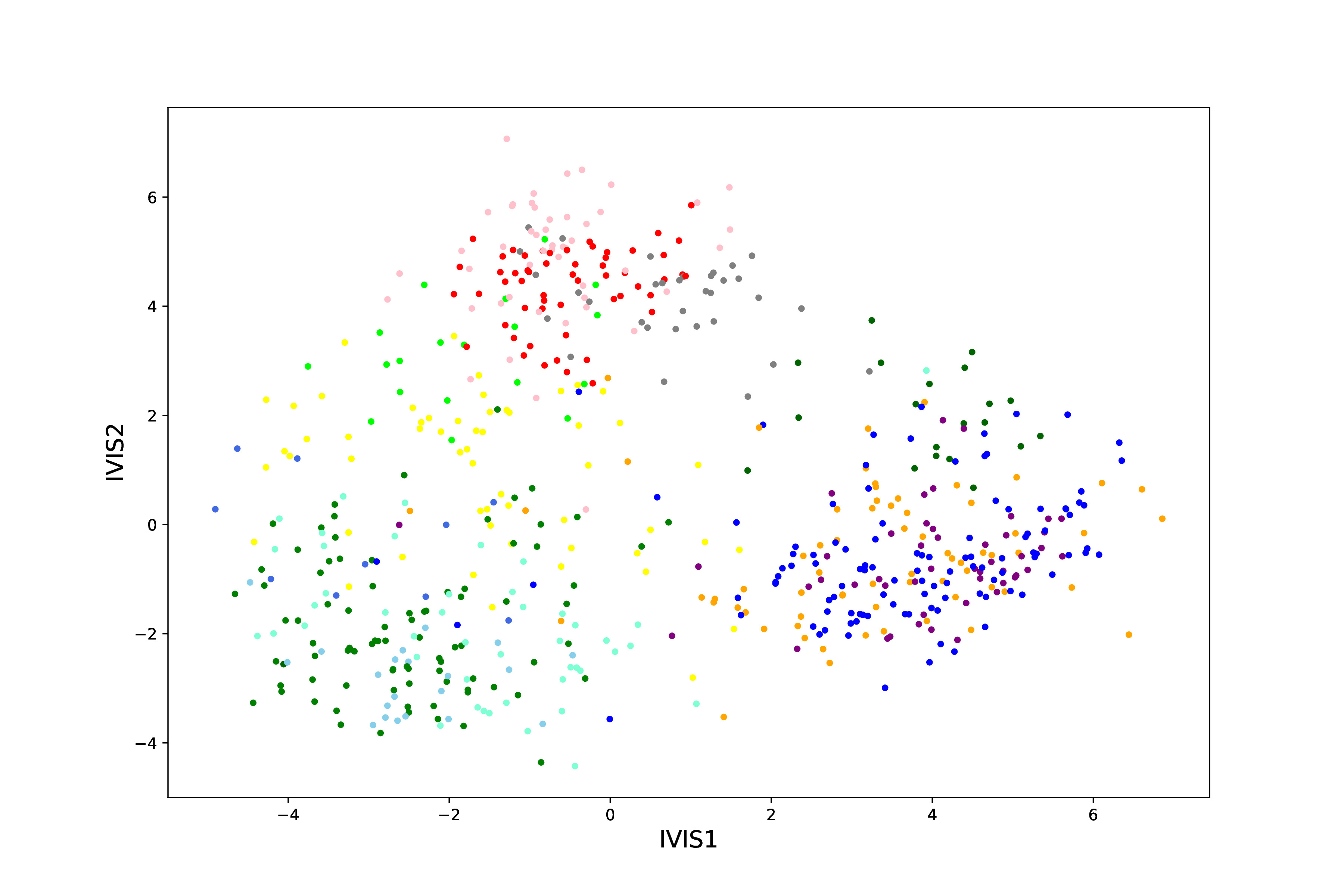}}
	
	\caption{$2$-dimensional embeddings of Usoskin data generated by NeuroDAVIS, t-SNE, UMAP, Fit-SNE, PHATE and IVIS respectively.}
	\label{fig:Usoskin}
\end{figure}

\subsubsection{Biological datasets}
\label{sec:results_biological}
Finally, we have explored the effectiveness of NeuroDAVIS on some biological datasets. We have used two scRNA-seq datasets for this purpose. Usoskin dataset \cite{usoskin2015unbiased} contains gene expression values for $622$ cells across $25334$ genes, while Jurkat dataset \cite{zheng2017massively} contains gene expression values for $3388$ cells across $32738$ genes. Both these datasets have been preprocessed using Scanpy \cite{wolf2018scanpy} following standard procedures as recommended in the tutorial \url{https://scanpy-tutorials.readthedocs.io/en/latest/pbmc3k.html}. Apart from t-SNE, UMAP and Fit-SNE, we have considered PHATE and IVIS for performance comparison on these datasets which are among the current state-of-the-art algorithms for scRNA-seq datasets besides t-SNE and UMAP. It may be mentioned here that in both these datasets, the number of samples is much lower than the number of dimensions, which escalates additional challenge in learning.

\begin{figure}
	\centering
	\subfloat[NeuroDAVIS]{
		\includegraphics[width=.3\columnwidth,height=5cm]{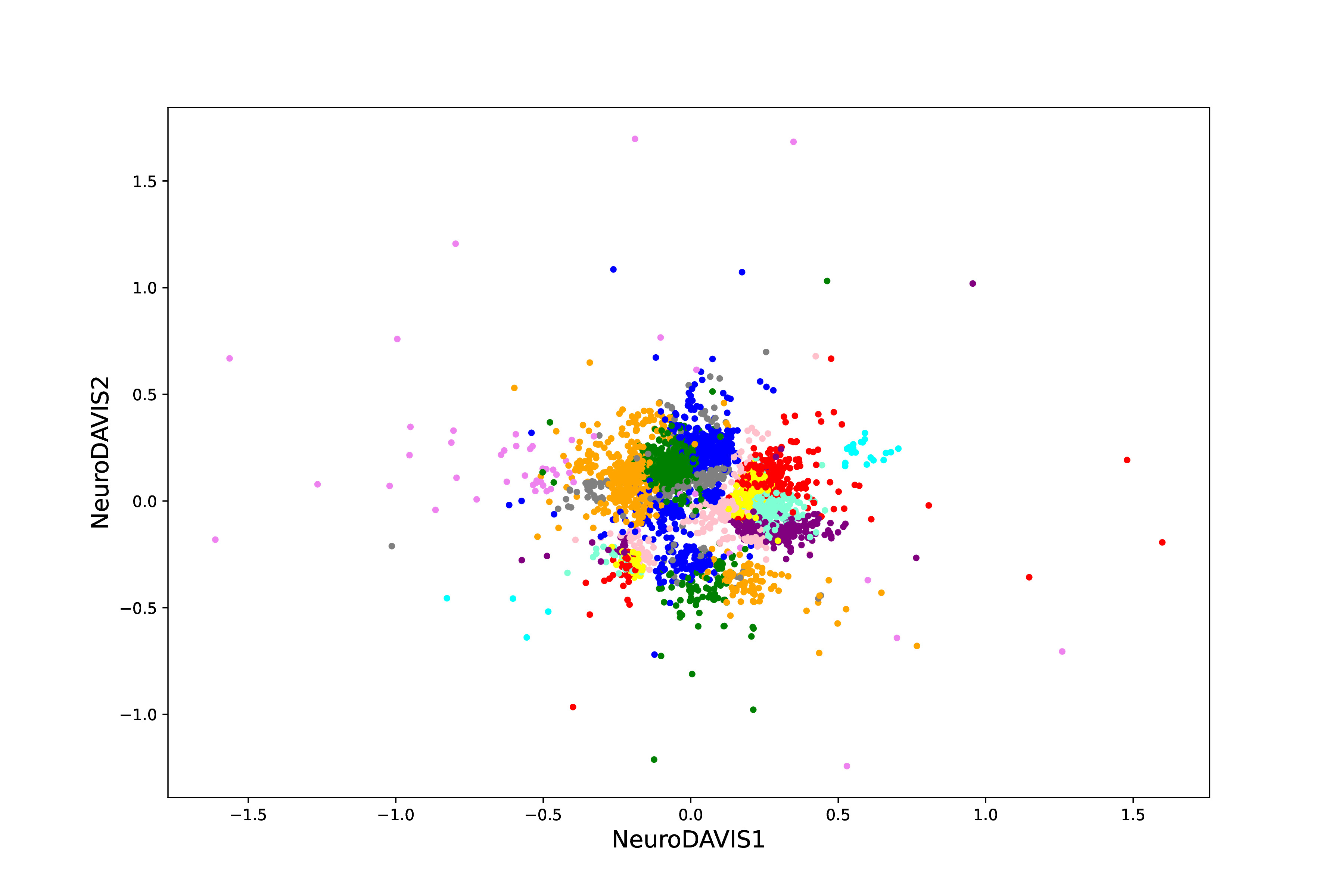}}\quad
	\subfloat[t-SNE]{
		\includegraphics[width=.3\columnwidth,height=5cm]{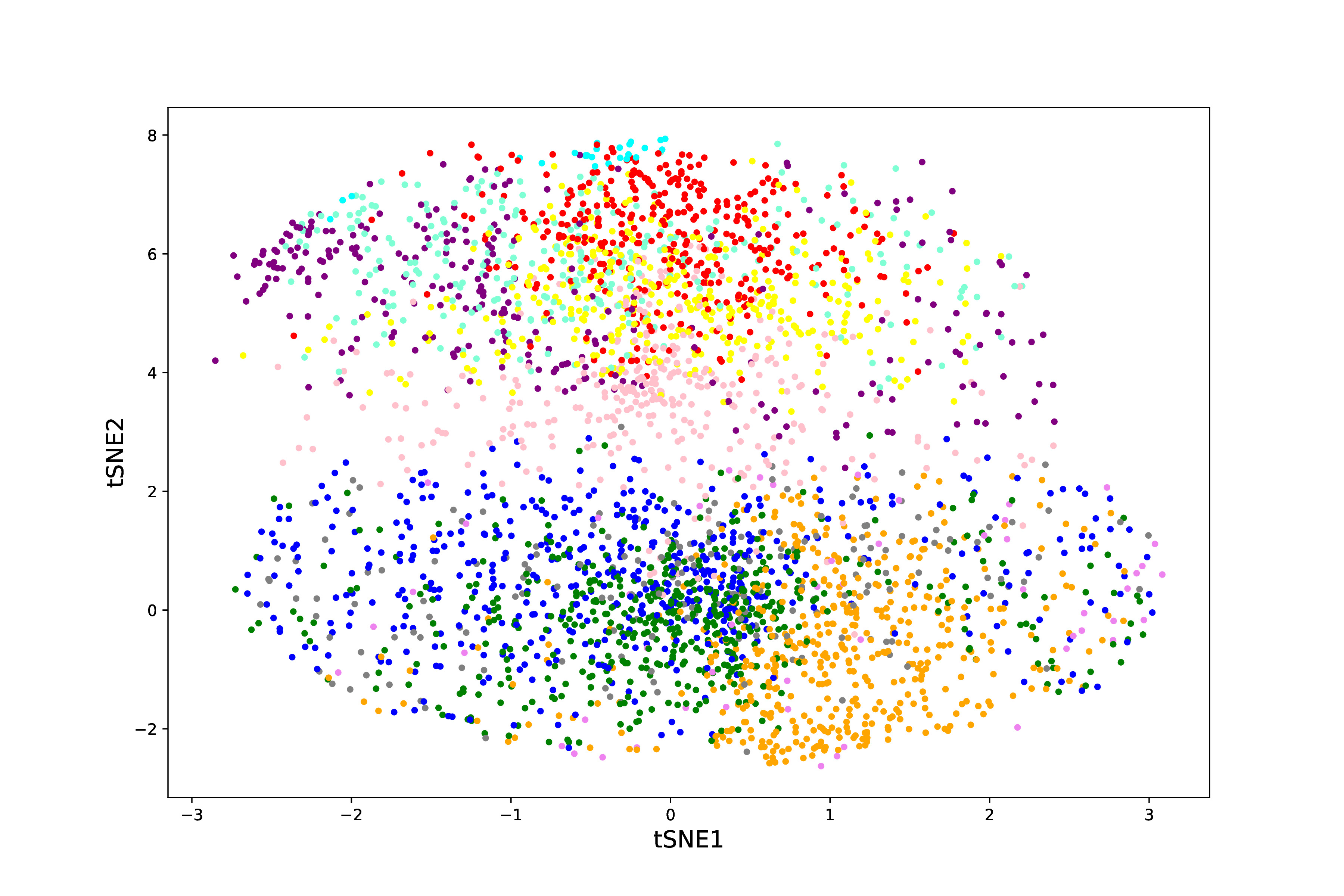}}\quad
	\subfloat[UMAP]{
		\includegraphics[width=.3\columnwidth,height=5cm]{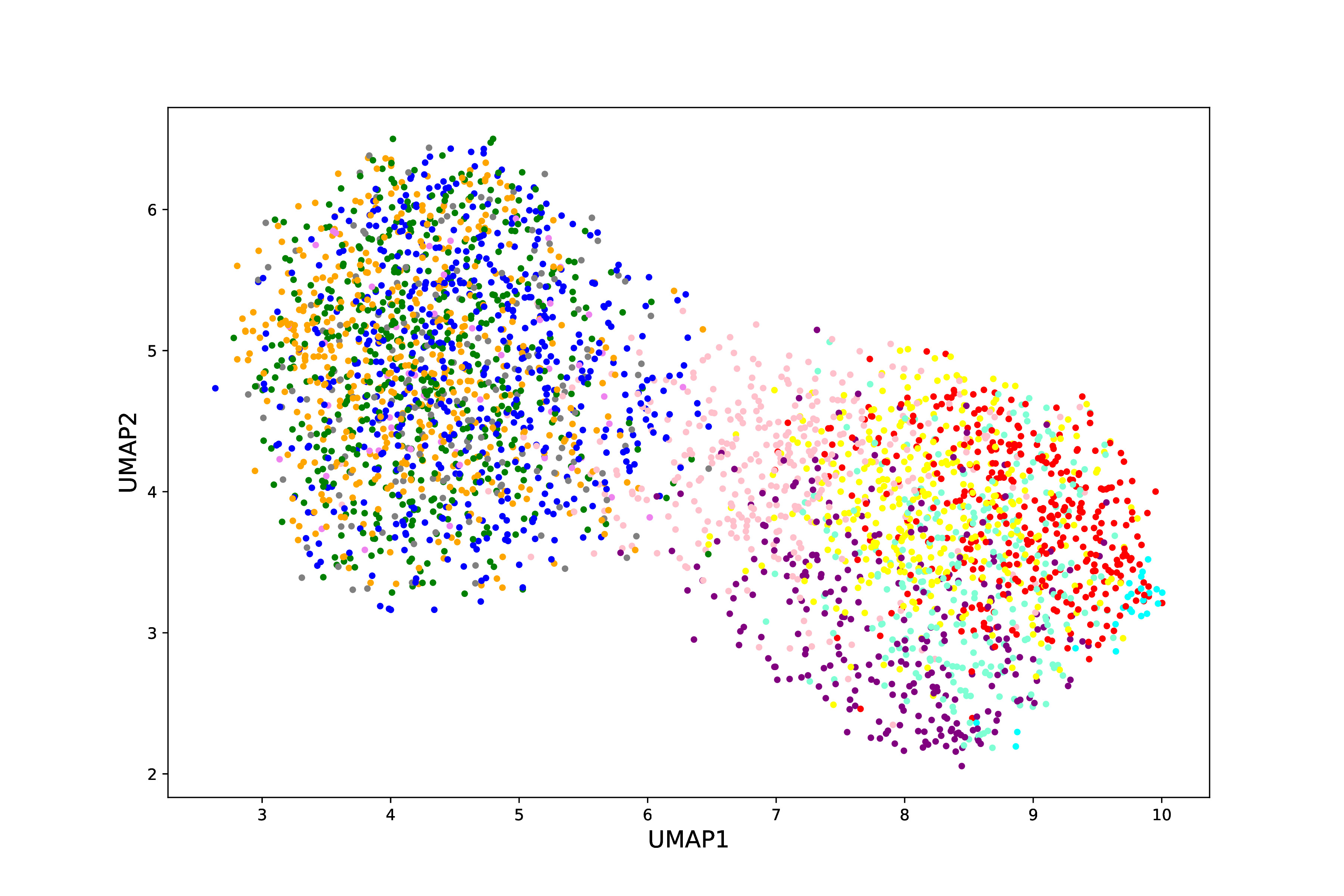}}
	
	\subfloat[Fit-SNE]{
		\includegraphics[width=.3\columnwidth,height=5cm]{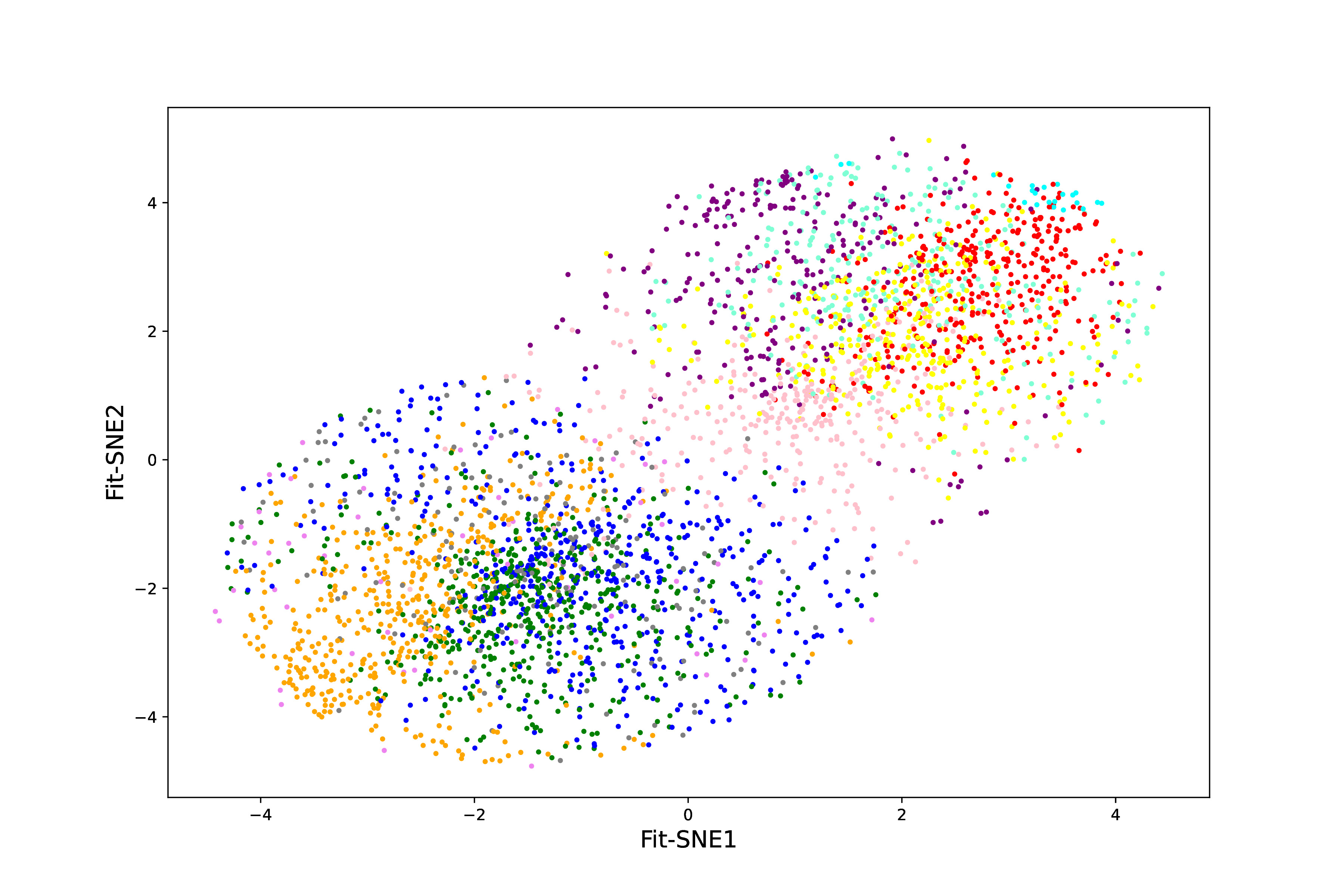}}
	\subfloat[PHATE]{
		\includegraphics[width=.3\columnwidth,height=5cm]{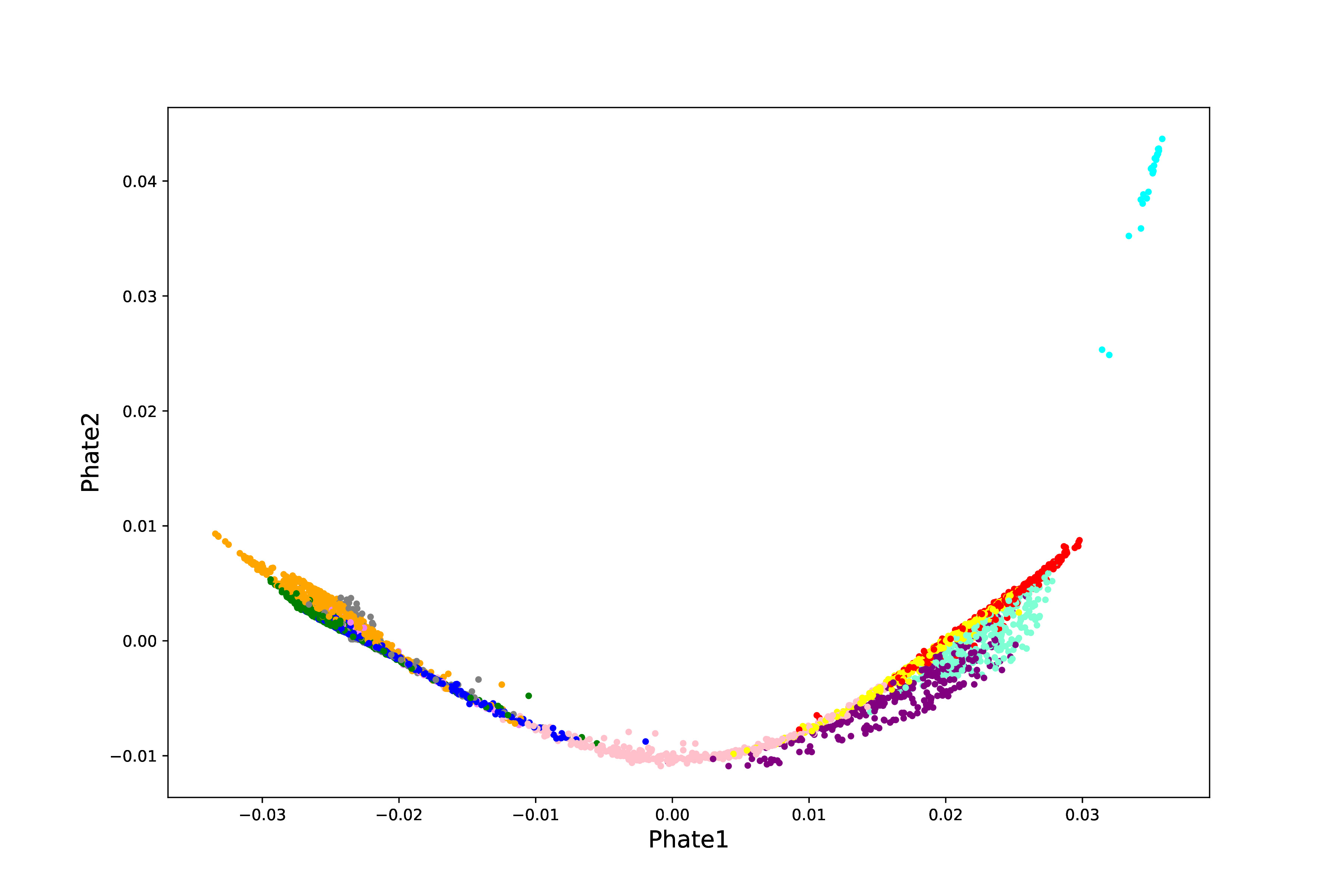}}\quad
	\subfloat[IVIS]{
		\includegraphics[width=.3\columnwidth,height=5cm]{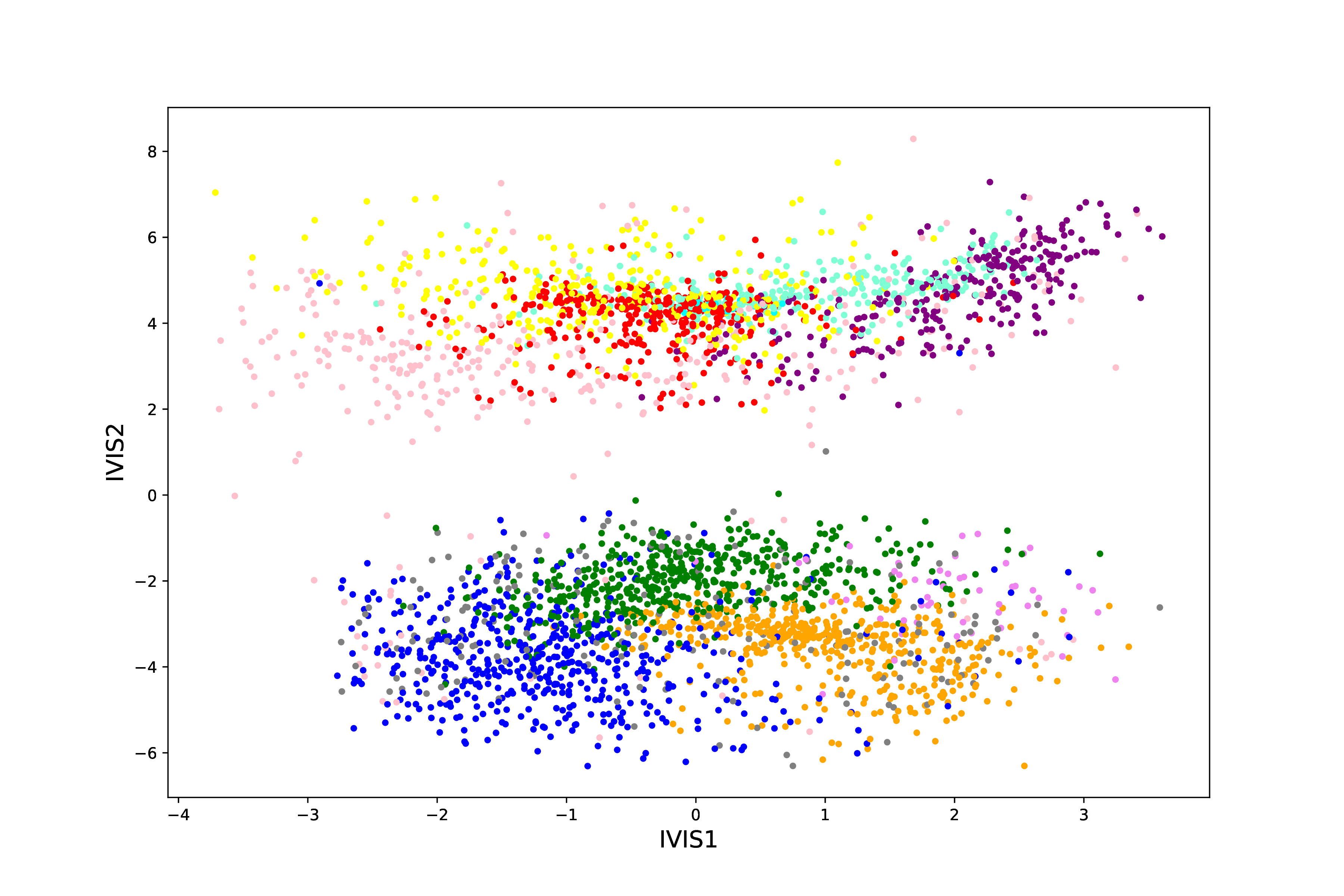}}
	
	\caption{$2$-dimensional embeddings of Jurkat data generated by NeuroDAVIS, t-SNE, UMAP, Fit-SNE, PHATE and IVIS respectively.}
	\label{fig:Jurkat}
\end{figure}

The two-dimensional embeddings obtained by NeuroDAVIS, t-SNE, UMAP, Fit-SNE, PHATE and IVIS for Usoskin and Jurkat datasets, as shown in Figures \ref{fig:Usoskin} and \ref{fig:Jurkat} respectively, have revealed that both t-SNE and UMAP have been unable to represent the clusters in the data clearly. For Usoskin dataset, NeuroDAVIS has been able to represent some of the clusters well, while for Jurkat dataset, clusters in the NeuroDAVIS projection has been much more dense and well-separated, as compared to the other methods. Overall, it can be said that NeuroDAVIS projections are better than t-SNE, UMAP and Fit-SNE projections, and are comparable to PHATE or IVIS projections only.

\begin{figure}
	\centering
	\includegraphics[width=\columnwidth,height=8cm]{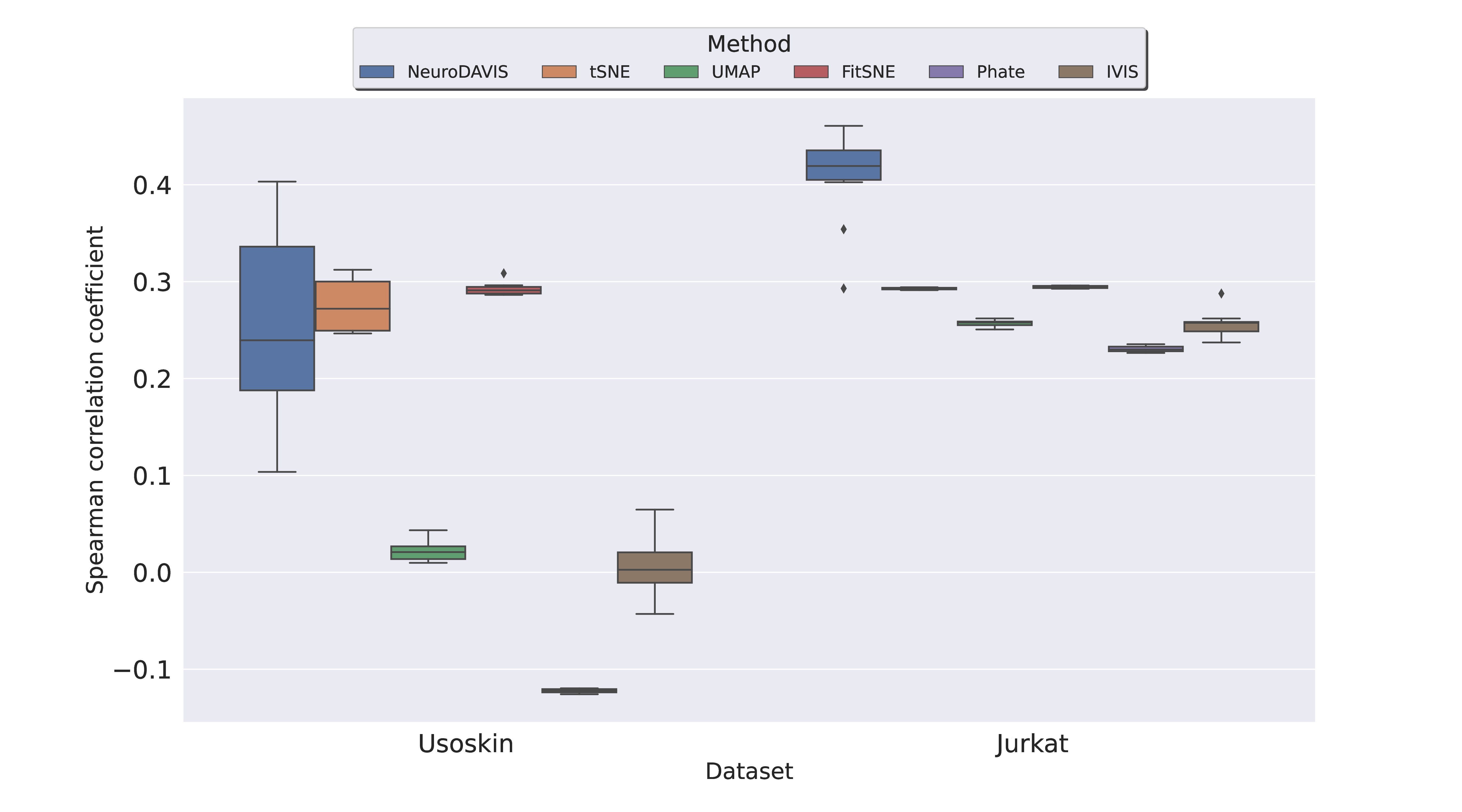}
	\caption{Spearman rank correlation between pairwise distances in original distribution and pairwise distances in NeuroDAVIS, t-SNE, UMAP, Fit-SNE, PHATE and IVIS-produced embeddings of Usoskin and Jurkat datasets. For Usoskin dataset, median correlation coefficient values obtained are $0.24$ (NeuroDAVIS), $0.27$ (t-SNE), $0.02$ (UMAP), $0.29$ (Fit-SNE), $-0.12$ (PHATE) and $0.002$ (IVIS). For Jurkat dataset, median correlation coefficient values obtained are $0.42$ (NeuroDAVIS), $0.29$ (t-SNE), $0.26$ (UMAP), $0.29$ (Fit-SNE), $0.23$ (PHATE) and $0.25$ (IVIS). The corresponding p-values obtained using Mann-Whitney U test for Usoskin dataset are $0.677$ (NeuroDAVIS-t-SNE), $0.472$ (NeuroDAVIS-Fit-SNE) $0.0001$ (NeuroDAVIS-UMAP, NeuroDAVIS-PHATE and NeuroDAVIS-IVIS), and $0.0001$ (NeuroDAVIS-t-SNE, NeuroDAVIS-UMAP, NeuroDAVIS-Fit-SNE, NeuroDAVIS-PHATE and NeuroDAVIS-IVIS) for Jurkat dataset.}
	\label{fig:SingleCell_Spearman}
\end{figure}

The correlation coefficient values measured between the pairwise distances in the original data and NeuroDAVIS-, t-SNE-, UMAP-, Fit-SNE-, PHATE- and IVIS-generated embeddings have been reported in Figure \ref{fig:SingleCell_Spearman}. We have once again observed that NeuroDAVIS has shown better correlation (median correlation coefficient = $0.41$) than all other methods for Jurkat dataset, while for Usoskin dataset, NeuroDAVIS has reported a median correlation coefficient of $0.23$, better than all other methods except t-SNE (median correlation coefficient = $0.27$) and Fit-SNE (median correlation coefficient = $0.29$). The correponding p-values obtained using Mann–Whitney U test are $0.0001$ (NeuroDAVIS-t-SNE, NeuroDAVIS-UMAP, NeuroDAVIS-Fit-SNE, NeuroDAVIS-PHATE and NeuroDAVIS-IVIS) for Jurkat dataset, while for Usoskin dataset, p-values obtained using Mann–Whitney U test are $0.677$ (NeuroDAVIS-t-SNE), $0.472$ (NeuroDAVIS-Fit-SNE) and $0.0001$ (NeuroDAVIS-UMAP, NeuroDAVIS-PHATE and NeuroDAVIS-IVIS).

As demonstrated in Figure S4 (in Supplementary Material), we have further observed that the clustering performance of k-means and agglomerative hierarchical clustering methods on NeuroDAVIS embedding of Usoskin data has been better than t-SNE and IVIS embeddings, but poorer than PHATE and Fit-SNE embeddings, while being closely comparable to that of UMAP embedding, in terms of ARI and FMI scores. For Jurkat dataset, NeuroDAVIS embedding has, however, resulted in the best clustering performance in terms of both ARI and FMI scores, which can only be challenged by IVIS embedding.

\section{Discussion and Conclusion}
\label{sec:conc}
In this work, we have developed a novel unsupervised deep learning model, called NeuroDAVIS, for data visualization. NeuroDAVIS can produce a low-dimensional embedding from the data, extracting relevant features useful for subsequent analysis pipelines. The effectiveness of NeuroDAVIS has been demonstrated on a wide variety of datasets including $2$D synthetic data, and high dimensional numeric, textual, image and biological data. It has shown excellent visualization capability supported by outstanding downstream analysis (clustering/classification), thus making it competitive against state-of-the-art methods, like t-SNE, UMAP and Fit-SNE. Results obtained are also statistically significant in favour of NeuroDAVIS. The strength of NeuroDAVIS lies in the fact that it is able to preserve cluster shape, size and structure within the data well. As demonstrated mathematically, the low-dimensional embedding produced by NeuroDAVIS has also been observed to preserve local relationships among observations in high dimension. Furthermore, it is capable of extracting the inherent non-linearity within the data. In addition, NeuroDAVIS does not presume data distributions, which makes the method more practical. The model can be termed as a general purpose dimension reduction and visualization system having no restrictions on embedding dimension. It is capable of producing interpretable visualizations independently without any preprocessing.

NeuroDAVIS, being a dimension reduction and visualization method, does not come without its weaknesses. It takes as input an identity matrix of order equal to number of observations in the input data. Thus, it consumes more space in memory than other existing methods. Moreover, hyper-parameters in NeuroDAVIS are sensitive to tiny modifications, and network tuning requires substantial amount of time and effort by an amateur.

Nevertheless, NeuroDAVIS serves as a competing method for visualization of high-dimensional datasets, producing a robust latent embedding better than some of the existing benchmarked methods, like t-SNE, UMAP and Fit-SNE, while also resolving the local-global shape preservation challenge. Thus, in this work, we have introduced a single solution to the long-standing problem of visualization of datasets belonging to different categories/domains/modalities. Adding a feature selection module to the existing architecture could be a future extension to NeuroDAVIS. It might also be extended to visualize high-dimensional multi-modal datasets, by producing low-dimensional embeddings that can capture significant features from the data.

\section*{Data and Code availability}
NeuroDAVIS has been implemented in Python 3. The codes to reproduce the results are available at \url{https://github.com/shallowlearner93/NeuroDAVIS}. The preprocessed datasets used in this work can be downloaded from \url{https://doi.org/10.5281/zenodo.7315674}.

\section*{Supplementary information}

The supplementary figures have been incorporated in the Supplementary file.

\section*{Authors' contributions}
Conceptualization of Methodology: CM, DBS, RKD. Data Curation, Data analysis, Formal analysis, Visualization, Investigation, Implementation, Validation, Original draft preparation: CM, DBS. Validation, Reviewing, Editing, Overall Supervision: DBS, RKD.

\section*{Acknowledgments}
	
This work is supported by DST-NSF grant provided to RKD through IDEAS-TIH, ISI Kolkata.

\section*{Declarations}
DBS works as an Associate Data Scientist at Tatras Data Services Pvt. Ltd. He has received no funds for this work.

\bibliographystyle{unsrt}	
\bibliography{NeuroDAVIS.bib}
	
	
\end{document}